\documentclass[epj,final]{svjour}
\usepackage{graphics}
\usepackage{epsfig}
\usepackage[latin1]{inputenc}
\usepackage{amsmath}
\usepackage{amssymb}

\newcommand{\gap}{\stackrel{>}{\sim}}
\newcommand{\lap}{\stackrel{<}{\sim}}
\newcommand{\dg}{\ensuremath{^{\circ}}}
\def\xbj{\ensuremath{x_{Bj}}}
\newcommand{\gev}{\,\mbox{GeV}}
\newcommand{\qsq}{\ensuremath{Q^2} }
\newcommand{\gevsq}{\ensuremath{\mathrm{GeV}^2} }
\newcommand{\ptpsi}{\ensuremath{p_{\prp,\psi}}}

\def\prp{\perp}

\def\sx{small-$x$}
\def\kt{\ensuremath{k_\prp}}
\def\kti#1{\ensuremath{k_{\prp #1}}}
\def\pt{\ensuremath{p_\prp}}
\def\pti#1{\ensuremath{p_{\prp #1}}}
\def\qt{\ensuremath{q_\prp}}

\def\qbi#1{\ensuremath{\bar{q}_{ #1}}}
\def\qb{\ensuremath{\bar{q}}}

\def\lepto{{\sc Lepto}}

\def\cascade{{\sc Cascade}}
\def\rapgap{{\sc Rapgap}}
\def\lepto{{\sc Lepto}}

\def\ldcmc{{\sc LDCMC}}

\def\as{\ensuremath{\alpha_s}}
\def\sub#1{\ensuremath{_{\mbox{\scriptsize #1}}}}
\def\alb{\ensuremath{\bar{\alpha}\sub{s}}}

\def\rb{{\bf r}}  
\def\kb{{\bf k}}
\def\bb{{\bf b}}
\def\zb{{\bf z}}
\def\xb{{\bf x}}
\def\yb{{\bf y}}
\newcommand{\CCFM}{CCFMa,CCFMb,CCFMc,CCFMd}
\newcommand{\BFKL}{BFKLa,BFKLb,BFKLc}
\newcommand{\LDCMC}{LDCa,LDCb,LDCc,LDCd}
\newcommand{\alphasb}{\alb}

\newcommand{\LEPTO}{Ingelman_LEPTO65}

\newcommand{\RAPGAPMC}{RAPGAP206}
\newcommand{\DGLAP}{DGLAPa,DGLAPb,DGLAPc,DGLAPd}
\newcommand{\asb}{\alb}

\newcommand{\CASCADEMC}{jung_salam_2000,CASCADEMC}
\newcommand{\ktalgo}{kt_1,kt_2}

\def\CASCADE{{\sc Cascade}}

\newcommand{\Pmax}{\bar{q}}

\begin{document}
\title{ Small \boldmath$x$ Phenomenology - Summary and status}
\subtitle{{\sc The Small $x$ Collaboration } }
\author{Jeppe~Andersen\inst{1}
  \and Serguei~Baranov \inst{2}
  \and John~Collins\inst{3}
  \and Yuri~Dokshitzer\inst{4}
  \and Lidia~Goerlich\inst{5}
  \and Günter~Grindhammer\inst{6}
  \and Gösta~Gustafson\inst{7}
  \and Leif~J\"onsson\inst{8}
  \and Hannes~Jung\inst{8}
  \and Jan~Kwieci\'nski\inst{5}$^{\dag}$
  \and Eugene~Levin\inst{9}
  \and Artem~Lipatov\inst{10}
  \and Leif~Lönnblad\inst{7}
  \and Misha~Lublinsky\inst{9}
  \and Martin~Maul\inst{7}
  \and Izabela~Milcewicz\inst{5}
  \and Gabriela~Miu\inst{7}
  \and Grazyna Nowak\inst{5}
  \and Torbj\"orn~Sj\"ostrand\inst{7}
  \and Anna~Sta\'sto\inst{11}
  \and Nicusor~T\^{\i}mneanu\inst{12}
  \and Jacek~Turnau\inst{5}
  \and Nikolai~Zotov\inst{13}
}                     
%
%
\institute{%
       DAMTP and Cavendish Laboratory, University of Cambridge, UK,
  \and Lebedev~Institute~of~Physics, Moscow, Russia
  \and Penn State Univ., 104 Davey Lab., University Park PA 16802, USA 
  \and LPTHE, Universit\'es P. \& M. Curie (Paris VI) et Denis Diderot
       (Paris VII), Paris, France    
  \and H. Niewodniczanski Institute of Nuclear Physics, Cracow, Poland
  \and Max Planck Institut, Munich, FRG.,
  \and Department of Theoretical Physics, Lund University, Sweden
  \and Department of Physics, Lund University, Sweden
  \and School of Physics, Tel Aviv University, Tel Aviv, Israel,
  \and Moscow~State~University, Moscow, Russia
  \and DESY, Hamburg, FRG and
       H. Niewodniczanski Institute of Nuclear Physics, Cracow, Poland,
  \and University of Uppsala, Sweden
  \and Skobeltsyn~Institute~of~Nuclear~Physics, Moscow~State~University, Moscow, Russia
\\ $^{\dag}$ deceased
   }
%
\date{23. Dec. 2003}
%
\abstract{
A second workshop on small $x$ physics, within the Small $x$   
Collaboration, was held in Lund
in June 2002 with the aim of over-viewing recent theoretical progress in 
this area and summarizing the experimental status.
\\
This paper is dedicated to the memory of Jan Kwieci\'nski, who died 
unexpectedly on August 29, 2003.
%
} 
\maketitle
\section{Introduction}
\label{sec:intro}
This paper is 
a summary of the 2nd workshop on \sx\ parton dynamics held in Lund in
the beginning of June 2002.  
During two days we went through a number
of theoretical, phenomenological as well as experimental 
aspects of \sx\ physics in short
talks and long discussions. Whereas our first workshop in 2001
and the resulting summary paper~\cite{smallx_2001} was dedicated to
a general survey and discussion of \sx\  physics
in order  to identify the most pending
questions, we concentrate here on those aspects, where progress has been made,
as well as on a more detailed discussion and some aspects
 of the experimental situation.
For a general introduction to \sx\ physics and the \sx\ evolution equations, as
well as tools for calculation in terms of Monte Carlo programs, we refer the
reader to~\cite{smallx_2001}.
\par
With the successful completion of the two full hadron level Monte Carlo programs
\ldcmc  ~\cite{\LDCMC} and \CASCADE ~\cite{\CASCADEMC}, the necessary tools
were provided for
detailed studies both on a theoretical and phenomenological level as well as for
detailed comparison with experimental data and the usage in the 
experimental groups at HERA and elsewhere. 
 Since then they have been used 
in very different areas, like jet
production and heavy flavor physics.   
The \sx\ improved
unintegrated parton densities obtained from CCFM evolution implemented in the
Monte Carlo generators have been proven to be a very powerful tool in describing
experimental data as well as for estimating the effect of higher order
corrections. 
For example only by also 
applying the \CASCADE\ Monte Carlo in the extraction of
$F_2^c$ and in the calculation of bottom production at HERA, it was 
recognized that the extrapolation from the measured visible range to the total
cross section is dangerous and  introduces large model dependencies. 
Now, in the area of bottom production, the visible cross sections
are in reasonable agreement both with calculations applying
$\kt$-factorization with CCFM evolved unintegrated gluon density as
well as with NLO calculations in the collinear approach.
This shows the importance of applying
alternative approaches even when extracting experimental measurements.
\par
This paper is organized as follows: 
First we discuss
in more detail the definition of unintegrated gluon densities, as well 
as the question on gauge invariance of parton densities in general and
especially of the \kt-factorization approach. In the following section 
we discuss results and problems in theoretical applications of the 
unintegrated parton distribution functions,  
different parameterizations, the scale in $\as$, the role of the 
non-singular terms in the $g\to gg$ splitting function, saturation and the 
effects of energy momentum conservation in the BFKL equation. The 
section also contains a discussion of polarized unintegrated 
distributions, polarization effects and color octet contributions 
in $J/\psi$ meson production.
The second part of this
paper deals with experimental investigations of \sx\ effects and with the
question, whether and where deviations from the collinear approach can be
established, and whether a sign for a new evolution scheme like 
BFKL/ CCFM/ LDC has
already been seen. We end this paper with an outlook and a definition of the
next steps and goals.
\section{\boldmath$\kt$ - factorization formalism}
\label{sec:kt-factorisation}
The DGLAP\cite{\DGLAP} evolution treats 
successive parton emissions which are
strongly ordered in virtuality and  resums the resulting large logarithms of
ratios of subsequent virtualities.
Because of the strong ordering of
virtualities, the virtuality of the parton entering the hard
scattering matrix element can be neglected (treated collinear with the
incoming hadron) compared to the large scale $Q^2$.
\par
At very high energies, it is believed that the
theoretically correct description is given by the BFKL~\cite{\BFKL}
evolution. 
Here, each emitted gluon is assumed to take a large fraction, 
$1-z|_{z\rightarrow 0}$
of the energy of the propagating gluon, and large
logarithms of $1/z$ are summed up to all orders.
\par
The CCFM~\cite{\CCFM}
 evolution equation resums also large logarithms of $1/(1-z)$
in addition to the $1/z$ ones. Furthermore it introduces angular
ordering of emissions to correctly treat gluon coherence effects. In
the limit of asymptotic energies, it is almost equivalent to 
BFKL~\cite{Forshaw:1998uq,Webber:1998we,Salam:1999ft}, but 
also similar to the DGLAP evolution for large $x$ and high
$Q^2$.  The cross section is \kt-factorized into an off-shell
matrix element convoluted with an unintegrated parton density (uPDF), which
now also contains a dependence on the maximum angle $\Xi$ allowed in
emissions.
This  maximum allowed angle $\Xi$
is defined by the hard scattering quark box, producing the 
(heavy) quark pair and also defines the scale for which parton emissions
are factorized into the uPDF.
\par
The original CCFM splitting function is given by:
\begin{equation}
\tilde{P}_g(z_i,\qbi{i}^2,\kti{i}^2)= \frac{\alphasb(\qbi{i}^2(1-z_i)^2)}{1-z_i} + 
\frac{\alphasb(\kti{i}^2)}{z_i} \Delta_{ns}(z_i,\qbi{i}^2,\kti{i}^2)
\label{Pgg}
\end{equation}
with $\alphasb=\frac{3 \as}{\pi}$ and 
the non-Sudakov form factor $\Delta_{ns}$ given by:
\begin{eqnarray}
\ln\Delta_{ns}(z_i,\qbi{i}^2,\kti{i}^2)& = & -
                  \int_{z_i}^1 \frac{dz'}{z'} 
                        \int \frac{d q^2}{q^2}\alphasb  \cdot \nonumber \\
           & &   \Theta(\kti{i}-q)\Theta(q-z'\qbi{i})
                  \label{non_sudakov}         \\
	    & = &  -
                  \int_{z_i}^1 \frac{dz'}{z'} 
                        \int^{\kti{i}^2}_{(z'\qbi{i})^2} \frac{d q^2}{q^2} 
             \alphasb  \label{non_sudakov_int}       
\end{eqnarray}
The angular ordering condition is given by:
\begin{equation}
z_{i-1} \qbi{i-1} < \qbi{i} 
\label{ang_ord}
\end{equation}
where the rescaled transverse momenta $\bar{q}_{i}$ of the emitted
gluons is defined by:
\begin{equation}
 \qbi{i} 
  = \frac{p_{\prp i}}{1-z_i}
\end{equation}
Here $z_i = x_i /x_{i-1}$ is the ratio of the energy fractions 
in the branching $(i-1) \to i$ and $p_{\prp i}$ is
the transverse momentum of the emitted gluon $i$. The transverse momentum
of the propagating gluon is given by $\kti{i}$.
It is interesting to note, that the angular ordering constraint, as given by
eq.(\ref{ang_ord}), reduces to ordering in transverse momenta 
$p_{\prp}$ for large $z$,
whereas for $z\to 0$, the transverse momenta are free to perform a 
so-called random walk.  
\par
In \cite{smallx_2001} it has been proposed to include also the non-singular
terms in the splitting function as well as to consistently 
use $\mu_r=\pt$ for the renormalization scale in $\alpha_s(\mu_r)$, 
everywhere. 
These changes, although formally sub-leading, have
significant influence for calculation performed at present collider energies.
\par
The inclusion of non-singular terms, as well as the evolution of
quarks, is straightforward in the LDC model \cite{\LDCMC}, which is a
reformulation of CCFM, where the separation between the initial- and
final-state emissions is redefined. In addition to the angular
ordering in eq.(\ref{ang_ord}), the gluons emitted in the initial-state are
required to have
\begin{equation}
  p_{\prp i} > \min(k_{\prp i},k_{\prp i-1}).
    \label{ldccut}
\end{equation}
In the double leading logarithmic approximation (DLLA), this requires
a reweighting of each splitting, completely canceling the non-Sudakov 
form factor, reducing the splitting function in (\ref{Pgg}) to the leading
singularities of the standard DGLAP one, making the inclusion of
non-singular terms as well as quark splittings a trivial exercise.
The constraint in (\ref{ldccut}) means that the $p_{\prp}$ of the
emitted gluon is always close to the highest scale in the splitting
and the argument in $\as$ is naturally taken to be
$p_{\prp}^2$. 
\par
While formally equivalent to the DLLA accuracy for the inclusive observable $F_2$, 
it is important to note that the  
sets of chains of initial-state splittings summed over, are
different in LDC and CCFM. 
Therefore results for exclusive final states agree only after addition
of final state radiation in the appropriate kinematical regions (which
are different in the two formalisms).
\par
We here also want to mention the formalisms developed in 
Refs.~\cite{Martin_Stasto,martin_kimber}.
An evolution equation for a single scale uPDF, which interpolates
between DGLAP and BFKL, is presented by Kwieci\'nski, Martin and
Sta\'sto in \cite{Martin_Stasto}. The
formalism for a two-scale uPDF by Kimber, Martin and 
Ryskin~\cite{martin_kimber} is
based on the same single scale evolution equation, but an angular cut is
applied for the last step in the chain.

\subsection{Unintegrated parton distributions}
\label{sec:unint}
In the following we discuss in detail the precise
definition of (integrated or unintegrated) 
parton density functions (PDFs)~\cite{pdf.defn}. 
\begin{enumerate}
\item A PDF is not a physical quantity in and of itself.  It merely is
   a useful tool.
\label{tool}
\item 
   A PDF is often given as a
   probability density of quarks or gluons within the framework of
   light-front quantization of QCD in the light-cone gauge.  Such a
   definition is useful to provide motivation, intuition and an
   initial candidate for a formal definition.  But this method does
   not necessarily provide a valid definition.
\item Whether or not some kind of consistent probability
   interpretation can be made with modified definitions is an open
   question. For many applications of PDFs, the answer to this question
   is irrelevant.
\item The physical significance of PDFs is that there are
   factorization formulae involving them\footnote{
      The status of a given factorization formula may be anywhere from
      being a completely proved theorem to merely being a conjecture.}.
   Factorization formulae (in their most general sense) give
   approximations to physical amplitudes or cross sections that are
   useful and predictive because:
   \begin{enumerate}
   \item The PDFs are universal -- the same in a range of different
      processes and situations.
   \item Some (not necessarily all) of the coefficients in a
      factorization formula may be estimated, for example in
      fixed-order perturbation theory with a weak coupling.
   \item Kernels of evolution equations (DGLAP etc) may similarly be
      estimated perturbatively.
   \end{enumerate}
\item Since a PDF will include non-perturbative physics, it is
   generally desirable that an explicit definition be given, for
   example in terms of some Green function or a matrix element of some
   (usually non-local) operator. 
\item Given point \ref{tool}, it is not necessary that a PDF's
   definition is explicitly gauge invariant.  However, if the
   definition is not gauge-invariant, the choice of gauge must be
   explicitly specified.  It should be possible to transform the PDF's
   definition into an explicitly gauge-invariant form.  But in general
   there should be extra parameter(s) for the parton density
   corresponding perhaps to
   a gauge-fixing vector or the direction of Wilson line factors in the
   operators.  It will also be necessary to obtain evolution equations
   with respect to the extra variable(s).  See the work of Dokshitzer,
   Diakonov and Troian \cite{DDT} and of Collins and Soper \cite{CS1,CS2} for
   example.
\item The most obvious candidate definition for a PDF is as a number
   density in the light-cone gauge, essentially
   \begin{equation}
     f(x,\kt) = \frac{\langle p | b_k^\dag b_k|p \rangle  }
                      {\langle p | p \rangle  }
   ,
   \end{equation}
   where $b_k^\dag$ and $b_k$ are creation and annihilation operators
   for a flavor of parton in the sense of light-front quantization in
   light-cone gauge.  However, as we will see below, such
   a definition is divergent beyond the lowest order of perturbation
   theory.  
\item The divergence arises from an integral over rapidity of emitted
   gluons and is present even if all IR and UV divergences are
   regulated.  
   The divergence is an endpoint divergence due to the 
   $1/k^+$ singularity in the gluon propagator in light-cone gauge: 
   $\int_0 d k_+/k_+$  
   Therefore it cannot be removed by a modification of the integration path,
   and in that way changing the analytic prescription of the singularity.
   
\item For an uPDF, 
   the divergence cannot be canceled
   between real final state 
   and virtual gluon emission:  Virtual gluon emission
   has an unrestricted transverse momentum integral, but real gluon
   emission is restricted by the transverse momentum of the 
   emitting parton (Fig.~\ref{updf-virt}).
   Hence a cancellation of real and virtual divergences cannot
   occur simultaneously for all values of the transverse momentum of
   the 
   emitted parton.
   This is a problem because without a cancellation between real and virtual
   divergencies the resulting parton density function diverges and becomes
   meaningless.
\begin{figure}[h]
\begin{center}
\resizebox{0.3\textwidth}{!}{\includegraphics{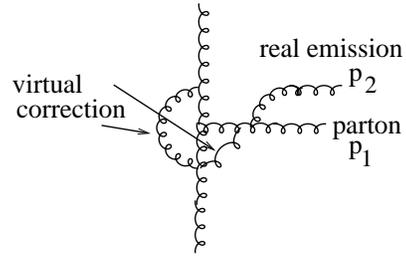}} 
 \caption{
 {\it A schematic drawing indicating the real and virtual corrections in a
 parton splitting process. The transverse momentum of the
 emitted parton $p_2$ is limited by the emitting parton $p_1$, whereas the
 transverse momenta in the virtual loops can reach any value.
 \label{updf-virt}}}
\end{center}
\end{figure}
\item The rapidity divergence ($\int dk_+/k_+ = \int_{-\infty} dy $)
   involves gluon momenta in a region that
   has no relevance to the process: The momenta have infinite rapidity
   relative to physical particles.  Any sensible definition of a PDF
   must have a cutoff.  A simple candidate would be obtained by taking
   the (incorrect) light-cone gauge definition 
    but
   with the use of a planar gauge: $n\cdot A = 0$, with $n^2 \ne 0$.
   The unintegrated parton density then has {\em two} extra
   parameters beyond the $x$ and $\kt$ kinematic variables.  These
   are the inevitable renormalization scale $\mu$ and the variable
   $(p\cdot n)^2/n^2p^2$.  
   The renormalization scale $\mu$ has the approximate interpretation of a 
   cut off on the transverse momentum or the virtuality of virtual 
   particles, and the last variable equals $\cosh^2 y$, where $y$ is the 
   rapidity difference between the target and the gauge fixing vector.
\item 
   In the CCFM formalism these two extra parameters are correlated. 
   Thus the CCFM parameter $\qb$ determines both the transverse momentum 
   cutoff and  the limiting rapidity 
   in the $p$-rest frame through the relation $y = 
   \ln(\qb /m_p x_g)$.
\item In Ref.~\cite{pdf.defn} Collins has proposed explicit
   gauge-invariant definitions of unintegrated PDFs that avoid the
   difficulties mentioned above.  The evolution equation would be that
   given by Collins, Soper and Sterman \cite{CS1,CS2}.  
   Details in this approach
   are being worked out.
\end{enumerate}   
\subsection{Further questions on gauge invariance } 
\label{sec:gauge}
The question of gauge invariance
\footnote{this section is based on some remarks by M. 
Ciafaloni (unpublished) during a discussion with John Collins and 
Yuri Dokshitzer}  
 is not only relevant in the discussion of PDFs, but also 
for \kt-factorization in general as well as for the
cross sections which are \kt-factorized~\cite{GLR,LRSS2,CCH,CE}
into an off-shell (\kt-dependent) partonic cross section
$\hat{\sigma}(\frac{x}{z},\kt^2) $
and a \kt-unintegrated parton 
density function\footnote{We use the 
  classification scheme introduced in Ref.\cite{smallx_2001}:
   $x{\cal G}(x,\kt^2)$ describes DGLAP type unintegrated
  gluon distributions, $x{\cal F}(x,\kt^2)$ is used for  pure BFKL and 
  $x{\cal A}(x,\kt^2,\Pmax^2)$ stands for a CCFM type or any other type having two
  scales involved.} ${\cal F}(z,\kt^2)$:
\begin{equation}
 \sigma  = \int 
\frac{dz}{z} d^2 \kt \hat{\sigma}\left(\frac{x}{z},\kt^2\right) {\cal F}\left(z,\kt^2\right)
\label{kt-factorisation}
\end{equation}
Here the partons generating a QCD hard process  are off-mass shell.
On-shell amplitudes in, say, dimensional regularization are
supposed to be gauge-invariant, if not yet physical.
The ensuing factorization
of mass singularities, introduces a scheme- and perhaps gauge-dependence, to be
canceled by (integrated) PDFs in a physical process involving hadrons.
Thus, any gauge-dependence introduced in the PDFs is in a sense an artifact of
the factorization procedure.
\par
A single off-shell gluon
is not gauge-invariant. However, experience with string 
theory~\cite{string-regge} suggests that
high-energy factorization
 could be a way of defining a physical off-shell
continuation, as residue of the Regge pole exchanged at the given (off-shell)
momentum transfer. In such a case the \kt cannot be assigned to a single gluon
(except in some approximation) because the Reggeon is a compound state of
(infinitely) many partons. Therefore, implementing such an idea in a formal
definition is hard and further complicated by the fact that gluon
Reggeization is infrared singular.
\par
The work on \kt-factorization by~\cite{CCH} provided a gauge-invariant
definition of off-shell matrix elements, 
based on the Regge-gluon factorization idea.
\footnote{Note, that there are certain issues on gauge invariance 
which the authors of \cite{CCH} and \cite{pdf.defn} 
have not been able to resolve completely, but which will be a 
topic of future work.}
The gluon Green function (and related uPDF)
was defined so as to satisfy the (gauge-independent) BFKL equation, and the
emphasis was on defining the corresponding off-shell matrix elements, given the
physical cross-section.
\par
At leading-log level, CCH~\cite{CCH} noticed that the LO 
off-shell matrix elements 
could be defined by
the high-energy limit of an on-shell six-point function (or an eight-point
function in the two-\kt case)
 whose expression was worked out in a
physical gauge first, and then translated to the Feynman gauge. 
Because of their definition, the LO matrix elements are gauge-invariant and
positive definite.
At next-to-leading parton level Ciafaloni~\cite{NLLCC}, and 
Ciafaloni and Colferai~\cite{CC98b}
noticed that one has to subtract, however, the leading kernel contribution
(including  gluon Reggeization) in order to avoid a rapidity divergence
related to the $\ln(s)$ term in the total cross-section.
This subtraction introduces a factorization-scheme dependence,
mostly on the choice of the scale of the process, but not a gauge dependence.
Recently the DESY group~\cite{nlofwdjet1,nlofwdjet2,nlofwdjet3}
 went a long way towards completing this approach
for DIS and jet production. The NLO matrix elements so defined are
gauge-invariant, while positivity has not yet been thoroughly investigated, 
and
is not guaranteed, because of the subtraction. The latter is devised so as
to put the whole energy dependence in the Green function.
\par
The CCFM equation employs a definition of unintegrated density
as a sum over physical final states, restricted to some angular region 
via angular ordering.
This definition with rapidity cutoff is consistent with the subsequent
analysis of matrix elements because the latter roughly provide upper and
lower bounds on the rapidity integration, due to the angular coherence
property. However, the relation of CCFM to BFKL was worked out at leading log (LL) only,
and no complete attempt has been made so far to match
this definition to exact next-to-leading log (NLL) calculations.
In this case the energy dependence of the physical cross-section is shared
between density and matrix elements, depending on the choice of the cutoff.
\par
The conclusion of the above considerations is, that any 
prediction for a physical process must be, obviously, gauge-invariant, 
however (unintegrated) PDF's are not guaranteed to be so. 
The formulation of \kt-factorization was meant to be gauge-invariant, and has
been carried through at LL by \cite{CCH,CE} and at NLL level 
by \cite{CC98b,BCV2001,bartels-nlo-qqbar-real_2}. It is not yet clear, whether
gauge invariance is restored  beyond that level.
However, gauge-dependent definitions of PDF's with the corresponding
matrix elements can be conceived also, provided their convolution reproduces 
the same (physical) cross-section.

\section{Theoretical applications}
\label{sec:theoretical-applications}
\subsection{Comparison of available parameterizations}
\label{sec:compar}
The original CCFM splitting function given in eq.(\ref{Pgg}) includes only the
singular terms as well as a simplified treatment of the scale in $\alpha_s$,
i.e. \kt was used as the scale in the $1/z$ term and the non-Sudakov factor
whereas \pt was used in the $1/(1-z)$ term and in the Sudakov form factor.
Due to the angular ordering 
a kind of random walk in the propagator gluon 
\kt can be performed, and therefore care has to be taken 
for small values of \kt. Even during the evolution the non-perturbative region
can be entered for $\kt < \kt^{cut}$.
In the region of small \kt ,
$\as$ and the parton density are large, 
and collective phenomena, like gluon recombination or saturation might play a
role. Thus, the fast increase of the parton density and the cross section is
tamed. 
However, for the calculation of the unintegrated gluon density
presented here, a simplified but
practical approach is taken:  
no emissions are allowed for  $\kt < \kt^{cut}$ and $\qt < Q_0$. The limitation
of \kt is necessary for the calculation of the non-Sudakov form factor
$\Delta_{ns}$ in Eq.(\ref{non_sudakov_int}) and it ensures a finite value
of $\as(\kt)$.
Different choices of $\kt^{cut}$ are discussed below.
\par
Following the arguments in \cite{smallx_2001}, 
the scale in $\alpha_s$ was changed to $\pt = q (1-z)$ everywhere, and 
the CCFM splitting function  was extended 
to include also the non-singular terms~\cite{jung-dis02,jung-dis03}.
The unintegrated gluon density at any $x$, $\kt$ and scale $\Pmax$ is obtained
by evolving numerically~\cite{jung_salam_2000}
a starting gluon distribution from the scale $Q_0$ according to CCFM to the
scale $\Pmax$. 
The normalization $N$ of
the input distribution as well as the starting scale $Q_0$, which also acts as a
collinear cutoff to define $z_{max} = 1- Q_0/q$, need to be specified. 
These parameters were  
fitted such that 
the structure function $F_2$ as measured at 
H1~\cite{H1_F2_1996,H1_F2_2001} and ZEUS~\cite{ZEUS_F2_1996,ZEUS_F2_2001}
can be described after convolution with the off-shell matrix element
in the region of $x < 5\cdot 10^{-3}$ and $Q^2 > 4.5$~GeV$^2$. 
Using 248 data points
a $\chi^2/ndf = 4.8, 1.29, 1.18, 1.83$ for {\bf JS, J2003 set 1,2,3 },
respectively, is obtained. The following sets of  
CCFM unintegrated gluon densities are obtained:
\begin{itemize}
\item[$\bullet$] {\bf JS} (Jung, Salam~\cite{jung_salam_2000})\\
  The splitting function $P_{gg}$ of eq.(\ref{Pgg}) is used.
  The soft region is defined by $\kt^{cut} = 0.25 $ GeV.
\item[$\bullet$] {\bf J2003 set 1} (Jung~\cite{jung-dis03})\\
  The splitting function $P_{gg}$ of eq.(\ref{Pgg}) is used,
  with $\kt^{cut}=Q_0$ fitted to $\kt^{cut} = Q_0=1.33$~GeV.
\item[$\bullet$] {\bf J2003 set 2} \\
  The CCFM splitting function containing also the non singular
  terms is used:
  \begin{eqnarray}
  P(z,q,k)& = &\asb \left(\kt^2\right)  \cdot\\
  & &\left( \frac{(1-z)}{z} + z(1-z)/2\right) \Delta_{ns}(z,q,k) \nonumber\\
  & &  + \asb\left((1-z)^2q^2\right) \left(\frac{z}{1-z} + z(1-z)/2\right)
  \nonumber
  \end{eqnarray}
  The Sudakov and non-Sudakov form
  factors were changed accordingly. 
  The collinear cut is fitted to $Q_0=\kt^{cut} = 1.18$ GeV.
\item[$\bullet$] {\bf J2003 set 3} \\
  CCFM splitting function containing only singular terms 
   but the scale in $\alpha_s$ is changed from \kt\ to \pt\ for the
  $1/z$ term.  The collinear cut is fitted to $Q_0=\kt^{cut} =1.35$ GeV.
  The problematic region in the non-Sudakov form factor in 
  eq.(\ref{non_sudakov_int}) is avoided by fixing $\alpha_s(\mu_r)$ for 
  $\mu_r<0.9$ GeV.
\end{itemize}
\begin{figure}[tb]
\begin{center}
\resizebox{0.57\textwidth}{!}{\includegraphics{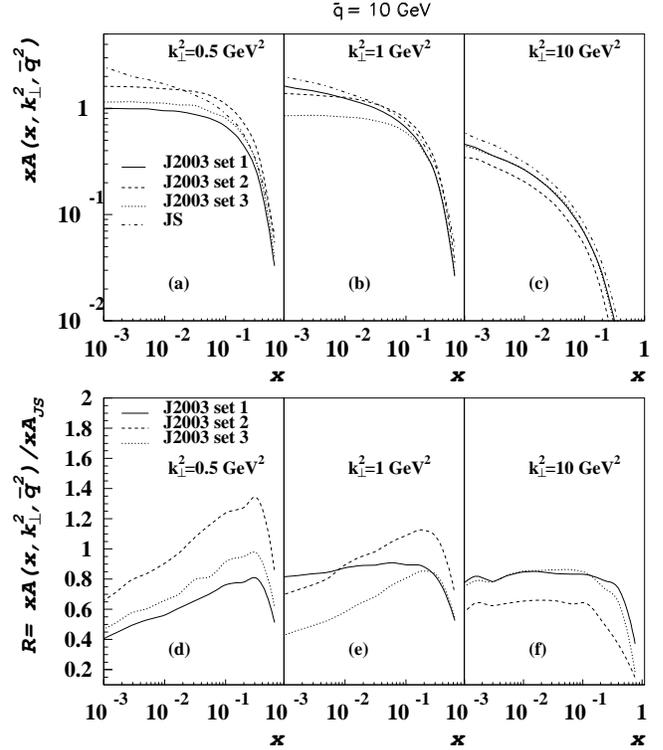}} 
 \caption{
 {\it Comparison of the different sets of unintegrated gluon densities obtained
 from the CCFM evolution as described in the text. In $(a-c)$ the unintegrated
 gluon density is shown as a function of $x$ for different values of \kt at a
 scale of $\bar{q}=10$ GeV. In $(d-f)$ the ratio 
 $R=\frac{x{\cal A}(x,\kt^2,\Pmax^2)}
 {x{\cal A}(x,\kt^2,\Pmax^2)_{\bf JS}}$ as a function of $x$ 
 for different values of \kt is shown.
 \label{ccfm-new}}}
\end{center}
\end{figure}
A comparison of the different sets of CCFM unintegrated gluon densities is shown
in Fig.~\ref{ccfm-new}. It is clearly seen, that the treatment of the soft
region, defined by $\kt < \kt^{cut}$ influences the behavior at small $x$ and
small \kt. 
\par
Also the LDC model describes $F_2$ satisfactorily, but the
corresponding unintegrated gluon densities are somewhat different.
\begin{figure}[tb]
\begin{center}
\resizebox{0.53\textwidth}{!}{\includegraphics{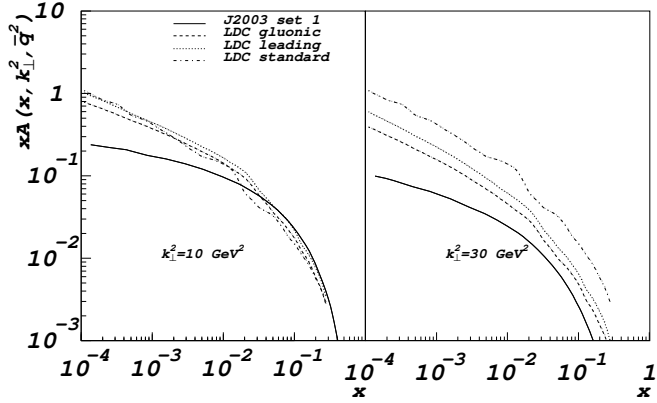}} 
 \caption{
 {\it Comparison of the different sets of unintegrated gluon densities obtained
 within LDC at
 scale of $\bar{q}=10$ GeV. \textrm{Standard} refers to the full
 LDC including quarks in the evolution and the full gluon splitting
 function. For \textrm{gluonic} and \textrm{leading} only gluon
 evolution is considered with only singular terms in the splitting 
 function for the latter. Also shown is the { J2003 set 1} for 
 comparison (divided by $\pi$).
 \label{ccfm-ldc}}}
\end{center}
\end{figure}
One major difference
as compared to CCFM is that LDC can also include quarks in the
evolution, and can therefore also reproduce $F_2$ in the valence
region of high $x$. In Fig.~\ref{ccfm-ldc} three different unintegrated
gluon densities for the LDC approach are presented.
The \textit{standard} set refers to the full
LDC including quarks in the evolution and the full gluon splitting
function, whereas for the \textit{gluonic} set and the 
\textit{leading} set only gluon
evolution is considered with only singular terms in the splitting 
function for the latter. 
All three alternatives have been
individually fitted to $F_2$ in the region $x<0.3$,
$Q^2>1.5$~GeV$^2$ for \textit{standard} and $x<0.013$ and
$Q^2>3.5$~GeV$^2$ for \textit{gluonic} and \textit{leading}.   
In LDC there is only one relevant infrared cutoff,
 $k_{\prp0}$, 
which limits the $p_\prp$ of emitted gluons. This has been fitted to
$0.99$, $1.80$ and $1.95$ for \textit{standard}, \textit{gluonic} and
\textit{leading} respectively.
No cut on the transverse momenta of the virtual gluons is applied and the
argument $\mu_r$ in $\as$ is set to $\mu_r = \pt$ which is then always larger
than the cutoff $k_{\prp\, 0}$.

\subsection{Semi analytical insight into the CCFM equation}
\label{sec:semi-analytical}
The CCFM equation interlocks in a rather complicated way the two 
relevant scales, i.e. the transverse momentum   
$\kt$ of the parton and the hard scale $\qb$ which is related to the 
maximal emission angle.  
Due to this complexity the existing analyses of the CCFM equation 
are based upon   numerical   
solutions.  After performing some approximations it is however possible 
to obtain   
semi analytical insight into the CCFM  equation and we would like to  
consider the following two cases:             
\begin{enumerate}    
\item The single loop approximation (SLA) \cite{Webber1990,SMALLXa,SMALLXb}, 
which corresponds to the DGLAP limit.      
\item The CCFM equation at small $x$ with consistency constraint (CC)   
\cite{LDCb,Martin_Sutton2}.   
\end{enumerate}    
\subsubsection{The single loop approximation}  
\begin{figure}[tb]
\begin{center}
\resizebox{0.53\textwidth}{!}{\includegraphics{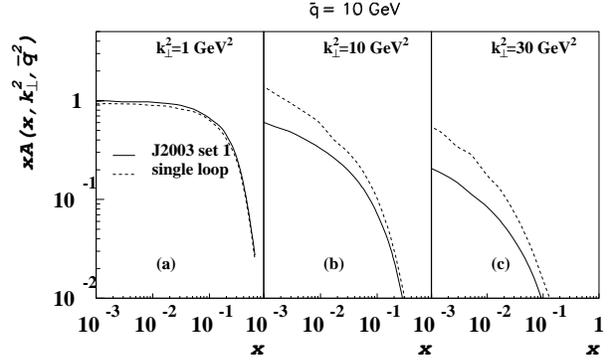}} 
\vskip -4.5cm
 \caption{
 {\it Comparison of the unintegrated gluon densities obtained
 in the CCFM and single loop approximation, using the same input distributions
 of J2003 set 1.
 \label{ccfm-dglap}}}
\end{center}
\end{figure}

This approximation  corresponds to setting the non-Sudakov form-factor 
equal to   
unity and to the replacement of the angular ordering by a $\qt$ ordering. 
It can be a reasonable approximation   
for large and moderately small values of $x$.
In Fig.~\ref{ccfm-dglap} we show a comparison of the unintegrated gluon density
obtained in the full CCFM and the single-loop approximation (using the same 
 input distributions).
\\  
In SLA the CCFM equation   
can be partially diagonalized by the Fourier-Bessel transform of the 
unintegrated   
gluon distribution ${\cal A}(x,\kt,\qb)$ \cite{JKB}\\  
\begin{equation}    
{\cal A}(x,\kt,\qb)=\int_0^{\infty} db b J_0(\kt,b) \bar {\cal A}(x,b,\qb)    
\label{fb1}  
\end{equation}   
\begin{equation}  
\bar {\cal A}(x,b,\qb)=\int_0^{\infty} d\kt \kt J_0(\kt,b)  {\cal A}(x,\kt,\qb)    
\label{fb2}  
\end{equation}  
with $b$ being the impact parameter and the integrated 
gluon distribution is given by:
\begin{equation}  
xg(x,Q^2)=2\bar {\cal A}(x,b=0,\qb=\sqrt{Q^2})  
\label{idvb}  
\end{equation}      
The transverse coordinate representation also partially diagonalizes 
the CCFM equation   
extended for polarized unintegrated parton distributions as will be 
discussed in a separate   
section below.  Due to the absence of the $1/z$ term in the polarized  
splitting function  it will be   
possible to utilize this representation {\it beyond}
the single loop case (see section 2.6).  
The tranverse coordinate representation has also been used for the  
analysis of the CCFM equation in SLA for the unintegrated gluon distributions  
in a photon \cite{AGJK}.    
\par
The CCFM equation in SLA takes the following form in the $b$ representation:    
$$  
\bar {\cal A}(x,b,\qb) = \bar {\cal A}^0(x,b) +    
\int_{q_0^2}^{\qb^2}{dq^2 \over q^2}{ \alpha_s(q^2)\over 2\pi}\int_0^1 
{dz\over z}zP_{gg}(z) \cdot   
$$    
\begin{equation}    
\left\{\Theta(z-x)    
 J_0[b(1-z)q]\bar {\cal A}\left({x\over z},b,q\right)-z    
\bar {\cal A}(x,b,q)\right\}    
\label{ccfmb}  
\end{equation}    
where for simplicity we have neglected the quark contribution. 
 At $b=0$ this equation reduces to the conventional   
DGLAP evolution equation in the integral form.    
Equation (\ref{ccfmb})  can be solved in a closed form using the moment 
function      
\begin{equation}  
\bar f_{\omega}(b,\qb)=\int_0^1 dx x^{\omega-1}\bar {\cal A}(x,b,\qb)    
\label{momb}  
\end{equation}  
Different approximations of eq.(\ref{ccfmb}) are related to formalisms used
e.g. in studies of Drell-Yan pairs, and can give more insight into the
properties of the solutions to the CCFM equation. Thus approximations in
the Bessel function and the Sudakov form-factor gives the relation \cite{JKB}:
$$    
{\cal A}(x,\kt,\qb) \simeq    
{T_g(b=1/\kt,\qb)\over \kt^2}\int_0^{1-\kt/\qb}dz  P_{gg}(z) \cdot    
$$   
\begin{equation}    
{\alpha_s(\kt^2)\over 2 \pi}    
\Theta(z-x){x\over z}    
g\left({x\over z},\kt^2 \right)    
\label{aconv}  
\end{equation}   
where the Sudakov form-factor $T_g(b,\qb)$ is defined by:     
\begin{equation}  
T_g(b,\qb)=\exp\left[-\int_{1/b^2}^{\qb^2} 
{dq^2\over q^2}{\alpha_s(q^2)\over 2\pi}\int_0^{1-1/(bq)}    
dz zP_{gg}(z)\right]    
\label{sudlike}  
\end{equation}  
Neglecting also contributions to the Sudakov form-factor for large $q^2$
gives:
\begin {equation}  
{\cal A}(x,\kt,\qb) \simeq  {\partial [T_g(b=1/\kt,\qb)xg(x,\kt^2)] 
\over \partial \kt^2}    
\label{aderiv}  
\end{equation}  
The approximate expressions (\ref{aderiv}) and (\ref{aconv})   
are similar to those discussed in \cite{DDT,Kimber:1999xc,martin_kimber}.    
It turns out that  expression (\ref{aconv}) gives a reasonable 
approximation of the   
exact solution of the CCFM equation in SLA while  expression (\ref{aderiv}) 
can generate   
negative result for large $\kt$ and large $x \sim 0.1$ \cite{JKB}.
\subsubsection{CCFM equation with consistency constraint} 
We shall consider now the CCFM equation in the small $x$ limit keeping 
only the singular   
$1/z$ part of the splitting function $P_{gg}(z)$ and neglecting the Sudakov   
form-factor. We shall also impose the consistency   
constraint \cite{LDCb,Martin_Sutton2} which is known to generate the dominant part 
of the sub-leading BFKL corrections.   
The integration limit(s) in the CCFM equation are now constrained by the 
following   
competing conditions:
\footnote{The single loop approximation is extended, since the \qb\ ordering is
replaced by angular ordering } 
\begin{enumerate}    
\item Angular ordering (AO)  $\leftrightarrow$ $z_{i-1} \qbi{i-1}< \qbi{i}$.     
\item Consistency constraint (CC) $\leftrightarrow$ $\qbi{i}^2<\kti{i}^2/z_{i}$.    
\end{enumerate}    
It can easily be observed that   
CC takes over AO for $\kt^2<\qb^2/z$.    
\par  
The structure of the CCFM equation at small $x$ with CC is different in the 
regions    
$\kt<\qb$ and $\kt>\qb$.  At $\kt<\qb$ the unintegrated distribution ${\cal A}(x,\kt,\qb)$ 
is independent of $\qb$, i.e.:       
\begin{equation}    
{\cal A}(x,\kt,\qb) \rightarrow {\cal F}(x,\kt)      
\label{fraf}  
\end{equation} 
after adopting the leading $\ln^2(\kt^2/\bar{q}^2)$ approximation
of the Sudakov form factor,  
while for $\kt>\qb$  we get the following expression after  
adopting the leading double $\ln^2(\kt^2/\qb^2)$ approximation:
\footnote{Double $\ln(k_t^2/\bar{q}^2)$ terms appear in the Sudakov form factor
for exclusive cross sections. They are not present in the inclusive cross
section, which is $ \propto F_2$}
\begin{equation}  
{\cal A}(x,\kt,\qb)={\cal F}(x,\kt)  
\exp\left[-{3\alpha_s\over 2\pi}\ln^2(\kt^2/\qb^2)\right]  
\label{fxqt}  
\end{equation}    
where    for simplicity we keep fixed   
$\alpha_s$.  The single scale   
function    
${\cal F}(x,\kt)$ satisfies the BFKL-like equation with sub-leading corrections.   
We found in this way that imposing the consistency constraint and the double   
$\ln^2(\kt^2/\qb^2)$ approximation in the region $\kt>\qb$ we reduce the two-scale
problem   
to the single-scale one and to the BFKL-like dynamics.    
The novel feature of the CCFM framework is however the exponential 
suppression   
of the unintegrated distribution in the region $\kt\gg \qb$ (cf. equation 
(\ref{fxqt}))  due to the double $\ln^2(\kt^2/\qb^2)$  
effects \cite{CCFMa}.  They are of course formally sub-leading at small $x$.      

\subsection{Effects of phase space constraints in BFKL}
\label{sec:momentum}
The leading logarithmic (LL) BFKL formalism  resums terms in the
perturbative series of the form $(\alpha_s\ln(\hat s/s_0))^n$, 
where $\hat s$ is the square of the center of
mass energy for the hard scattering and $s_0$ some
perturbative scale separating the evolution of the
$t$--channel exchange from the matrix elements.
These logarithms arise due to the emission of
gluons from the $t$--channel exchange. For the
scattering of two particles $p_A p_B\to k_a k_bk_i$
where $k_i$ are the momenta of the gluons emitted from
the BFKL evolution, we 
have $\hat s=2 p_A p_B$ and $s_0$ is often chosen as
$s_0=k_{a\prp}k_{b\prp}$ with $k_{a\prp}$ ($k_{b\prp}$) the transverse
part of $k_a$ ($k_b$ respectively).
 In deep inelastic scattering large $\hat s$ 
corresponds to  small $x$ of the probed parton. For
hadronic dijet production, large $\hat s$ corresponds
to large separation in rapidity between the leading
jets, and therefore to moderate values of $x$, where
normal DGLAP evolution of the partons is valid (and
therefore the standard PDFs can be used~\cite{Mueller-Nav}).
 The next--to--leading logarithmic
corrections~\cite{Fadin:1998py,NLLCC} consist
of terms proportional to $\as(\as\ln(\hat s/s_0))^n$,
i.e.~suppressed by one power of $\as$ compared to the LL component.  The
logarithms resummed in the BFKL approach correspond to the enhanced terms in
scattering processes for large center of mass energies and also the enhanced
terms in the description of the small $x$ 
behavior of the gluon distribution function. 
\par
When confronting BFKL predictions with data, several points are worth
observing. First of all, present day colliders do not operate at ``asymptotic
energies'' where the high energy exponent
dominates
the BFKL prediction under the assumption that the coupling can be held
fixed and small,
leading to a
prediction of an exponential rise in cross section with an intercept of
$\asb 4\ln 2$, with $\asb=\frac{3 \as}{\pi}$. For example at HERA, the separation between the struck
quark and the forward jet can reach up to about four units of rapidity,
whereas the measurable jet separation at the Tevatron is up to six units.
This is not asymptotically large. Secondly, it should be remembered that the
logarithms resummed are kinematically generated, and in the derivation of the
standard analytic solution to the BFKL equation, the transverse momentum of
the gluons emitted from the BFKL evolution has been integrated to infinity.
It is therefore apparent that any limits on the phase space probed in an
experiment can have a crucial impact on the theoretical prediction. Such
limits can either be the cuts implemented in the measurement or 
the limits on the available energy at a collider. The total
available energy will affect the impact factors, while taking into
account also detailed energy-momentum conservation in each gluon
emission will in addition affect the BFKL exponent.
Taking hadronic dijet production
as an example, the energy constraint will obviously not just limit the
possible rapidity separation of the leading dijets, but also the amount of
possible radiation from the BFKL evolution, especially when the leading
dijets are close to the kinematical boundary. For a multi--particle final
state described by two leading dijets with transverse momentum and rapidity
$(p_{a/b\prp},y_{a/b})$ and $n$ gluons described by $(\kti{i},y_{i})$,
the total energy of the event is given by $\hat s=x_a x_b s$ where ${s}$
is the square of the total energy of the hadron collider and
\begin{align}
  \begin{split}
    \label{eq:fullbjorkenx}
    x_a=&\frac{p_{a\prp}}{\sqrt s}e^{y_a} +
    \sum_{i=1}^{n}\frac{k_{i\prp}}{\sqrt s}e^{y_i}+\frac{p_{b\prp}}{\sqrt
s}e^{y_b}\\
    x_b=&\frac{p_{a\prp}}{\sqrt s}e^{-y_a} +
    \sum_{i=1}^{n}\frac{\kti{i}}{\sqrt s}e^{-y_i}+\frac{p_{b\prp}}{\sqrt
      s}e^{-y_b}.
\end{split}
\end{align}
While it can be argued that the contribution to $\hat s$ from the gluons
emitted from the BFKL evolution is subleading compared to the contribution
from the leading dijets, it is not obvious that the effect on the cross
section is small, simply for the reasons mentioned above: ignoring the
contribution to the parton momentum fractions will resum logarithmically
enhanced contributions from regions of phase space that lie outside what can
be probed at a given collider. 
Here we have taken as an example dijet
production at a hadron collider, but a similar effect will be found for any
BFKL evolution, whether it describes $\gamma^*\gamma^*$, $ep$, or $pp$
physics.
\par
The iterative approach of Ref.~\cite{Orr:1997im,Schmidt:1997fg}
 to solving
the BFKL equation at leading logarithmic accuracy allows not only a study on
the partonic level of BFKL in processes with complicated impact factors,
where it might be difficult to get analytic expressions for the cross
section. The method also allows for the reconstruction of the full final
state configurations contributing to the BFKL evolution, and therefore it is
possible to study quantities such as multiplicities and distribution in
transverse momentum of the emitted gluons~\cite{Andersen:2003gs}. Only this
reconstruction of the full final state allows for the observation of energy
and momentum conservation. The effects of energy and momentum conservation
have been studied in several
processes~\cite{Orr:1998hc,Andersen:2001kt,Andersen:2001ja}. When no phase
space constraints are imposed, the iterative solution reproduces the known
analytic solution to the BFKL equation. This iterative approach has recently
been generalized~\cite{Andersen:2003an,Andersen:2003wy} to solving the NLL
BFKL equation thereby joining other approaches 
\cite{Ciafaloni:2003rd,Ciafaloni:2003ek,Altarelli:2003hk,Thorne:2001nr,Thorne:1999rb} in studying effects of
NLL corrections.
\par
\begin{figure}[bp]
  \centering
  \resizebox{0.5\textwidth}{!}{\includegraphics{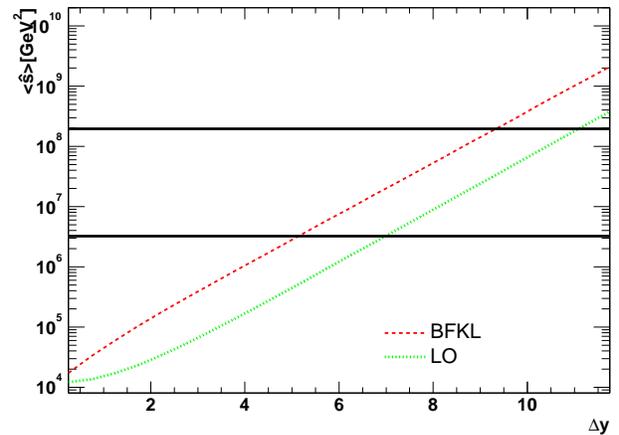}} 
  \caption{{\it The average center
    of mass energy in $gg\to gg$ scattering with (red/dashed) and without
    (green/dotted) 
    BFKL evolution of the $t$ channel gluon, with
    $p_{\prp\mathrm{min}}=20$~GeV for the dijets and $\alpha_s=0.1635$. Also
    plotted is the hadronic center of mass energy squared for the Tevatron
    ($(1.8\mathrm{TeV})^2$) and the LHC($(14\mathrm{TeV})^2$).}}
  \label{fig:avghats}
\end{figure}

The effects on the total center of mass energy of considering the full
multi--gluon BFKL final state in gluon--gluon scattering is seen in
Fig.~\ref{fig:avghats}. We have plotted the result of considering only the
two leading dijets (i.e.~ignoring the sum in Eq.~(\ref{eq:fullbjorkenx})),
and from considering the full BFKL final state (i.e.~using the full 
expression in
Eq.~(\ref{eq:fullbjorkenx})) (see Ref.~\cite{Andersen:2003gs} for more
details).
Fig.~\ref{fig:avghats} shows the average energy for a BFKL dijet event as a
function of the rapidity separation of the leading dijets, when the BFKL
gluon phase space is unconstrained. The standard analytic solution to the
BFKL equation implicitly assumes that all of this phase space is available.
It is clear from Fig.~\ref{fig:avghats} that the energy taken up by BFKL
radiation is significant compared to the center of mass energy at present and
planned colliders. 
For example at four units of rapidity, which is the upper limit of HERA 
at present, the average energy
of a BFKL dijet event is about $\sqrt{\hat{s}} = 1$~TeV according to 
Fig.~\ref{fig:avghats}, which is far beyond the maximum energy
available. At the HERA center-of-mass energy $\sqrt{\hat{s}} = 300$~GeV 
the rapidity range would
be less than one unit and thus leave very little phase space for 
additional emissions.
Therefore, any constraint on the BFKL radiation from
e.g.~overall energy conservation will have an impact on BFKL phenomenology
predictions at such colliders. In fact, it is found that if energy and
momentum conservation is satisfied,
by using the full Eq.~(\ref{eq:fullbjorkenx})
when calculating hadronic dijet production at the
LHC, then the exponential rise in cross section as a function of the rapidity
separation found for gluon--gluon scattering  
(when the BFKL gluon phase space
is integrated to infinity)
is moderated 
to an almost no--change situation
compared to the fixed leading order QCD prediction. Other BFKL signatures,
like the increasing dijet angular de-correlation with increasing rapidity
separation~\cite{Stirling:1994zs,DelDuca:1994mn,DelDuca}, 
are still present.
\par
With the iterative method of solving the LL BFKL equation it is of course
also possible to calculate jet rates and transverse momentum distributions
(since full information of the final state configuration is obtained) arising
from the BFKL dynamics. Below we present results on the jet rates in gluonic
dijet production (on the partonic level, i.e.~with the full gluonic phase
space assumed in the standard analytic solution of the BFKL equation) with a
BFKL chain spanning 5 units of rapidity using a very simple jet definition.
We simply let any gluon with a transverse momentum greater than some cut off
$\mu_R$ define a jet (this is a reasonable jet definition since at leading
logarithmic accuracy the emitted gluons are well separated in rapidity).
\begin{figure}[tbp]
  \centering
  \resizebox{0.5\textwidth}{!}{\includegraphics{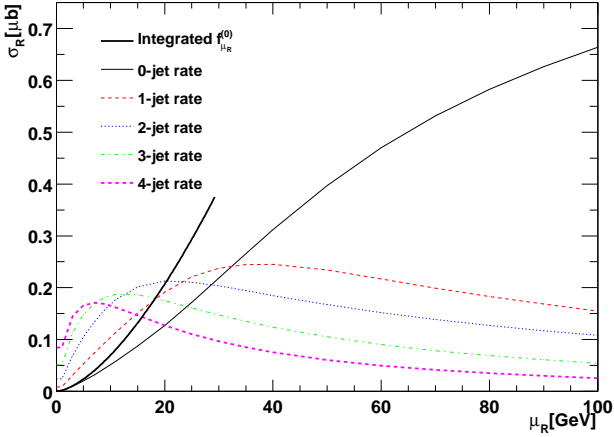}} 
  \caption{{\it The $0$--, $1$, $2$--, $3$--
    and $4$--jet parton--level cross sections as a function of 
    the cutoff $\mu_R$, for a
    rapidity span of $\Delta y = 5$ and $p_{\prp\mathrm{min}}=20$~GeV for
    the leading dijets.  Also shown is the analytic $0$--jet prediction valid
    for small $\mu_R$. }}
  \label{fig:zeroonejet}
\end{figure}
These jet rate predictions will change, once the partonic cross section is
convoluted with parton density functions. The jet rates of
Fig.~\ref{fig:zeroonejet} are the ones responsible for the increase in the
center of mass energy of a BFKL event over a simple LO configuration seen in
Fig.~\ref{fig:avghats}, but they are also responsible for the rise in cross
section predicted from the BFKL dynamics when the BFKL gluonic emission is
unbounded. In Ref.~\cite{Andersen:2003gs} it is found that for gluon--gluon
scattering (with a minimum transverse momentum of the leading dijets of
$20$~GeV) with a BFKL exchange, one can expect a BFKL gluon emission density
of about one hard ($\kti{i}>20$~GeV) gluon for every two units of rapidity
spanned by the BFKL evolution, when the energy of the event is unconstrained.
This amount of radiation is implicitly assumed in the standard analytic
solution of the LL BFKL equation.
\par
In conclusion, carefully taking energy-momentum conservation into account
dramatically modifies the strong increase for small $x$, predicted in
the leading log BFKL approach.
\subsection{The saturation scale}
\label{sec:satscale}
\begin{figure}[htb] 
\begin{center}
\resizebox{0.48\textwidth}{!}{\includegraphics{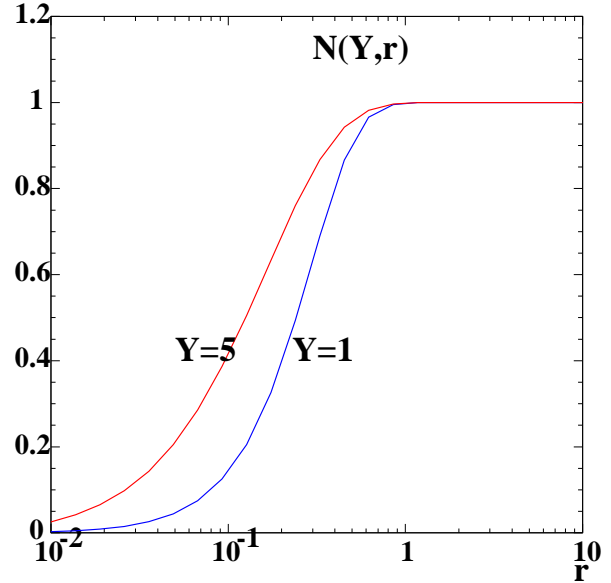}}
\caption{{\it 
The amplitude $N(r,Y)$ as a function of the dipole size $r$ for two
different rapidities $Y=1,5$} }
\label{fig:NrY}
\end{center} 
\end{figure}

The  parton saturation idea is realized with the help of a nonlinear evolution  
equation
in which the gluon splitting is described by a linear term while the negative  
nonlinear term results from 
the competing gluon recombination (see also \cite{GLR}).
The Balitsky-Kovchegov (BK) equation
\cite{Balitsky:96,Kovchegov:99} 
was derived for deep inelastic scattering 
of a virtual photon on a large nucleus by the resummation of multiple pomeron 
exchanges in the leading logarithmic approximation
(when  $\alpha_s \ln(1/x)\sim 1$ and
$x\simeq Q^2/s$)  in  the large $N_c$ limit.
It is an equation for the dipole--proton forward scattering amplitude 
$N(\xb,\yb,Y)$ where $\xb,\yb$ are the end points of the $q\bar{q}$ dipole and 
$Y=\ln 1/x$ is the rapidity of the process. The BK equation has the following
integro-differential form~\cite{Kovchegov:99}
\begin{multline}
\frac{\partial N(\xb,\yb,Y)}{\partial Y} \,=\, 
\asb\  \int \frac{d^2 \zb (\xb-\yb)^2}{(\xb-\zb)^2(\yb-\zb)^2} 
\left[N(\xb,\zb,Y)\, +\, \right.  \\ N(\yb,\zb,Y) 
\left. -\, N(\xb,\yb,Y)\, -\, N(\xb,\zb,Y)N(\yb,\zb,Y) \right]
\label{eq:kov} 
\end{multline}
where $\asb=3\as/\pi$ and is fixed in the leading $\ln
1/x$ approximation. The linear term in (\ref{eq:kov})
 is the dipole version of the BFKL equation whereas the quadratic term
describes the gluon recombination.
Instead of $\xb,\yb$ one often uses their linear combinations: 
$\rb=\xb-\yb$ which is the size of the dipole  and $\bb=\frac{1}{2}(\xb+\yb)$
the impact parameter. Equation (\ref{eq:kov}) can be easily solved when using
the approximation of the infinitely large nucleus, i.e. assuming that the
amplitude $N(\xb,\yb,Y)\equiv N(|\rb|,Y)$ depends only on the size of the
dipole but not on the impact parameter \cite{BRAUN:00,LEVLUB:01,GBMS:02}. The
more complicated case with the full impact parameter dependence has been
analyzed recently \cite{GBS:03}. 
For the $b$-independent and cylindrically symmetric
solution, $N(\rb,Y)=N(r,Y)$,  Eq.~(\ref{eq:kov}) can be rewritten
in momentum space in a much simpler form after
performing the following Fourier transform 
\begin{eqnarray}
\phi(k,Y)
&=&
\int \frac{d^2\rb}{2\pi} \exp(-i\kb\cdot \rb)\, 
\frac{N(r,Y)}{r^2} \nonumber \\
&=& 
\int_0^\infty 
\frac{dr}{r}\, J_0(k\/r)\,N(r,Y), 
\end{eqnarray}
where $J_0$ is the Bessel function. In this case the following equation   
is obtained 
\begin{eqnarray}
\label{eq:newkov}
\frac{\partial \phi(k,Y)}{\partial Y} 
& = 
\asb\, (K\otimes \phi)(k,Y)  
\,-\ 
\asb\, \phi^{\,2}(k,Y), 
\end{eqnarray}
Here expression $(K\otimes \phi)(k,Y)$ means the action of the usual BFKL
kernel in the momentum space onto the function $\phi(k,Y)$.
Let us briefly analyze the basic features of $N(r,Y)$ and $\phi(k,Y)$.
In Fig.~\ref{fig:NrY} we plot the amplitude $N(r,Y)$ as a function of the
dipole size $r$ for different values of rapidity $Y$. The amplitude $N(r,Y)$
 is small 
for small values of the dipole size. It is governed in this regime by the
linear part of equation (\ref{eq:kov}). For larger values of dipole sizes,
$r>1/Q_s(Y)$ the amplitude
 grows and saturates eventually to~$1$. This is the regime
where the nonlinear effects are important. As it is clear from the
Fig.~\ref{fig:NrY} the saturation scale grows with rapidity $Q_s(Y)$. It means
that with increasing rapidities the saturation occurs for smaller dipoles. It
has been shown that  the growth of the saturation momentum is exponential in
rapidity 
$Q_s(Y) \, = \, Q_0 \, \exp(\lambda  Y)$ with $\lambda \simeq 2 \asb$
being a universal coefficient and governed by the equation.
 The normalization $Q_0$ on the other hand is dependent on the 
initial condition $N(r,Y=0)$.
We note however,  that when the rapidity $Y$ is not too large the initial
conditions are still important. In this region  the coefficient $\lambda$ can
still depend on the rapidity \cite{LEVLUB:01}.
\par
It has to be stressed that in the leading-log $x$ approximation the strong
coupling constant is fixed.  The running of the coupling, although being 
a next-to-leading-log $x$ effect, is obviously more physical. In this case the 
rapidity 
dependence of the saturation scale is changed.
We adopt a natural approximation that 
the local exponent of the saturation scale $\lambda(Y)= d
\ln(Q_s(Y)/\Lambda)\; /dY$  
takes the form $\lambda(Y) = 2 \bar\alpha_s (Q^2_s(Y))$ where
$\Lambda=\Lambda_{QCD}$. 
The above form is motivated by the leading logarithmic result  
with the fixed coupling as discussed before, i.e. $Q_s(Y)=Q_0 \exp(\lambda 
Y)$.  
\par
Thus, we have 
\begin{equation}  
{d\ln (Q_s(Y)/\Lambda) \over dY} = {12 \over b_0 \ln ( Q_s(Y) /\Lambda)}\,,  
\end{equation} 
with the initial condition $Q_s(Y_0) = Q_0$  and  $Y_0$ chosen in the region
where 
scaling sets in. The solution takes the form  
\begin{equation}  
Q_s(Y) = \Lambda\; \exp\left( \sqrt{{24\over b_0}\, (Y-Y_0) + L_0 ^2}
\right)\,,  
\qquad Y>Y_0\,,  
\label{eq:satscalerc} 
\end{equation}  
where $L_0 = \ln (Q_0/\Lambda)$.
Thus, the exponential dependence on the rapidity $Y$ of the saturation scale
is changed in the running coupling case   to the milder behavior
\cite{GBMS:02}. In the phenomenological analysis of the HERA data based on the
BK equation \cite{GLLM}, the running of the $\alpha_s$ has been included.
It has been shown that, in the limited range of $x$, the effective power
governing the behavior of the saturation scale
can be still fitted using the simple form $Q_s(Y) \sim \exp(\lambda_0 Y)$ with
$\lambda_0 \simeq 0.18$,
 which is  close to the value used in the Golec-Biernat
and W\"usthoff saturation model \cite{GBW_1998,GBW_1999}.
\par
\begin{figure}[htb] 
\begin{center}
\resizebox{0.5\textwidth}{!}{\includegraphics{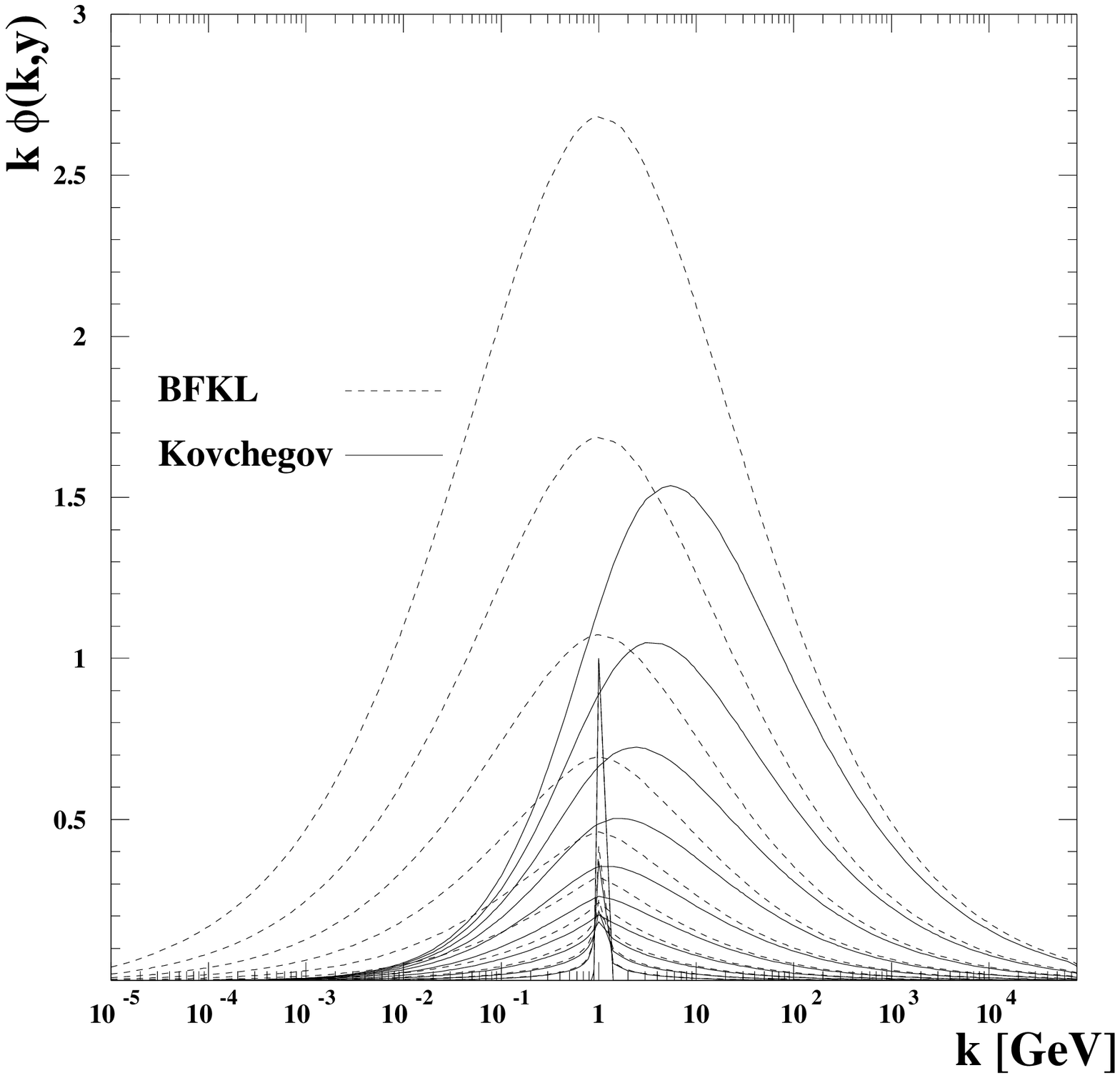}} 
\caption{{\it The Fourier transform of the amplitude $N(r,Y)$ as a function of the
momentum $k$ for different values of rapidities $Y$ increasing from $1$ to
$10$. Dashed lines correspond to the solution to the BFKL equation whereas
solid lines to the full nonlinear BK equation.}}  
\label{fig:FkY} 
\end{center} 
\end{figure}

In what follows we consider the  solution to the Balitsky-Kovchegov equation
in the  leading-log $x$ fixed coupling case.
In Fig.~\ref{fig:FkY} we plot the solution $k\phi(k,Y)$ in momentum space
as a function of $k$ for increasing rapidities.
We have imposed the initial condition of the form of the delta function,
localized at some $k=k_0=1$~GeV. We compare the solution of the BK equation with
that of the linear BFKL which exhibits the  strong unlimited diffusion into
both infrared and ultraviolet regimes. The BFKL solution  is symmetric and
peaked around the initial value $k=k_0$. The solution to the BK equation shows
on the other hand a suppression of the diffusion into the low momenta. We
clearly see that the peak of the distribution moves with increasing
rapidity to the higher momenta. 
One can identify the value of the momentum $k$ at which the maximum occurs as
the saturation scale $Q_s(Y)$. At large momenta $k$ where the
nonlinearity in the BK equation does not play a role,  the two solutions BFKL
and BK are close to each other. The overall height of the distribution is
strongly damped with respect to the linear case.
The solution to the BK equation shows also another interesting feature, namely 
the property of scaling (e.g. \cite{GBKS:01,GBMS:02,LUB:01}). In the
nonlinear regime $r>1/Q_s(Y)$ the amplitude
depends only on one combined variable instead of $r$ and $Y$ separately.
\begin{equation}
N(r,Y) \; \equiv \; N(r \, Q_s(Y))
\label{eq:Gscal}
\end{equation}
This is also property of the GBW saturation model, though in the latter case
the scaling was present for all values of $r$. In the case of BK equation,
scaling only occurs in the saturation domain that is  for large values of
dipole sizes.
\subsection{Non-linear evolution versus HERA data}
A new approach to a global QCD analysis based 
on the non-linear QCD evolution by Balitsky-Kovchegov (BK)
is presented in~\cite{LEVLUB:01,GLLM}. 
The BK equation improved by the DGLAP
corrections for small dipole sizes
(independently of impact parameter) 
is in fact very successful in describing
the low $x$ part of the structure function $F_2$ at HERA.
In the  following a brief summary of the results in Ref.~\cite{GLLM} is given.
\par
\begin{figure*}[htb]
\begin{tabular}{c c}
(a) & (b) \\
 \epsfig{file=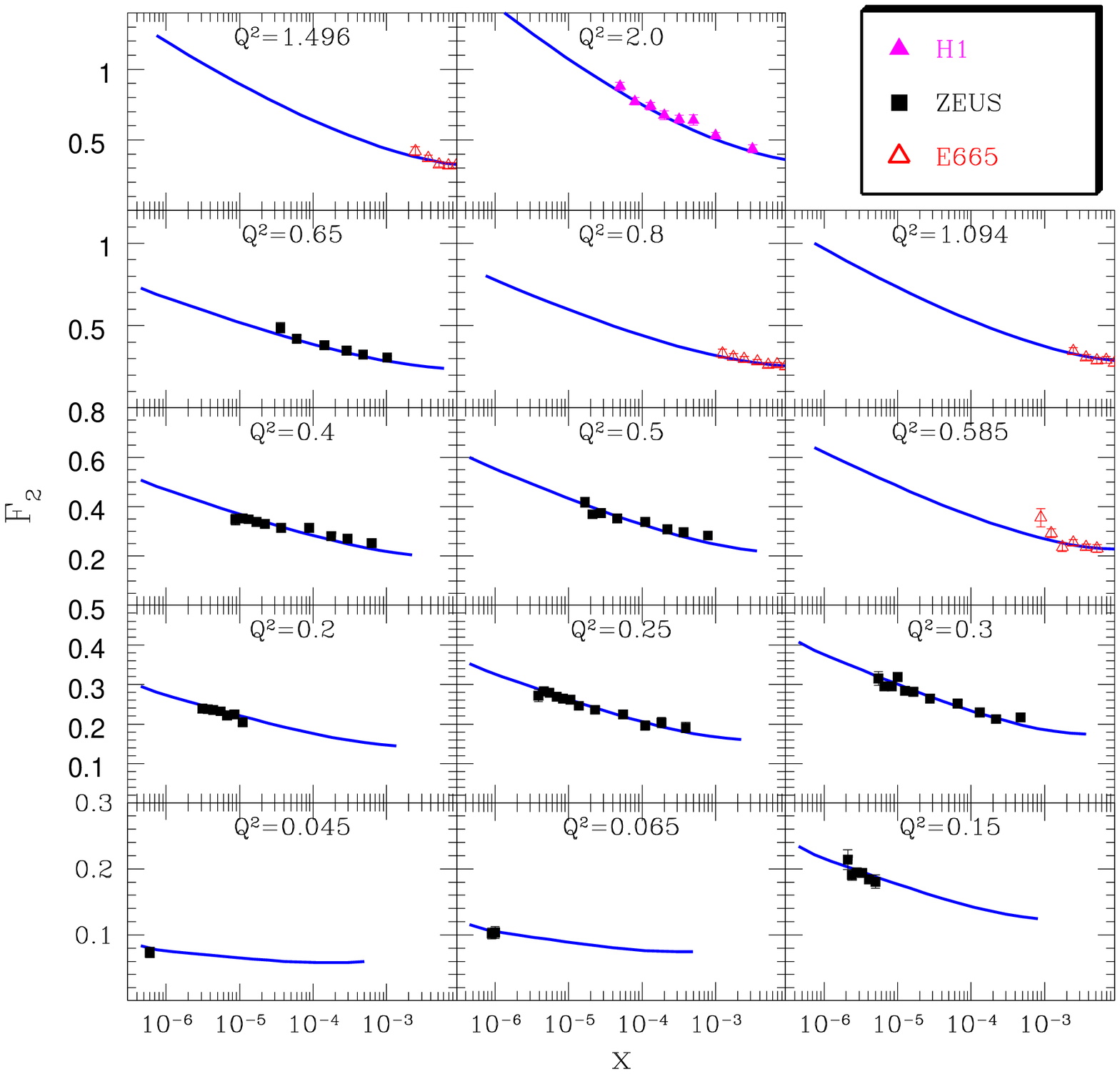,width=85mm, height=95mm}&
 \epsfig{file=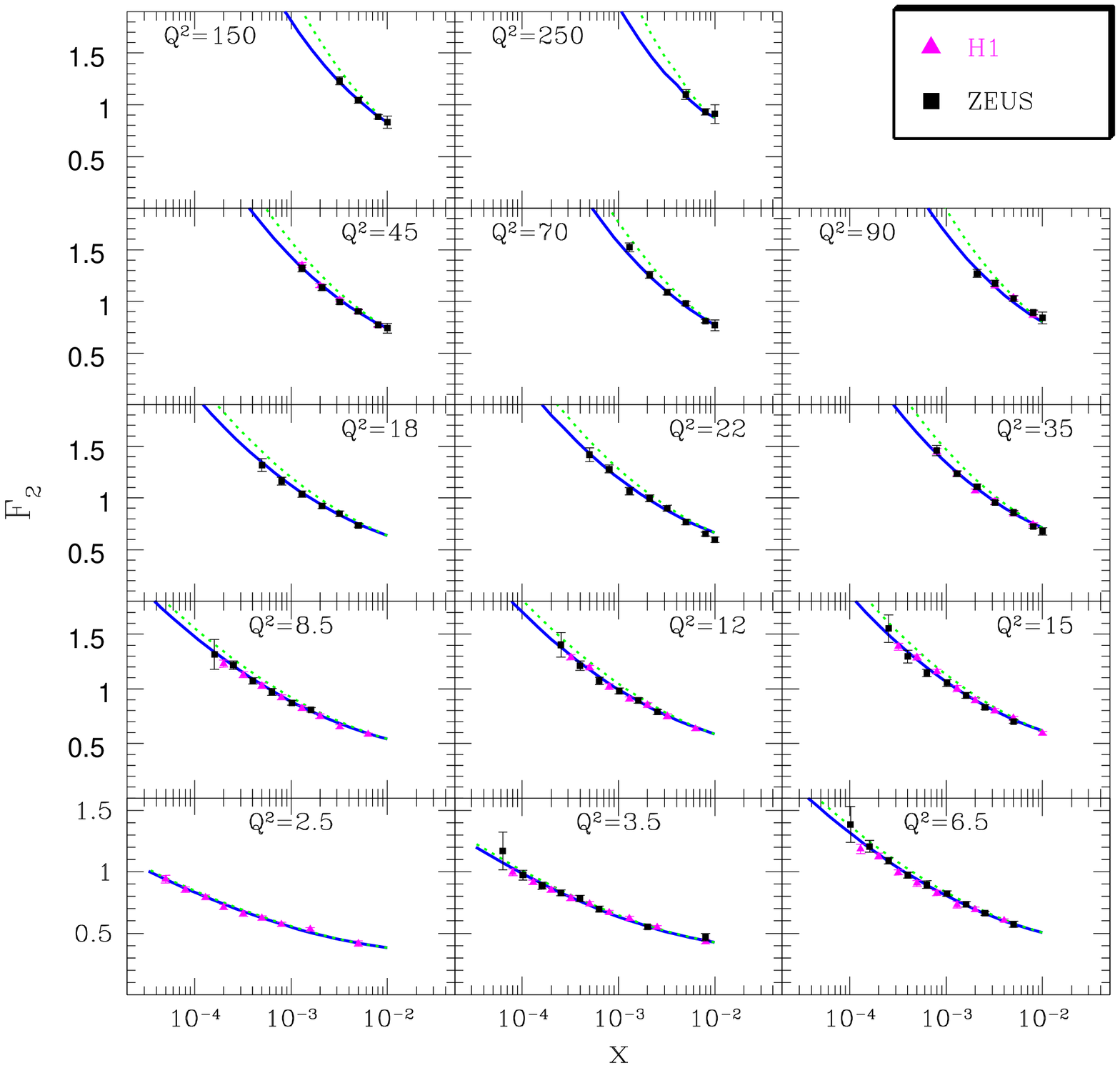,width=85mm, height=95mm}  \\
\end{tabular}
  \caption{{\it Fit to the $F_2$ structure function. The dashed line
 is a result obtained without the DGLAP corrections. The data are from
Refs. \protect\cite{ZEUSF2L,ZEUS_F2_2001,H1F2,E665F2}}}
    \label{F2plot}
\end{figure*}

With the initial conditions specified at $x_0\,=\,10^{-2}$ the BK equation
(without impact parameter dependence) is solved numerically towards 
smaller $x$. The impact parameter dependence is restored using a 
rescattering ansatz of the Glauber-type.
 All existing low $x$  ($x \leq 0.01$) data on the $F_2$ structure function are
reproduced  with  resulting $\chi^2/ndf\,\simeq 1$ 
(Fig. \ref{F2plot}). 
Only two parameters and a few fixed ones 
(associated with the initial conditions) 
 are used for the fit.
The fitted parameters are the effective proton radius,  
entering the Gaussian impact parameter distribution, and the scale at which
the DGLAP corrections are switched on (${\cal O}(1\mbox{ GeV})$). The DGLAP 
corrections are important for  large photon virtualities only and reach up to
$15 $\%. 
The low $Q^2$ (of the order of a few GeV$^2$ and below) data are described 
solely by the BK equation.  
\par
In DIS the Pomeron intercept is obtained by a measurement of
 $\lambda\,\equiv d \ln F_2/d\ln (1/x)$. For large photon virtualities
the fit based on the BK equation reproduces the HERA data with
$\lambda\,\simeq\,0.3\,-\,0.4$, the hard BFKL intercept. 
\begin{figure}[htb]
\resizebox{0.48\textwidth}{!}{\includegraphics{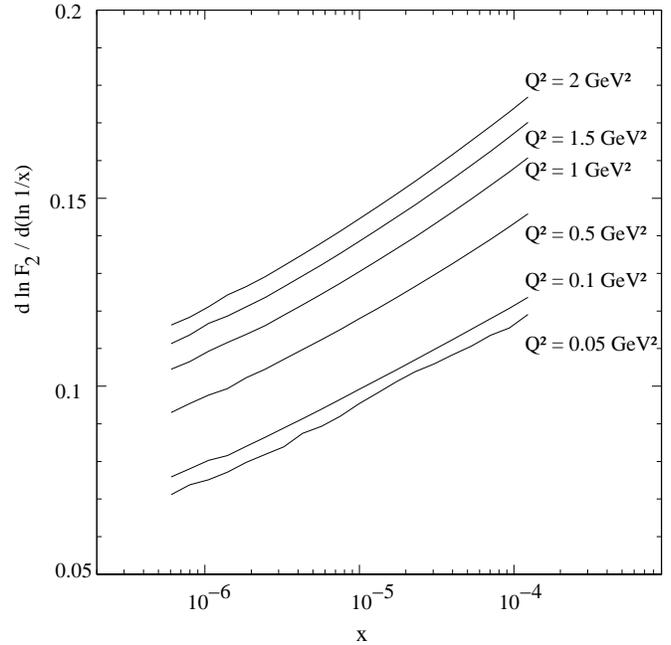}}
\caption{ \it The logarithmic derivative $\lambda\,=\,\partial \ln
F_2/\partial \ln 1/x$ plotted at low $Q^2$ and very low $x$.}
\label{lamdalow}
\end{figure}
In the small $Q^2$ region the non-linear terms in the BK equation are reflected
in the smaller  $\lambda$ values at smaller $Q^2$.  
Fig. \ref{lamdalow}
 presents a prediction for $\lambda$  at 
at smaller $Q^2$ and smaller $x$. Fig.~\ref{lamdalow-new} presents results for 
$x \gap 10^{-4}$. In this region $\lambda$ decreases strongly for small $Q^2$, 
but varies relatively slowly with $x$. In fig. 10 we see, however, that 
for smaller values of $x$, $\lambda$ decreases more strongly with $x$, 
for fixed $Q^2$, tending to zero in agreement
with
the unitarity constrain. 
\begin{figure}[htb]
\resizebox{0.48\textwidth}{!}{\includegraphics{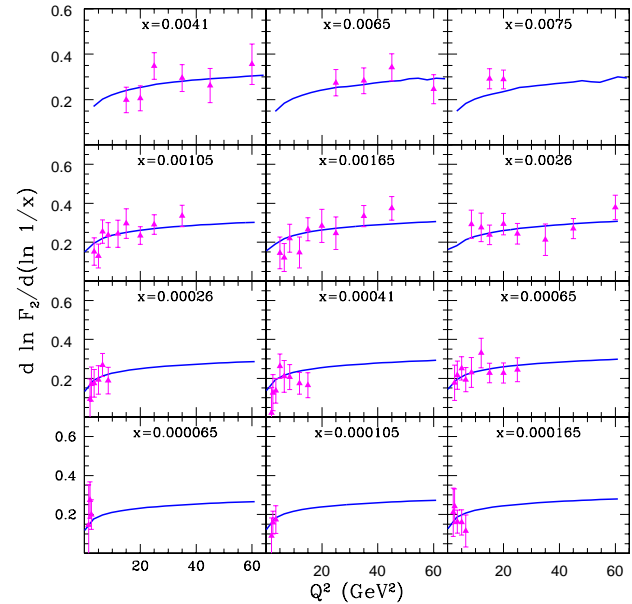}}
\caption{ \it The logarithmic derivative $\lambda\,=\,\partial \ln
F_2/\partial \ln 1/x$ as a function of$Q^2$ for different values of $x$.}
\label{lamdalow-new}
\end{figure}
At $Q^2$ well below $1\,GeV^2 $ and
$x\simeq 10^{-6}$, $\lambda \simeq 0.08 - 0.1$. This value of
$\lambda $ coincides with the "soft Pomeron" intercept. Thus 
the nonlinear evolution provides a solution to the problem
of  hard-soft Pomeron transition.
\par
The main fitting parameter used for the fit is an effective
proton radius $R$. The optimal fit is achieved at  
$R\,\simeq\, 0.3 \,fm$, the radius
which is much smaller than the electro-magnetic
radius of proton. On one hand, 
this small proton radius might be an artifact of the approximations used.
On the other hand, it may indeed indicate a small size dense gluon spot
inside proton. Such scenario arises  in several other models for high 
energy scattering off proton \cite{SZ,KT}.
\par
The approach based on non-linear QCD evolution
allows the extrapolation of the parton distributions to very high
energies available at the LHC as well as very low photon
virtualities, $Q^2\ll 1\, {\rm GeV^2}$.
\subsection{Multiple Interactions in non-ordered cascades}
At high energies the perturbative jet cross section in $pp$ collisions 
becomes larger than the total cross section.  This implies that there are
often several hard sub-collisions in a single event. Therefore 
correlations become important,
and the observed ``pedestal effect'' implies that the hard sub-collisions 
are not independent~\cite{sjostrand_zijl_1987}, indicating an impact parameter dependence such that 
central collisions have many mini-jets, while peripheral collisions have fewer
mini-jets~\cite{sjostrand_zijl_1987}.
Also at HERA the final state properties in photoproduction cannot be
reproduced without assuming multiple hard 
scattering~\cite{mult-scatt-h1,mult-scatt-zeus,mult-scatt-zeus2}. 
At higher $Q^2$ the
indications for multiple scattering are reduced, and thus HERA offers a 
unique possibility to study how, with decreasing $Q^2$, multiple interactions 
become
more and more important, until eventually a situation similar to $pp$
collisions is reached for $Q^2 =0$.
\par
In a non-$\kt$-ordered BFKL ladder, it is possible to have two (or  more)
local $\kt$-maxima, which then correspond to two different hard
sub-collisions.
Thus there are two different sources for multiple interactions: It is possible
to have two hard scatterings in the same chain, and there may be more than one
chain in a single event. The BFKL, CCFM or LDC
formalism can be used to estimate 
multiple collisions in a single chain. The 
symmetric properties of the LDC model
 for DIS makes it especially suited to be applied to $pp$ collisions, and in
Ref.~\cite{LDC-mult-scatt}
it is demonstrated that it is possible to deduce the
average number of chains in $pp$ scattering from data on deep inelastic 
$ep$ scattering.
\par
The LDC model can, 
however, not determine the correlations between the chains. Uncorrelated
chains would be described by a Poissonian distribution, but the observed
pedestal effect, mentioned above, makes it more likely that central
collisions have more, and peripheral collisions fewer, chains. The analysis
by Sj\"ostrand and von Zijl \cite{sjostrand_zijl_1987} favors an impact
parameter dependence described by a double Gaussian distribution. It turns
out that this distribution leads to a geometric distribution in the 
number of
sub-collisions, with the tail suppressed by energy conservation.
Some predictions for mini-jet multiplicity and the pedestal effect in 
$pp$ collisions are presented in Ref.~\cite{sjostrand_zijl_1987}, assuming
such a geometric distribution for the number of chains in a single $pp$ 
event. Further work is in progress, and
it would be very interesting to test these ideas, not only in $pp$ or
$p\bar{p}$ collisions, but also in $ep$ scattering, varying $Q^2$ from 
the DIS region to photoproduction.

\subsection{Spin dependent unintegrated parton distributions}
\label{sec:spin}
The basic, universal quantities which describe the inclusive cross-sections   
of hard processes within the QCD improved parton model are the scale 
dependent   
parton distributions.  These parton distributions or distribution amplitudes  
describe how the momentum of the nucleon is distributed among its 
constituents,  
i.e. quarks and gluons.
\par
Polarized parton distributions are   
a probabilistic measure for the distribution of the nucleon's longitudinal 
spin  
(helicity) among its constituents. More precisely, one defines polarized 
parton  
distributions as the difference of the probability density to find a parton 
$f$  
with its longitudinal spin parallel aligned minus the probability density  
to find the same parton with its longitudinal spin antiparallel aligned 
relative  
to the spin of the nucleon:  
\begin{equation}  
\Delta f = f_{\uparrow \uparrow} - f_{\uparrow \downarrow}\;.  
\end{equation}   
These parton distributions conventionally
only depend on $x$ and $Q^2$, but just as for the spin-independent case
it may be beneficial to also consider \kt -unintegrated polarized
parton distributions.
An evolution equation analogous to CCFM has been
derived for the unintegrated gluon distribution in~\cite{Maul:2001uz} and the
result is quite similar, although contrary to the unpolarized case
there is no non-Sudakov form factor since the polarized splitting
function does not have a $1/z$ pole.
\par
One can show that the principles discussed in \cite{Maul:2001uz} with slight  
modifications also apply for the case of including quarks in the 
evolution.  
Thus one arrives at a complete set of evolution equations along 
the lines  
of CCFM for the unpolarized case. The CCFM evolution for polarized gluons is called 
{ pCCFM} evolution equation~\cite{Kwiecinski:2003wp}.
\par
It can easily  be shown that in the small $x$ limit the pCCFM equation 
formulated in   
\cite{Maul:2001uz} generate the 
double\footnote{The double $\ln^2(1/x)$ terms come from 
non-ladder bremsstrahlung terms.}
$\ln^2(1/x)$ for distributions integrated   
over transverse momentum of the partons.  Their detailed structure is
however different from the  collinear QCD expectations   
\cite{Kwiecinski:1999sk,Bartels:1995iu}.
One can modify the pCCFM equations in order to incorporate those 
expectations  and make  contact with the evolution  
equations in the integrated case containing 
Altarelli-Parisi + ladder
contributions which have  been discussed in \cite{Kwiecinski:1999sk}.    
These modifications  contain the following steps~\cite{Kwiecinski:2003wp}: 
\begin{enumerate}  
\item  
In order to get the expected double logarithmic limit of the integrated
distributions  it is sufficient to replace the angular ordering constraint   
$\Theta(Q- z  |{\bf q_\prp}|)$ by   
the stronger constraint $\Theta(Q^2-zq_{\prp}^2)$ in the corresponding
evolution  equations for  integrated distributions.   
\item The argument of $\alpha_s$ will  be set equal to $q_{\prp}^2$ instead   
of $q^2 \equiv q_{\prp}^2(1-z)^2$. 
\item The non-singular parts of the splitting function(s)   
will be included in the definition of the Sudakov form-factor(s).
\item Following Ref. \cite{Kwiecinski:1999sk} we include 
the complete splitting   
functions $P_{ab}(z)$ and not only their singular parts at $z=1$ and 
constant  contributions at $z=0$.    
\item We represent the splitting functions $\Delta P_{ab}(z)$ as:\\   
$\Delta P_{ab}(z)=\Delta P_{ab}(0)+ \Delta \bar P_{ab}(z)$ where   
$\Delta \bar P_{ab}(0)=0$.  \\
Following \cite{Kwiecinski:1999sk} we shall   
multiply $\Delta P_{ab}(0)$ and $\Delta \bar P_{ab}(z)$ by   
$\Theta(Q^2-zq_{\prp}^2)$ and $\Theta(Q^2-q_{\prp}^2)$ respectively in 
the integrands   
of the corresponding integral equations. Following    
the terminology of Ref. \cite{Kwiecinski:1999sk} we call the corresponding 
contributions   
to the evolution kernels    
the 'ladder' and 'Altarelli-Parisi' contributions respectively.       
\item We shall 'unfold'  the eikonal  form factors in order to treat    
real emission and virtual correction terms on equal footing.   
\end{enumerate}  
After those modifications one arrives at an evolution equation for the  
unintegrated polarized parton distributions which includes the complete LO 
Altarelli-Parisi  and the double $\ln^2(1/x)$   
effect generated by ladder diagrams in   
a consistent way, i.e. if one integrates the evolved unintegrated parton 
distributions  
over the transverse momentum \kt the result will be the same as if 
one had done  
an evolution with the integrated parton distributions using 
'Altarelli-Parisi+ladder' evolution equation.  
It means that the corresponding diagram between evolution and transverse
 momentum integration  
commutes. 
\par
One can utilize the fact that the pCCFM equation can be (partially) 
diagonalized   
 by the Fourier-Bessel transform.     
 It turns out that the difference between the integrated  
and the unintegrated evolution equation in Fourier-space is only a single 
factor  
$J_0(b_\prp q_\prp (1-z))$, where $b_\prp$ is the transverse impact 
parameter conjugate to the   
transverse momentum of the parton, $q_\prp$ the   
transverse evolution scale, $z$ the momentum fraction and $J_0$ the Bessel 
function of order 0.  
The evolution equation for the integrated case is simply restored by setting
 $b_\prp =0$.
\par
There is a third contribution to the evolution of unintegrated polarized parton 
distributions which is  
not covered by the 'Altarelli-Parisi + ladder' approximation of the modified 
pCCFM equation,   
these are the non-ladder bremsstrahlung  
contributions. 
A general method of implementing the non-ladder bremsstrahlung
corrections 
into the double logarithmic resummation was proposed by Kirschner and 
Lipatov   
\cite{Kirschner:di,Kirschner:1994vc}. For unintegrated polarized 
parton distributions  
 they have been implemented in Ref.~\cite{Kwiecinski:1999sk}. In the 
 unintegrated  
case one can simply add them by analogy  
to the Altarelli-Parisi  and ladder contribution by inserting the 
factor  $J_0(b_\prp q_\prp (1-z))$  
because then again one obtains perfect commutativity of the diagram 
between the integrated and unintegrated  
parton distributions. 
\par
Putting all three contributions together (Altarelli-Parisi, 
ladder + non-ladder) one obtains a set of   
linear  
integral equations for unintegrated polarized quark and gluon distributions. 
\begin{figure}[tb]  
\resizebox{0.5\textwidth}{!}{\includegraphics{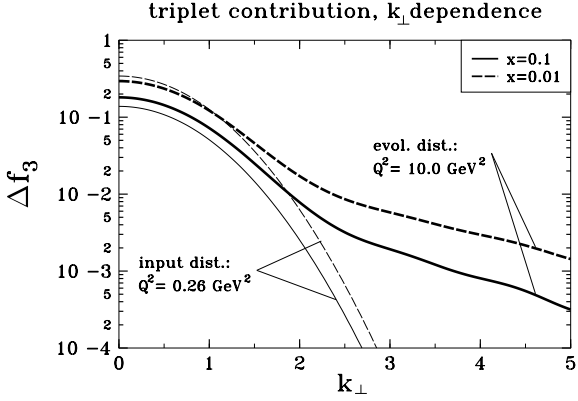}} 
\caption{{\it Transverse momentum dependence  
for the triplet contribution   
$\Delta f_3$ = $\frac{1}{6} \left(\Delta u + \Delta \bar u-\Delta d - 
\Delta \bar d \right)$  
using pCCFM evolution 
 including non-ladder contributions.   
The thin lines show the input distributions  
for $Q^2_0 = 0.26\; {\rm GeV}^2$ for $x=0.1$ (solid) and $x=0.01$ (dashed), 
while the bold  
lines show the same distribution evolved to $Q^2= 10\;{\rm GeV}^2$. 
}  }
\label{ccfmk}  
\end{figure}  
In Fig.~\ref{ccfmk}  
we show the evolution of the \kt dependence for  the triplet 
contribution : 
\begin{equation}
\Delta f_3 = \frac{1}{6} \left(\Delta u + \Delta \bar u-\Delta d - 
\Delta \bar d \right). 
\end{equation} 
The input distributions at $Q^2_0 = 0.26 \;{\rm GeV}^2$ are taken 
from \cite{Gluck:2000dy,GRV98} and are  
compared to the evolved distributions  
at $Q^2_0 = 10.0 \;{\rm GeV}^2$. The width of the initial transverse 
momentum dependence   
$\sigma$ has been chosen to be 1 GeV. For the simulation   
the Altarelli-Parisi, ladder and non-ladder contributions all have been 
included.  
It is seen that due to the evolution the \kt dependence  
is broadening away from a Gaussian behavior to a more exponential 
decay.   
  
\subsection{$J/\psi$-production and polarization effects}
In the following
we consider $J/\psi$ meson production in $ep$ deep inelastic scattering
in the color singlet model using \kt-factorization.   
It should be noted 
that heavy quark and quarkonium cross section calculations 
within the collinear factorization  
in  fixed order pQCD show 
a  large discrepancy (by more than an order of 
magnitude)~\cite{zotov26a,zotov26b,zotov27a,zotov27b} 
to measurements at the Tevatron for 
hadroproduction of $J/\psi$  and $\Upsilon$ mesons.
This fact has resulted in intensive theoretical investigations of such 
processes. In particular, it was suggested to add an additional transition 
mechanism from $c\bar c$-pairs to $J/\psi$ mesons, the so-called color 
octet (CO) model~\cite{zotov28a}, 
where a $c\bar c$-pair is produced in a color octet state and 
transforms into the final color singlet (CS) state by the help of very soft 
gluon radiation. 
The CO model is based on the general principle of the non-relativistic
QCD factorization (see \cite{zotov28a}). The physics behind this factorization
is simple: the heavy quark-antiquark meson is produced at distances
which are not so short as the distances for heavy quark-antiquark
production (which are of the of $1/2m_Q$, where $m_Q$ is the mass
of heavy quark). Indeed, we can easily estimate that the typical
distances for e.g. $J/\psi$ production is about $1/\alpha_s(m_Q)m_Q$.
These distances are much longer than the distances of the typical hard
process but they are still much shorter than the hadronization distances.
Therefore $J/\psi$ production is still under control of perturbative QCD
but on the other hand it could be accompanied by a highly non-perturbative
production of soft gluons. 
By adding the contribution from the CO model and  fitting the free parameters
one was able to describe the existing data on the production of $J/\psi$
production at the Tevatron.
However, in recent years, we have seen a lot of difficulties of the
CO model. The first and the most disturbing is the fact that the fit with
the CO model gives values of wave functions at the origin which are in
contradiction
with the non-relativistic (NR) 
QCD hierarchy where each non-perturbative CO matrix element
has a definite order of magnitude in the relative heavy quark velocity.
The qualitative prediction for the CO model is the transverse polarization
of the produced $J/\psi$ since the main contribution of the CO model to
$J/\psi$ production in $p\bar p$-collisions comes from gluon-gluon fusion
with transverse polarized gluons.
 The second important question is about the NR QCD factorization itself.
Is the heavy quark mass really large enough to have well separated
scales, $1/2m_Q$ and $1/2\alpha_s(m_Q)m_Q$, or is a special selection
needed as suggested in 
Ref.\cite{likhoded}
The CO model has been applied earlier in the analysis of inelastic 
$J/\psi$ production~\cite{zotov33,zotov34} at HERA. However, 
as noted in Ref.~\cite{zotov34}, the results from Ref.~\cite{zotov33} and 
\cite{zotov34}  do not agree with each other.
Also
 the  results 
obtained within the usual collinear approach and the CS 
model~\cite{zotov35,zotov36,zotov37,zotov38} 
underestimate the experimental data by factor about two.
\par
First attempts to investigate the $J/\psi$ polarization problem
in $ep$-interactions at HERA  and in $p\bar{p}$-interactions
at the Tevatron were made in \cite{zotov24aa,zotov24a,zotov24b,zotov39,zotov40}
using the \kt-factorization approach. An extensive analysis of the
production of $J/\psi$, $\chi_c$ and $\Upsilon$ mesons (including
the polarization properties) in $p\bar{p}$-collisions has
been recently presented in \cite{Baranov:2002cf}.
\par
The matrix element for DIS electro-production of $J/\psi$ mesons has been
calculated in~\cite{zotov_jpsi}, keeping the full $Q^2$ dependence as well as
the full polarization state of the $J/\psi$ meson.
For studying $J/\psi$ meson polarization properties we calculate 
the $\ptpsi$- and $Q^2$-dependences of the spin alignment parameter 
$\alpha$~[19, 61]:
\begin{equation}
\alpha (\omega) = {d\sigma/d\omega - 3\,d\sigma_L/d\omega\over d\sigma/d\omega + d\sigma_L/d\omega},
\end{equation}
\noindent
where $\sigma $ ($\sigma_L$) is the production cross section for 
(longitudinally polarized) 
$J/\psi$ mesons and with 
$\ptpsi$
 or $Q^2$ substituting 
 $\omega$.
The parameter $\alpha$ controls the angular distribution for leptons
in the decay $J/\psi \to \mu^{+}\,\mu^{-}$ (in the $J/\psi$ meson 
rest frame):
\begin{equation}
{d\Gamma(J/\psi \to \mu^{+}\,\mu^{-})\over d\cos\theta} \sim 1 + \alpha\,\cos^2\theta.
\end{equation}
\noindent
\begin{figure}[tb]
\begin{center}
\resizebox{0.5\textwidth}{!}{\includegraphics{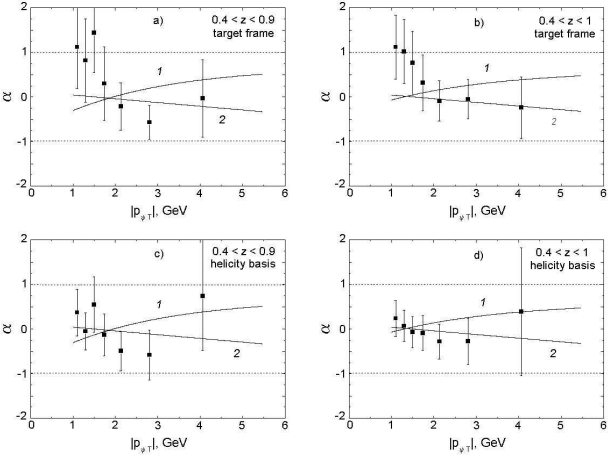}} 
 \caption{
 {\it The spin alignment parameter $\alpha(\ptpsi)$, which is
calculated in the region $0.4 < z < 0.9$ (a,c) and in the
region $0.4 < z < 1$ (b,d) at $\sqrt s = 314\,{\rm GeV}$, 
$m_c = 1.4\,{\rm GeV}$ and $\Lambda_{{\rm QCD}} = 250\,{\rm MeV}$. 
Curves {\sl 1} corresponds to  calculations in the collinear approach 
at  leading order with the GRV gluon density,
curves {\sl 2} corresponds to the \kt-factorization calculations
with the {\bf JB} unintegrated gluon distribution.
}}
\label{zotov-jpsi}
\end{center}
\end{figure}
Fig.~\ref{zotov-jpsi} shows the spin alignment parameter  
$\alpha(\ptpsi)$
calculated in the region $0.4 < z < 0.9$ $(a)$ and  $0.4 < z < 1$ $(b)$,
with $z=E_{\psi}/E_{\gamma}$ in the $p$-rest frame,  
in comparison with experimental data taken by the 
ZEUS collaboration at HERA~\cite{Chekanov:2002at}. 
Curves {\sl 1} corresponds to 
calculations in the collinear approach at 
leading order using the GRV gluon density, curves {\sl 2} corresponds 
to the \kt-factorization calculations
with the BFKL {\bf JB }~\cite{Bluemlein} unintegrated gluon distribution 
with $Q_0^2 = 1\, {\rm GeV}^2$.
\par
We note that it is impossible to make definite conclusions about
the \kt-factorization approach considering the polarized $J/\psi$
production cross section because of the large uncertainties in the 
experimental data. 
The large additional contribution from the initial 
longitudinal polarization of virtual photons weakens the effect 
of initial gluon off-shellness even more.
However at low $Q^2 < 1\,{\rm GeV}^2$ such contributions are negligible.
This fact should result in observable spin effects of final $J/\psi$ mesons.
As an example, we have performed calculations for the spin alignment 
parameter $\alpha$ as a 
function of $\ptpsi^2$ at fixed values of $Q^2$ for
$40 \,{\rm GeV} \le W \le 180 \,{\rm GeV}$, 
$z > 0.2$, $M_X \ge 10\,{\rm GeV}$ at HERA.

\begin{figure}[tb]
\begin{center}
\resizebox{0.5\textwidth}{!}{\includegraphics{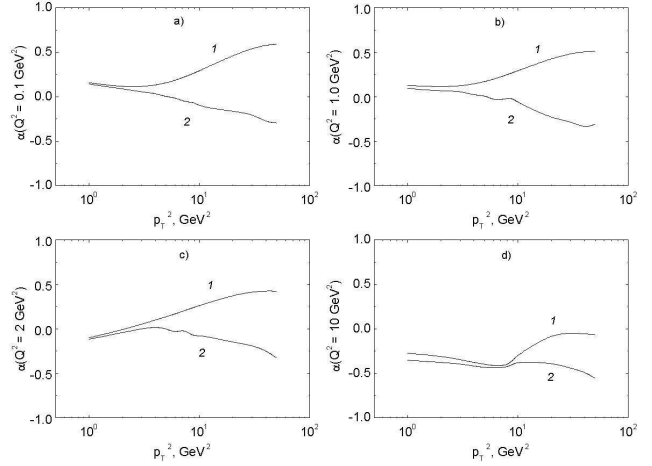}} 
 \caption{{\it The spin alignment parameter $\alpha(\ptpsi^2)$
at fixed values of $Q^2$ for
$40 \,{\rm GeV} \le W \le 180 \,{\rm GeV}$, 
$z > 0.2$, $M_X \ge 10\,{\rm GeV}$ at $\sqrt s = 314\,{\rm GeV}$, 
$m_c = 1.4\,{\rm GeV}$ and $\Lambda_{{\rm QCD}} = 250\,{\rm MeV}$.
Curves {\sl 1} and {\sl 2} correspond to the calculations as in Fig. 12. 
}}
\label{zotov-jpsi2}
\end{center}
\end{figure}
The results of our 
calculations at fixed values of $Q^2 = 0.1,\,1,\, 5,\,10\,{\rm GeV}^2$
are shown in Fig.~\ref{zotov-jpsi2}.
Curves {\sl 1} correspond to calculations in the collinear approach at leading
order using the GRV gluon density and curves {\sl 2} correspond to the 
\kt-factorization calculations with the {\bf JB } unintegrated  
at $Q_0^2 = 1\, {\rm GeV}^2$.
We observe large differences between predictions 
of the leading order of 
the color singlet model with the GRV gluon density 
 and the \kt-factorization approach at low 
$Q^2 < 1\,{\rm GeV}^2$ (Fig.~\ref{zotov-jpsi2}).
\par
Therefore more accurate  measurements of 
polarization properties of the $J/\psi$ mesons will be an
interesting test of the \kt-factorization approach.
\par
 For the production of $J/\psi$ particles in the framework of the 
CS model the relevant partonic subprocess is
\begin{eqnarray}
\gamma + g &\to& ~^3S_1[1]+g. \label{psig}
\end{eqnarray}
When the CO $q \bar{q}$ states are allowed, there 
appear additional contributions from  the following partonic 
subprocesses:
\begin{eqnarray}
\gamma + g &\to& ~^1S_0[8]+g,\;~^3S_1[8]+g,\;~^3P_J[8]+g. \label{1S0g}
\end{eqnarray}
The CO matrix elements responsible for the non-perturbative       
transitions in eq.(\ref{psig}) are related to the fictitious      
CO wave functions, that are used in calculations in place of     
the ordinary CS wave functions.
\begin{figure}[thb]
\begin{center}
\resizebox{0.48\textwidth}{!}{\includegraphics[bb=25 90 539 717]{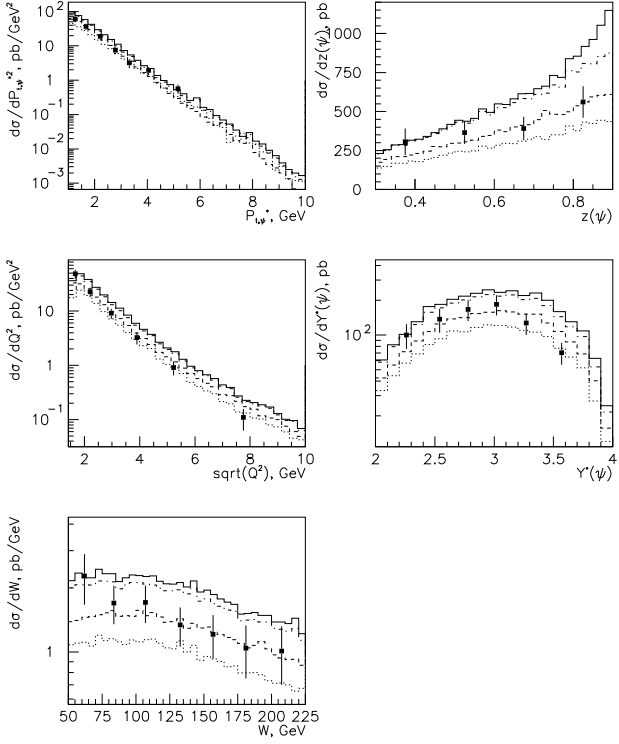}} 
\caption{{\it A comparison between the theoretical predictions and
experimental data \protect\cite{newH1jpsi} for inelastic $J/\psi$ production.
Dash-dotted histogram, the CS contribution
with {\bf JB} gluon density and $\as(\kt^2)$;
dashed histogram, the same
with ``derivative of GRV" and $\as(\kt^2)$;
dotted histogram, the same
with {\bf JB} gluon density and $\as(m_{\psi \prp}^2)$;
solid histogram, the sum of the CS and CO
contributions, with {\bf JB} gluon density, $\as(\kt^2)$.}}
\label{new-nikolai}
\end{center}
\end{figure}
In Fig.~\ref{new-nikolai} we present a comparison between our theoretical
calculations and experimental data collected by the H1 collaboration at 
HERA~\cite{newH1jpsi} in the kinematic range $2$ GeV$^2<Q^2<100$ GeV$^2$,
50 GeV $<W<$ 225 GeV, $0.3<z<0.9$, $p_{T,\psi}^{*\,2}>1$ GeV$^2$.
\par
The effect of the different evolution equations 
(BFKL {\bf JB }
 or DGLAP ``derivative of GRV", for a detailed description 
 see~\cite{smallx_2001}) which govern the
evolution of gluon densities is found to be as large as a factor of 2 in 
the production cross section. This is illustrated by a comparison of 
dash-dotted and dashed histograms in Fig.~\ref{new-nikolai}. 
A similar effect is connected with the renormalization scale $\mu_r^2$ in 
the running coupling constant $\as(\mu_r^2)$. The calculations made 
with $\mu_r^2=\kt^2$ and $\mu_r^2=m_{\psi \prp}^2$ are represented by the 
dash-dotted and dotted histograms in Fig.~\ref{new-nikolai}. 
Note that the setting 
$\mu_r^2=\kt^2$ is only possible in the  \kt-factorization approach.
In this case, $\alpha_s(k_\prp)$ was fixed at $\alpha_s$=1 if the formal
running value was greater than 1, and it was set to zero if
$k_\prp < \Lambda_{QCD}$.
The contributions from the $2\to 1$ CO subprocesses are cut away
by the experimental restriction $z<0.9$.
Turning to the $2\to 2$ CO contributions, one has to take care
about the infrared instability of the relevant matrix elements. In
order to restrict the $2\to 2$ subprocesses to the perturbative domain, we
introduce the regularization parameter $q^2_{\mbox{reg}}$.
The numerical results shown in Fig.~\ref{new-nikolai}  
are obtained with setting
$q^2_{\mbox{reg}}=1$~GeV$^2$ and $m_c = m_{\psi}/2 = 1.55$ GeV.
\par
The results shown in Fig.~\ref{new-nikolai} are 
obtained with the non-perturbative
CO matrix elements of Ref. \cite{zotov27b}. If the values extracted
from the analysis \cite{Baranov:2002cf} were used instead, the contribution
from the CO states would be a factor of 5 lower.
One can see that, irrespective to the particular choice of the 
non-perturbative matrix elements,the production of $J/\psi$ mesons at the HERA
 is reasonably described within the
color-singlet production mechanism (with \kt factorization) and the
color-octet contributions are not needed. 
 
\section{Selected topics on the current experimental status}
\label{sec:curr-exper-stat}
\begin{table*}[htb]
  \begin{center}
{\scriptsize
    \begin{tabular}{|l|l|l|l|l|l|}
      \hline
      & \multicolumn{4}{|c|}{ collinear factorization}&  $\kt-$\\
      &\multicolumn{2}{|c|}{direct } &  \multicolumn{2}{|c|}{resolved} & factorization\\
      \hline	
      & LO+PS & higher order   &   LO+PS & higher order  & LO+PS  \\ 
	&       & NLO (dijet)    &         &  NLO (dijet)   &       \\
      \hline
      \hline
      HERA observables & & & & &  \\
      \hline
      \hline
      DIS D$^*$ production &                                      & ok~\protect\cite{ZEUS_F2charm,H1_f2charm_2000} & ? & ?  
                           & ok~\protect\cite{H1_f2charm_2000,BJJPZ_2002}  \\
      photoprod. of D$^*$  & ok & ok~\protect\cite{ZEUS_dstar,MNR_2002}    & ok~\protect\cite{ZEUS_dstar} & no~\protect\cite{ZEUS_dstar}
                           & ok~\protect\cite{jung_salam_2000,baranov_zotov_1999,Charm,baranov_zotov_2000,BJJPZ_2002} \\
      \hline
      DIS B production (visible) &  ok~\protect\cite{ZEUS_bbar_dis}&  ok~\protect\cite{ZEUS_bbar_dis}& --- & ---  
                           & ok~\protect\cite{ZEUS_bbar_dis}  \\
      DIS B production (total)  &  no~\protect\cite{H1_bbar_dis} &  ok~\protect\cite{H1_bbar_dis}    & --- & ---
                           & no~\protect\cite{H1_bbar_dis}   \\
      photoprod. of B (visible) & ok~\protect\cite{H1_bbar-eps03,ZEUS_bbar} & ?  &  &  
                          & ok~\protect\cite{jung_ringberg2001,jung-hq-2001}  \\
      photoprod. of B (total)& no~\protect\cite{H1_bbar,ZEUS_bbar} &   no~\protect\cite{H1_bbar,ZEUS_bbar,MNR_2002}  & ? & ? 
                          &ok~\protect\cite{jung_ringberg2001,jung-hq-2001}  \\
     \hline
      high $Q^2$ di-jets & ? &  ok~\protect\cite{ZEUS_resgamma_dis,H1_2+1jets_data}  & ? & ?
                         & ?   \\
      low $Q^2$ di-jets (cross sec.) & ? & ok~\protect\cite{H1dijet}  & ? &  no~\protect\cite{ZEUS_resgamma_dis,H1_2+1jets_data,ZEUS_resgamma_99}
                         & ?   \\
      low $Q^2$ di-jets (azim.corr.)
	&no~\protect\cite{H1dijet}&  no~\protect\cite{H1dijet}& ok~\protect\cite{H1dijet} & ? 
	                   & ok~\protect\cite{H1dijet}  \\
				 & &{\scriptsize NLO 3-jet no~\protect\cite{H1dijet}}& & & \\ 
       photoprod. of di-jets & ? &  ok~\protect\cite{H1incphojet} & ? &  no~\protect\cite{ZEUS_resgamma_99,H1_resgamma_00}
	                   & ? \\
				     &   &                                       &   &  ok~\protect\cite{H1incphojet}
	                   &   \\
      \hline
       particle spectra &  no~\protect\cite{H1_energyflow,H1_chargedpar} & --- & ok~\protect\cite{resolvedJJK}  & --- 
                         & ok~\protect\cite{jung_salam_2000}  \\
       energy flow      &no~\protect\cite{H1_energyflow,H1_energyflow2,H1_energyflow3} & --- & ok~\protect\cite{H1_energyflow3,resolvedJJK} & ---
                        &  ?  \\
      \hline
      \hline
      HERA small-$x$ observables & & & & & \\
      \hline
      \hline
 DIS     forward jet production & no~\protect\cite{H1_fjets_data,H1fwdjet2,ZEUSfwdjet1,H1fwdjet1,ZEUSfwdjet2}
	&  no~\protect\cite{H1fwdjet1,ZEUSfwdjet2,KPfwdjet}
	& 
                ok~\protect\cite{H1fwdjet2,ZEUSfwdjet1,H1fwdjet1,ZEUSfwdjet2,JJK3} 
		    &  ok~\protect\cite{ZEUSfwdjet2,KPfwdjet}    
		    & ok~\protect\cite{jung_salam_2000} \\
 DIS     forward $\pi$ production &    no~\cite{H1fwdpi0,H1pi0new} & ? & ok \cite{H1fwdpi0,H1pi0new} & ? 
	                         & 1/2 \cite{H1pi0new} \\
      \hline
      DIS $J/\psi$ prod.     &  ?                     &                      & ? & ? 
                               & ok \cite{BZ_2003,zotov_jpsi}  \\
      photoprod. of $J/\psi$ & no~\protect\cite{H1_jpsi} &  ok
	\cite{Kraemer_1996} &   & ok \cite{Kraemer_2001} 
	                       & ok \cite{Saleev_1994,LSZ_2000,zotov_jpsi}  \\
      $J/\psi$ polarization  &                           &                     
	&   & low.stat. \cite{Kraemer_2001} 
	                       & low.stat. \cite{zotov19,zotov_jpsi}  \\
      \hline  
      \hline
      Tevatron observables & \multicolumn{2}{|c|}{direct }  &
	\multicolumn{2}{|c|}{heavy quark excitation } & \\
      \hline
      \hline
      $D$ meson  prod.       &    & 
	& no & ? 
                             & ok~\protect\cite{zotov-loennblad,jung-mpla} \\
      \hline
      $J/\psi$ prod.         &  ok \cite{zotov26a,zotov27a,zotov27b,Sanchis_pl,Sanchis_np}  &    &    &
                             &    ok \cite{zotov24a,zotov24b,j_sha}      \\
      $\chi_c$ prod.         &  ok \cite{zotov27a,zotov27b}  &    &    & 
                             &    ok \cite{zotov40,j_sha}   \\ 
      $J/\psi$ polarization  &  low.stat.\cite{Lee_2000}                        &  no & no & no
                               &  ok \cite{zotov24b,j_sha}   \\ 
      \hline 
      high-$p_\prp$  $B$ prod. & no~\protect\cite{rfield} & ok~\cite{MNR_1994} & ok~\protect\cite{rfield} & ? 
                             &
				    ok~\protect\cite{jung-hq-2001,zotov-loennblad,Hagler_bbar,LSZ_2003,BLZ_2003} \\
      $b\bar{b}$ (azim.corr.)  &    & ok~\cite{MNR_1994}  &    & 
                               & ok \cite{BLZ_2003}  \\
      \hline
      $\Upsilon$ prod.       &  ok \cite{zotov27a,zotov27b}  &    &    & 
                               & ok \cite{j_sha}   \\ 
      \hline
      high-$p_\prp$ jets at large $\vert\Delta\eta^{\star}\vert$  & no &  ? &  & ?
	                       & ?\\
      \hline
    \end{tabular}
    }
    \caption{{\it Summary of the ability of the collinear and
        \kt-factorization approaches to reproduce 
          the current measurements of some observables: $ok$
        means a satisfactory description; 
   1/2 means a not perfect but also a not too bad description, 
    or in part of the phase space an acceptable description;
            $no$ means that the 
        description is bad; and ? means that no thorough comparison has
        been made. The label {\rm NLO-dijet} means the calculation was performed in next-to-leading order for a  
	  dijet configuration available for example in the {\rm DISENT}, {\rm MEPJET} or {\rm DISASTER++} programs. 
        LO + PS is a short for LO Matrix Element + parton shower calculations as implemented 
	  in {\rm LEPTO }~\protect\cite{\LEPTO}
        or {\rm RAPGAP}~\protect\cite{\RAPGAPMC} in the collinear approach  
	  and {\rm LDC }~\protect\cite{\LDCMC} or {\rm CASCADE}~\protect\cite{\CASCADEMC} 
	  in the \kt-factorization approach.
  }}
    \label{tab:collfac}
  \end{center}
\end{table*}

In spite of the fact that QCD has been
extremely successful in describing the physics at $Q^2 \gg \Lambda_{QCD}$, 
the total
cross section in deep inelastic scattering  (DIS)
is dominated by soft and semi-hard processes which can not be
described by perturbative QCD. It is thus of fundamental importance to
provide experimental measurements which may give hints to how these
processes can be described within the QCD framework. 
\par
\begin{figure}[tb]
\vspace{-2.0cm}
\begin{center}
\resizebox{0.45\textwidth}{!}{\includegraphics{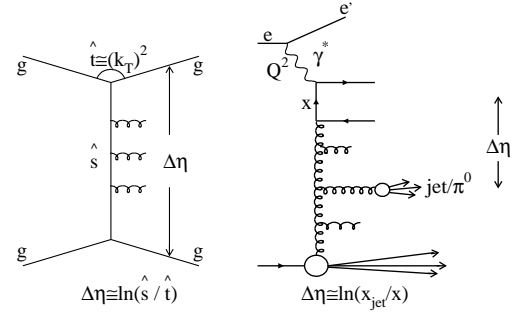}} 
\caption{{\it Kinematics of hard emissions for di-jets with large 
rapidity separation 
in hadron-hadron collisions (left) and for  forward jets/particles
at HERA. The maximal measurable jet separation at the Tevatron is about 
six rapidity units.
}}
\label{fig:deleta}
\end{center}
\end{figure}

One of the long standing questions in high energy collisions is whether 
significant deviations 
from the successful DGLAP \cite{\DGLAP}
evolution equations can be observed at the HERA and/or Tevatron colliders.
A fundamental question is where DGLAP evolution breaks down and emissions,
not ordered in virtuality, play a significant role.
In deep inelastic scattering processes (DIS) at low values of Bjorken $x$ 
it is assumed that
the struck parton results from a long cascade of parton branchings. 
Similarly, in hadron-hadron 
collision processes where two jets are separated by a large rapidity 
interval $\Delta\eta$
one expects a long partonic cascade between them (see Fig.~\ref{fig:deleta}).
\footnote{Large rapidity interval events in hadron-hadron collisions 
correspond to a subclass of DIS events at 
small $x$ characterized by
presence of the forward jet with $x_{jet}=p_{jet}/p_{proton} \gg x$ and 
transverse momentum 
$p_T \approx Q$. For such events
rapidity interval between forward jet and struck quark 
$\Delta\eta \approx ln(x_{jet}/x) \gg 1$}    
At sufficiently low values of $x$ 
(high values $\Delta\eta$)
 the DGLAP 
approximation 
should fail while the BFKL and CCFM approximations should be applicable. 
\par
Calculations of inclusive quantities like the structure function 
$F_2(x,Q^2)$ at HERA, 
performed in NLO DGLAP,  are  in very good agreement
with the measurements~\cite{H1F2,ZEUSF2}.
\footnote{The longitudinal structure function $F_L(x,Q^2)$ at small values of
$Q^2$ is an exception, see section~\protect\ref{flong}} 
However the interplay of non-perturbative (input starting distribution) 
and the perturbative (NLO DGLAP evolution) elements in
this calculation makes it impossible to decide if parton cascades with strongly ordered 
transverse momenta are the dominant mechanism
leading to scaling violations. 
\par
When exclusive quantities are investigated, 
the agreement
between NLO coefficient functions convoluted with NLO DGLAP parton densities 
and the data is 
less satisfactory, and for some processes the
DGLAP based theory fails completely. 
One example is the cross section of forward going jets at HERA, which will be
discussed below.
The question therefore is, 
for which observables the next order in the perturbative
expansion is enough, and for which a
resummation to all orders is needed.
\par
The forward jet production cross section
at small Bjorken $x$ at HERA  and the cross section for jet production
with large rapidity separation in  
high energy hadron-hadron  collisions (Tevatron) have since long been 
advertised as an ideal test of the perturbative 
dynamics~\cite{Mueller_fjets1,Mueller_fjets2,Bartels_fjets,Mueller-Nav,DelDuca}. 
More refined theoretical and phenomenological analyses have shown that
these tests are not so decisive and straightforward, however they remain
in the center of experimental activity of small $x$ physics.
Several measurements of highly energetic jets 
($x_{jet} = E_{jet}/E_p \gg x$) at 
large pseudo-rapidities\footnote{The pseudorapidity $\eta_{lab}$ 
is defined as
$\eta_{lab} = - \ln \tan(\theta/2)$, with polar angle $\theta$ 
being measured with respect to 
positive $z$-axis, which is given by the proton
beam (or forward) direction} 
$\eta_{lab}$  with transverse energies squared $E_{T,jet}^2$ of the order 
of $Q^2$ 
have been made by both the H1 \cite{H1fwdjet2,H1fwdjet1} and the 
ZEUS collaborations \cite{ZEUSfwdjet1,ZEUSfwdjet2}. 
This kind of measurement, originally proposed by Mueller and Navelet
(so called {\it Mueller-Navelet 
forward jets}~\cite{Mueller_fjets1,Mueller_fjets2,Mueller-Nav}) 
is  designed such  
that DGLAP evolution in transverse
momentum space is suppressed ($E_{T,jet} \approx Q$) while BFKL evolution 
in $x$-space ($x_{jet} \gg x$) is enhanced. 
The measurements showed 
large discrepancies to
DGLAP NLO-dijet calculations at low values of Bjorken-$x$, 
which  was taken as an indication of the breakdown of the DGLAP approximation
and the onset of BFKL effects. 
However, in the NLO-dijet calculations 
the scale uncertainties are very large.
It was also shown \cite{JJK3} that a 
good description of the data can be obtained by considering
the partonic structure of the virtual photon, which is expected to be 
important for $E_{T,jet}^2 > Q^2$ and $Q^2$ not too large. In this approach
we are back to the classic DGLAP approximation, with simultaneous evolution 
in transverse momentum space from both the photon and the proton sides towards
the hard scatter and this approach should be relevant if chains with at most one
local maximum in tranverse momentum dominate.
H1 \cite{H1fwdjet1,H1pi0new} also measured single 
forward particles ($\pi^0$) as 
opposed to jets, allowing 
the study of angles closer to the proton
direction and smaller Bjorken-$x$, at the price of a stronger dependence 
on the fragmentation process. Studies of the
inclusive cross section for the production of forward particles led 
essentially to the same conclusions 
as the study of forward jets. 
\par
Studies of the transverse energy flow provide a complementary means of 
investigating QCD processes. Compared with jet and leading particle studies,
measurements of transverse energy flow are sensitive to parton 
emissions of lower transverse momentum and to the modeling of both
the perturbative QCD evolution and the soft hadronization process.
In Ref.~\cite{H1pi0new} both types of measurement 
(forward particle and energy flow) have been merged. The transverse energy 
flow measured for events with  forward $\pi^0$ reveals the range over
which the transverse momentum of the forward parton is compensated. We may 
expect that different models of parton evolution lead to
different radiation patterns.
\par
At first look the measurements of the Mueller-Navelet jets and jet azimuthal 
de-correlation at the
Tevatron should be more promising as a test of non-DGLAP dynamics 
compared to HERA due to 
higher energy and
therefore larger phase space open to gluon emissions.
 While at HERA,
the separation between the struck
quark and forward jet can reach up to about four units of rapidity, the 
measurable jet 
separation at the Tevatron is up to six units. 
Although the BFKL calculations are expected to be more reliable at high
energies, it should be kept in mind that energy-momentum conservation is not
fulfilled. Thus, effects due to the conservation of energy and momentum (consistency
constraint) will be significant.
\par
In Tab.~\ref{tab:collfac} we present a collection of various experimental 
results from HERA and the
Tevatron which relate to low $x$ physics and 
we state the result of the comparison of 
these data with NLO DGLAP 
theory and BFKL and/or CCFM evolution schemes. 
The aim of this section is to review some of the items 
in Tab.~\ref{tab:collfac} in more detail. In the
next two subsections we review measurements   
which can be well described by the DGLAP approximation
and then  discuss measurements which were especially 
designed to extract a BFKL signal i.e. where the DGLAP evolution was suppressed 
by experimental cuts.
\subsection{Where NLO DGLAP (almost) works}
\label{coll-nlo}
Several  programs exist for the numerical NLO calculation 
of jet observables at the parton level in collinear factorization. 
They are known to give comparable results. All of them
calculate the direct photon contributions to the cross sections.  Only the 
JetVip program \cite{JetVip1,JetVip2} provides the additional possibility to 
calculate
a cross section consisting of both direct and resolved photon contributions, 
however in the DIS case conceptual difficulties are encountered
\cite{Ringberg} which lead to ambiguous results.
\par
All the parton level calculations in the collinear approach presented in
this paper were performed using the DISENT program~\cite{DISENT} and 
the CTEQ6M~\cite{cteq6} set of parton distribution functions. 
The renormalization scale was set to
$\mu_r = \sum \pti{i}$ and the factorization scale $\mu_f = Q$.
\footnote{For technical reasons DISENT allows only $\mu_f = Q$ or 
$\mu_f = const.$}
\begin{figure*}[tb]
\begin{center}
\resizebox{0.9\textwidth}{!}{\includegraphics{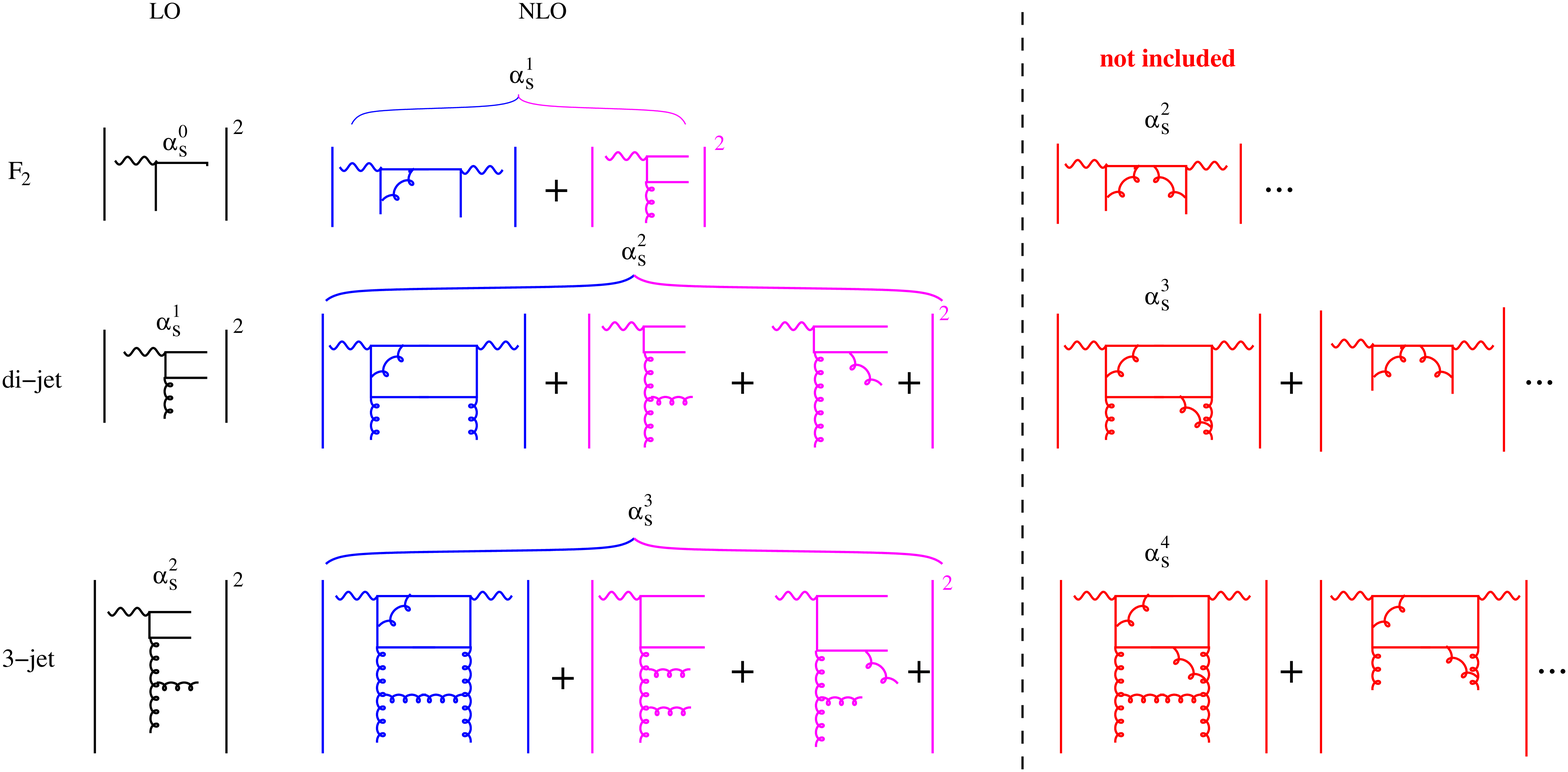}} 
\caption{{\it Schematic picture of the diagrams which are included in NLO calculations
for single-jet ($F_2$), di-jet and 3-jet processes.
 Also indicated are example of diagrams which are
not included (since of higher order in $\alpha_s$ or do not contribute to the
$n$-jet observable)
}
\label{nlofig}}
\end{center}
\end{figure*}
\par
There has been some confusion in the literature concerning the   
concept of NLO. Formally, whether a calculation is leading or
next-to-leading depends on the observable. LO is then the lowest
order in $\alpha_S$ in which the observable obtains a non-zero
value, and NLO is one order higher in $\as$. 
However, sometimes it is difficult to define the order in $\as$ 
appropriate for a specific measurement.
Therefore, in this paper we state clearly to which order
in $\as$ a process has been calculated.
In Fig.~\ref{nlofig} representative diagrams are shown for NLO
calculations of single-jet ($F_2$), di-jet and 3-jet processes. It is obvious from 
Fig.~\ref{nlofig} that NLO dijet calculations do not include all diagrams necessary for
a NNLO single-jet calculation, as indicated in the right column of Fig.~\ref{nlofig}.
\par
The calculations summarized above use integrated parton
distributions, convoluted with LO or NLO coefficient functions.
As is usual in the standard formulation of factorization, the
coefficient functions are on-shell partonic cross sections with
subtractions of the singular collinear regions. The parton
distributions are typically in the $\overline{\rm MS}$ scheme,
where the partons are integrated over \emph{all} transverse
momentum with the resulting ultra-violet divergences being
canceled by renormalization in the $\overline{\rm MS}$ scheme.
There is an evident mismatch of approximated and exact parton
kinematics in such calculations.  For the infra-red-safe jet cross
sections that are the domain of validity of the calculations, the
factorization theorem ensures that the calculations are consistent
and valid.
\par
The transverse momenta of the partons entering the hard scattering can be seen
as being generated by two sources: the intrinsic (primordial) transverse
momentum, which reflects the Fermi motion of the partons in the hadron,
typically of the order of one GeV, and the transverse momentum generated by the
QCD evolution (DGLAP or BFKL/CCFM/LDC), which can reach large values, even in
DGLAP up to the factorization scale. 
Therefore, for more exclusive components of the cross section, it is
better to use suitably defined unintegrated distributions and
off-shell parton kinematics.  For one treatment along these lines
that is specifically designed to treat NLO corrections in the
context of showering Monte-Carlo event generators, see the paper of
Collins and Zu~\cite{Collins-collfac}.   
\subsubsection{The longitudinal structure function $F_L(x,Q^2)$}
\label{flong}
\begin{figure}[tb]
\begin{center}
\vspace*{-2cm}
\resizebox{0.45\textwidth}{!}{\includegraphics{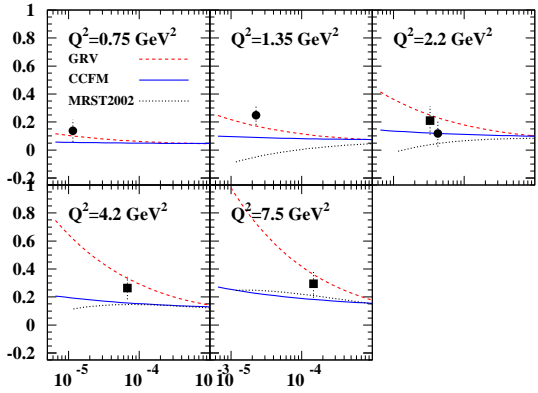}} 
\vspace*{-0.5cm}
 \caption{
 {\it The longitudinal structure function $F_L(x,Q^2)$ as measured by H1 and
 ZEUS~\protect\cite{fl-hera} compared to different calculations, including also
 the collinear DGLAP approach.
}}
\label{h1-fl}
\end{center}
\end{figure}

The longitudinal structure function $F_L(x,Q^2)$ is dominated by the gluon
density at large enough $Q^2$. 
In the limit of small $Q^2$ there is no phase space for strong ordering in virtuality
(DGLAP will not work) and  unphysically negative values for $F_L(x,Q^2)$ are
obtained in some calculations for $Q^2 < 1$~GeV$^2$~\cite{ZEUS-qcdfit}.
However in the \kt-factorization approach there is no strong ordering in virtuality and
therefore the parton evolution may generate arbitrarily small \kt 
(down to an artificial
cutoff) which means that the parameterization is valid over the full range in \kt.
In Fig.~\ref{h1-fl} the structure
function $F_L(x,Q^2)$ as measured by H1 and ZEUS~\cite{fl-hera} is compared to 
calculations using \kt-factorization in the
framework of Ref.~\cite{BKS}. 
The unintegrated gluon density was taken
from CCFM ({\it J2003 set 1}). Shown for comparison is  
another unintegrated gluon density obtained from  
the derivative of the integrated gluon density (here GRV~\cite{GRV98} is used)
and the prediction for $F_L$ obtained in the collinear DGLAP approach using the
MRST2002~\cite{MRST2002} parton densities.

\subsubsection{Single inclusive jets at HERA}
\begin{figure}[tb]
\begin{center}
\resizebox{0.52\textwidth}{!}{\includegraphics{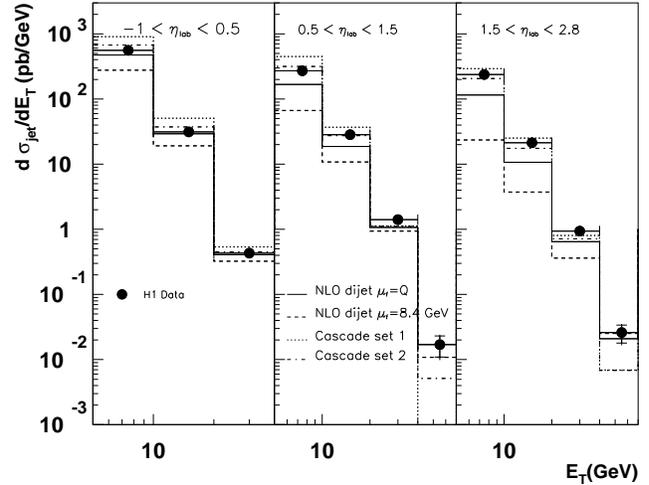}} 
 \caption{
 {\it Inclusive jet cross section ${\rm d}\sigma_{Jet}/{\rm d}E_T$ in 
 different 
  ranges of $\eta_{lab}$ ,integrated over the region 
  $5 < \qsq < 100 ~\gevsq$ and 
  $0.2 < y < 0.6$. 
  The data are compared to NLO dijet  calculation (DISENT) and 
  to the predictions from \cascade. 
}}
\label{fig1}
\end{center}
\end{figure}
Jets have been studied extensively at HERA and other colliders. 
These measurements have shown
that at sufficiently large transverse momenta and/or momentum transfers 
the NLO QCD theory based
on the DGLAP approximation is in excellent agreement with the data. To judge
 how well this approximation
works let us mention that the determination of the strong coupling constant 
$\alpha_s$ from recent 
H1~\cite{H1alpha} and ZEUS~\cite{ZEUSalpha} jet measurements are not only 
in perfect 
agreement with the world average value but are also in precision comparable to 
LEP measurements.
Another example of a jet measurement fully compatible with NLO  
theory in the collinear approach is the  measurement 
of dijet angular distributions~\cite{D0comp}
performed by the D0 Collaboration. The result of the
data-theory comparison is an exclusion limit on the quark substructure which is 
competitive with many LEP results.
In spite of this spectacular success of the QCD theory in the 
collinear approach 
one should keep in mind that there are regions of phase space, 
where the description of the data is less than
satisfactory. It is the aim of this subsection to localize these regions and 
observe patterns characteristic for a possible failure to  
describe the data by NLO theory in the collinear approximation.   
\begin{figure*}[tb]
\begin{center}
\resizebox{0.9\textwidth}{!}{\includegraphics{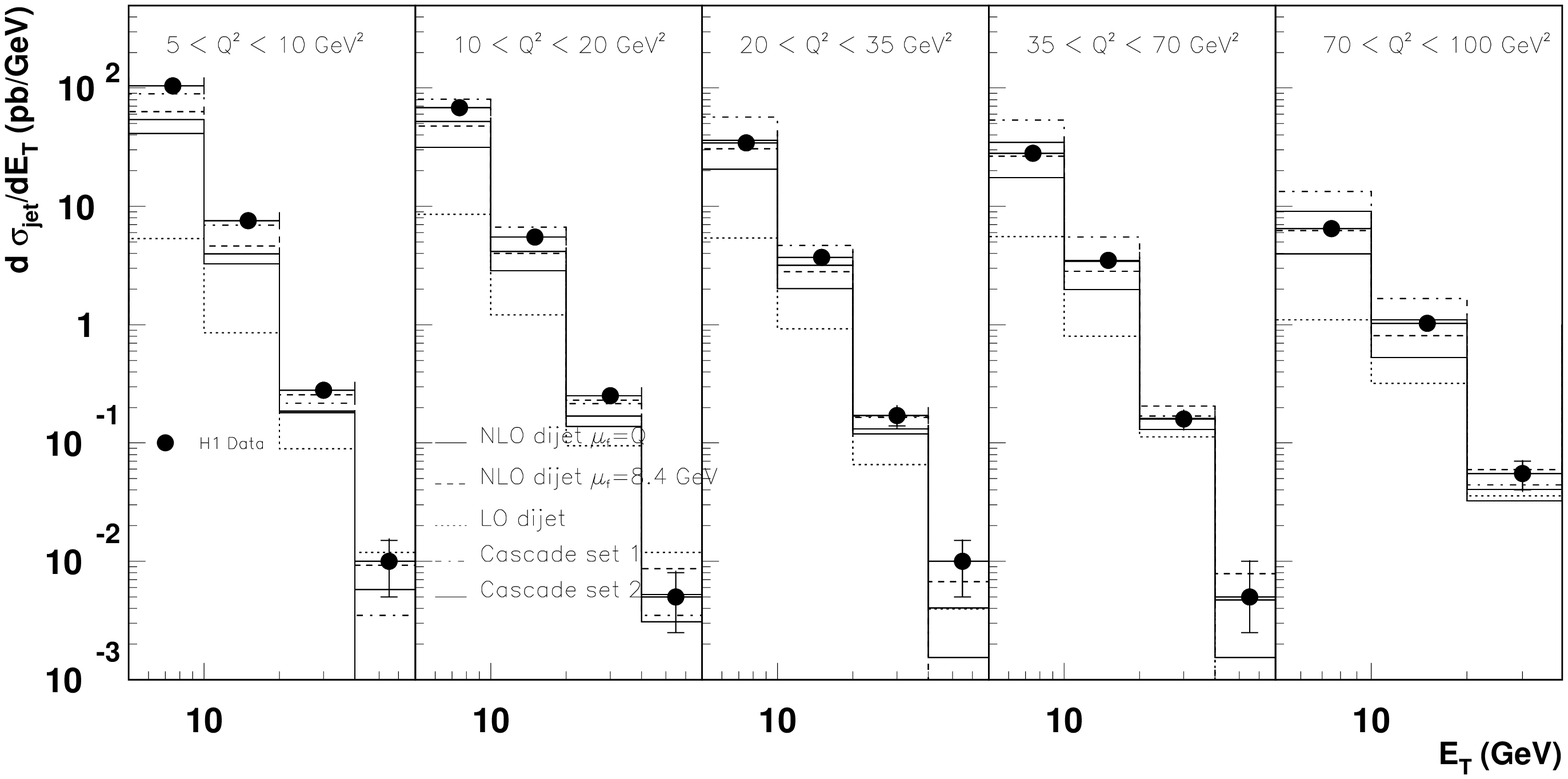}} 
\caption{{\it Inclusive jet cross section ${\rm d}\sigma_{Jet}/{\rm d}E_T$ 
for the forward
region $1.5 < \eta_{lab} < 2.8$ in different ranges of $Q^2$.
The  NLO-dijet (DISENT) and the LO dijet calculation as well as the
prediction \cascade\  are shown.}}
\label{fig2}
\end{center}
\end{figure*}
\par
In a recent H1 paper \cite{H1incjet}, NLO calculations of the inclusive jet 
cross sections, using the DISENT program \cite{DISENT}, 
 were confronted with high statistics data.
The kinematic range considered in this analysis was constrained by the 
conditions $5 < \qsq < 100 ~\gevsq$ and $0.2 < y < 0.6$ 
(the latter condition leads to a reduction of photoproduction background). 
Jets are defined using the inclusive $\kt$ cluster algorithm~\cite{\ktalgo} 
in the
Breit frame\footnote{The Breit frame is defined by $2x \vec{p}+\vec{q}=0$, 
where $x$ is Bjorken scaling variable, and $\vec{p}$ and  $\vec{q}$ 
are the proton and the virtual photon momenta, respectively} and selected 
by the requirement $E_{T,jet} >  5$~\gev. 
\par
Fig.~\ref{fig1} shows the inclusive jet cross section as a function of 
the transverse jet energy $E_{T,jet}$ in different regions of the 
pseudorapidity $\eta_{lab}$: 
in the backward region $ -1 < \eta_{lab} < 0.5$, the central region 
$ 0.5 < \eta_{lab} < 1.5$ and the forward region $ 1.5 < \eta_{lab} < 2.8$. 
The measured cross sections, which extend over four orders of magnitude, 
are compared to the calculations obtained in the collinear and 
\kt-factorization approaches, respectively. 
\par
While there is 
good agreement between the data and the NLO-dijet calculation in the 
backward region for all $E_{T,jet}$ values, discrepancies are observed 
for more
forward jets with low $E_{T,jet}$. In the lowest $E_{T,jet}$ range 
($5 < E_{T,jet} < 20$ \gev, for $\eta_{lab}> 1.5$), the assumed renormalization 
scale uncertainty
($E_{T,jet}/2 < \mu_r < 2E_{T,jet}$) does not cover the large difference 
between the data and the calculation. 
In Fig.~\ref{fig2} the forward region from Fig.~\ref{fig1} is studied in 
bins of $Q^2$, showing that 
discrepancies to NLO-dijet calculations are most significant
at  small \qsq and small $E_{T,jet}$ values. 
The factorization scale uncertainty is estimated by changing $\mu_f =
\sqrt{Q^2}$ to $\mu_f = 8.4$~GeV, which is the average jet transverse momentum.
The correlation of large NLO/LO corrections and high sensitivity to 
renormalization scale variations 
with poor agreement between data and QCD predictions
strongly suggests that the inclusion of higher order (e.g. NNLO or resolved 
photon component) terms in the QCD calculations is necessary in order to 
improve the description of the data.
\par
The predictions obtained in the \kt-factorization approach, supplemented with
the CCFM unintegrated gluon densities, as implemented in \cascade , are in
reasonable agreement with the data. Especially the forward region 
(Fig.~\ref{fig2}) is reasonably well described. The quality of the various
approaches to describe the data can be seen in Tab.~\ref{chisquared}, where we
quote the $\chi^2/ndf$, both for the NLO-dijet and the \cascade\ calculations. 
\begin{table*}
{\scriptsize
\begin{tabular}{| l | c | c | c | c | c | c | c |}
\hline 
\hline 
           & \multicolumn{2}{c|}{NLO-dijet} & \multicolumn{2}{c|}{CASCADE} & \multicolumn{3}{c|}{RAPGAP}\\
           &               &                &  \multicolumn{2}{c|}{J2003}& dir& \multicolumn{2}{c|}{dir + res} \\
parton density             & CTEQ6M     &  CTEQ6M      & set 1 &  set2 &  CTEQ6M &  \multicolumn{2}{c|}{CTEQ6M + SaS}\\  
  factorization scale $\mu_f^2 $    &  $ 70  $ GeV$^2$   & $  Q^2$ &       &   &    & $Q^2 + p_\prp^2$ & $ Q^2 + 4 p_\prp^2$\\
\hline    
$d \sigma /d E_t$ (in bins of $\eta$) 
(cf. Fig.~\protect\ref{fig1}) & 12.8  & 13.2 & 25.5   & 4.0 &23.7 & 1.3& 8.6\\ 
$d \sigma /d E_t$ (in bins of $Q^2$ for $1.5<\eta<2.8$) 
(cf. Fig.~\protect\ref{fig2})& 3.9  & 13.6 & 17.3   & 6.0 & 13.2 & 2.1 & 13.2\\ 
$d \sigma /d \Delta \eta$               
(cf. Fig.~\protect\ref{fig4})& 40.1 & 40.9&  116.8 & 37.7 & 66.9 & 22.6& 46.7\\ 
$ S = \frac{\int^{\alpha}N_{2-jet}(\Delta\phi^{*},x,Q^2)d\Delta\phi^{*}}
            {\int
		N_{2-jet}(\Delta\phi^{*},x,Q^2)d\Delta\phi^{*}} $
(cf. Fig.~\protect\ref{fig5})		
                 &  17.8 & 15.7 &   23.2 & 3.9  & 3.3 & 2.6 & 1.7 \\ \hline
forward jets H1  $p_\prp > 3.5$ GeV  
(cf. Fig.~\protect\ref{fig:jets94}$(a)$)	&  8.9  & 17.0 & 2.7    & 4.7  & 10.8 & 4.4  & 0.3  \\
forward jets H1  $p_\prp > 5$ GeV    
(cf. Fig.~\protect\ref{fig:jets94}$(b)$)&  5.7  & 11.2 & 1.9    & 2.3  & 6.7 & 2.6 & 0.7 \\
forward jets H1 prel $p_\prp > 3.5$ GeV 
(cf. Fig.~\protect\ref{fig:jets97})&  1.7  & 6.4  & 1.3    & 1.1 & 4.4  & 0.9 & 1.7 \\
forward jets ZEUS   $p_\prp > 5$ GeV 
(cf. Fig.~\protect\ref{fig:zeusfwdjet})& 28.9  & 38.4 & 19.2   & 9.5  & 27.1 & 20.1 & 16.8 \\ \hline
\hline 
\end{tabular}}
\caption{{\it Comparison of $\chi^2 /ndf$ obtained from comparing different
predictions to the data. For the NLO-dijet calculation with the DISENT program
the renormalization scale
was set to $\mu_r =\sum \kt$, 
the CTEQ6M~\protect\cite{cteq6} and SaS~\protect\cite{Sasgam} 
parton distribution functions of the proton and photon, respectively, are used.
\label{chisquared}}}
\end{table*}
\par
\begin{figure}[tb]
\begin{center}
\resizebox{0.5\textwidth}{!}{\includegraphics{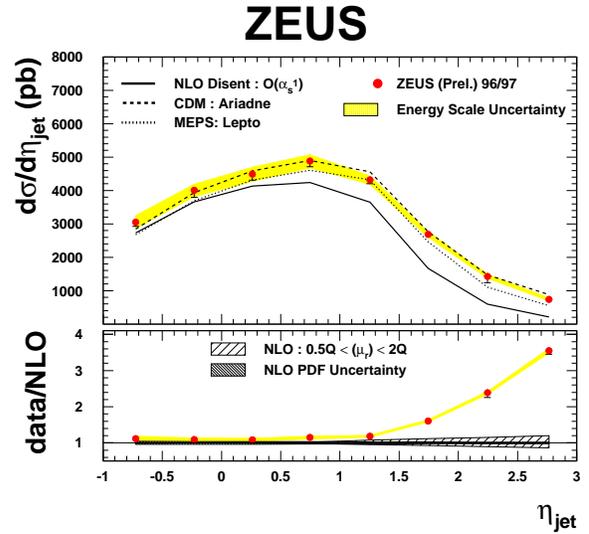}}
\vspace{-2.0cm} 
\caption{{ \it
Upper part: Measured differential inclusive jet cross section 
$d\sigma/d\eta_{jet}$ for the inclusive
phase space compared to ARIADNE (CDM), LEPTO(MEPS) and NLO Disent in order ${\cal O}(\alpha_s^1)$
Lower part: Relative difference of the measured inclusive jet cross section $d\sigma/d\eta_{jet}$ to the
NLO DISENT calculation with renormalization scale $\mu_r^2 = Q^2$  }}
\label{fig:zeusetadep}
\end{center}
\end{figure}

It is interesting to quote in the above context the recent ZEUS measurement on
inclusive jets \cite{ZEUSincjets} presented in Fig.~\ref{fig:zeusetadep}. 
The cross section
for jets reconstructed in the laboratory frame with the inclusive $\kt$ 
algorithm is compared to
a  calculation in   
NLO (here ${\cal O}(\as)$). 
When going  from small towards
large values of $\eta_{jet}$ the 
description of the data by the NLO calculation becomes worse. 
The reason for this is that  
 the $\as^0$
contribution goes to zero and the ${\cal O}(\as)$ calculation becomes
essentially the LO contribution, since    
for a fixed $Q^2$ and $x$ (or $y$),
$\eta_{jet}$ is fixed in an $\as^0$ calculation, simply given  
by $\frac{1}{2}\ln\frac{Q^2(1-y)}{4E_e^2y^2}$. The range used in the
measurement is $Q^2>25$~GeV$^2$, $y>0.04$ which, for the lowest
$Q^2$ gives a maximum $\eta_{jet}$ of about $0.8$. For larger $Q^2$
the maximum $\eta_{jet}$ is a bit larger, but the suppression of
the $\as^0$ contribution is still visible in 
Fig.~\ref{fig:zeusetadep} around
$\eta_{jet}=1$. Beyond this, the lowest order contribution is dominated by 
 $\as^1$ and the "NLO" calculation in the figure becomes
leading order. 
As can be seen from Fig.~\ref{fig:zeusetadep},  
even DGLAP type Monte Carlo models (here \lepto ) give a
rather reasonable description, if further parton radiation is included via parton
showers. Thus this comparison shows the need for higher order corrections
but not necessarily a need for any BFKL contribution.
\subsubsection{Inclusive di-jets at HERA}
\label{sec:incl-jets}
\begin{figure}[tb]
\begin{center}
\resizebox{0.3\textwidth}{!}{\includegraphics{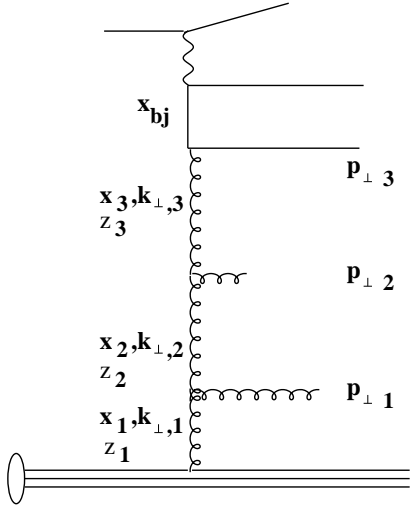}} 
\caption{{\it Generic leading order diagrams for di-jet production 
in $ep$ scattering: 
the variables
$\kti{i}$, $x_i$ and $z_i$ denote the transverse momenta, the longitudinal
energy fractions and the fractional energy in the splitting, respectively,
and  \pti{i} are the transverse momentum of the 
radiated gluons.
}}
\label{fig3}
\end{center}
\end{figure}
The measurement of di-jet production, which is less inclusive 
compared to  the
measurement described before, might provide a stronger 
test of the  NLO QCD calculations in the collinear factorization approach,
as it involves more observables.
Experimentally, possible deviations
from the DGLAP approach can best be observed by selecting events 
in a phase space 
regime, where the main assumption, the strong ordering in \kt of the 
exchanged parton cascade (Fig.~\ref{fig3}), is no longer
strictly fulfilled. This is the case at small $x$. 
The parton configurations not included
in DGLAP-based calculations 
might contribute significantly to the cross section.
Moreover, with respect to the photon-proton center-of-mass system 
(hcms), the two partons produced in a hard 
scattering process (Fig.~\ref{fig3}) are no longer balanced in 
transverse momentum. 
Events coming from calculations
beyond ${\cal O}(\as)$ will lead to a situation where the two hard jets are no
longer back-to-back. The excess of such events is expected to be higher for a BFKL
scenario compared to DGLAP, due to the possibility of hard emissions in the parton
evolution provided by the non-ordering in \kt.
\par
Di-jet production in deep inelastic $ep$ scattering was  
investigated in the region
of low $x$ ($10^{-4} < x < 10^{-2}$) and 
low  $Q^2$
($5 < Q^2 < 100 ~\gevsq$)~\cite{H1dijet}. Jets were reconstructed in the 
hcms using the \kt-algorithm.
 The minimum transverse jet energy 
$E_T^{\star}$ of 5 \gev\  was required and 
an additional requirement on the most energetic jet 
$E_{T,{\rm max}}^{\star} > 7 \gev$ (in the hcms)
was added to
avoid a scenario for which NLO-dijet predictions become 
unreliable~\cite{klasen-kramer,frixione-ridolfi}. 
\begin{figure}[tb]
\begin{center}
\resizebox{0.5\textwidth}{!}{\includegraphics{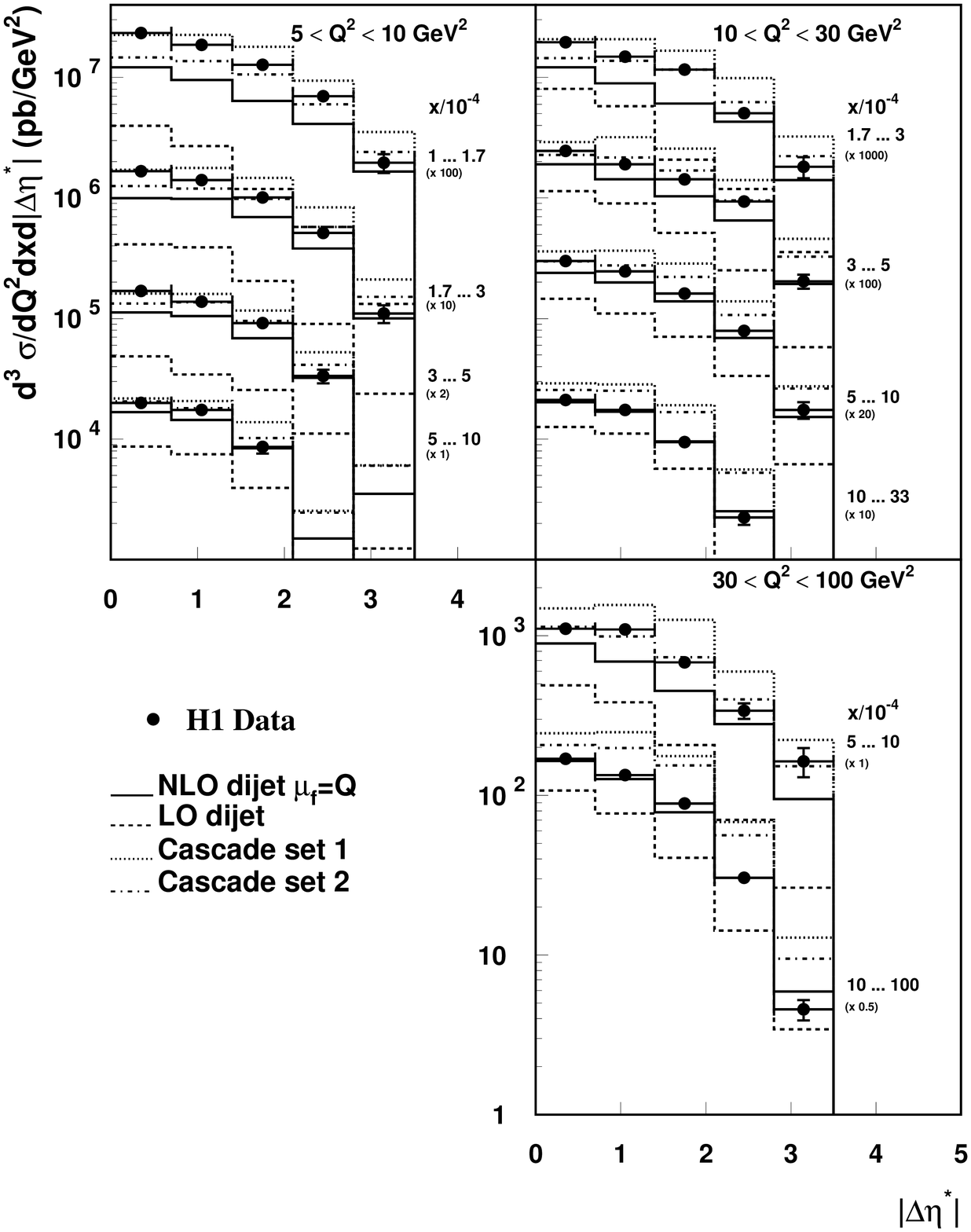}} 
\caption{{\it The triple differential inclusive di-jet cross section 
in bins of Bjorken-$x$ and
\qsq as a function of the distance \mbox{$|\Delta\eta^{*}|$} 
between the di-jets compared to 
NLO-dijet calculation (DISENT, solid line) and predictions from \cascade (dashed
and dotted line).
}}
\label{fig4}
\end{center}
\end{figure}
In Fig.~\ref{fig4} the triple differential inclusive
di-jet cross section in bins of Bjorken-$x$ and $Q^2$ as a function of 
the distance $\mid\Delta\eta^{*}\mid$
between jets is presented. The data are compared to NLO-dijet predictions. 
The NLO-dijet calculation with $\mu_f^2 = Q^2$ falls well below the data. A
better description over the full phase,
including the regime of very low $x$,
is obtained using $\mu_f^2 = 70 $ GeV$^2$.
It should be noted however, that even with $\mu_f^2 = 70 $ GeV$^2$ 
the theoretical 
uncertainty due to scale dependence of the NLO calculation is rather 
large so again no strong statement about  the DGLAP 
approximation for  dijet production at 
low $x$ can be made on the basis of the 
cross section measurement alone. 
In Fig.~\ref{fig4} the data are also compared with the predictions using the 
\kt-factorization approach in \cascade . The quality of the data description is
again quoted as $\chi^2/ndf$ for both approaches in Tab.~\ref{chisquared}.
\par
\begin{figure}[tb]
\begin{center}
\resizebox{0.5\textwidth}{!}{\includegraphics{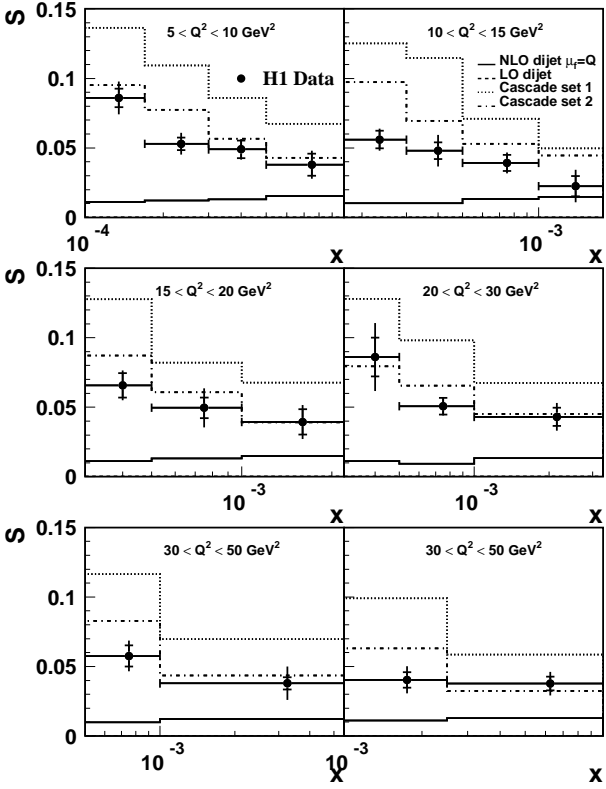}} 
\caption{{\it Ratio $S$ of the number of events with a small azimuthal 
jet separation ($\alpha < 120^{\dg}$)
 of the two most energetic jets with respect to the total number of 
 inclusive di-jet events, given as
a function of Bjorken-$x$ and $Q^2$. The data are compared to NLO-dijet
calculations (DISENT, solid line) and predictions from \cascade .
}}
\label{fig5}
\end{center}
\end{figure}

Further insight into small-$x$ dynamics may be gained from inclusive di-jet 
data by studying the
behavior of events with a small  separation
in the azimuthal angle, $\Delta\phi^{*}$, 
of the two hardest jets as
measured in the hcms, as proposed in 
\cite{Askew,Forshaw-phi,Kwiecinski-phi,Szczurek}. Partons entering the 
hard scattering process with negligible transverse momenta, as assumed in 
the DGLAP formalism, lead mainly
to back-to-back configurations of the two outgoing jets with 
$\Delta\phi^{*}=\pi$. Higher order QCD
processes lead to azimuthal jet separations different from $\pi$, 
however, the effect might be smaller than
in the case of the BFKL and CCFM evolution schemes. In the above quoted 
di-jet analysis~\cite{H1dijet} the jet azimuthal 
de-correlation was studied using a variable which has been proposed by 
Szczurek {\it et al.}~\cite{Szczurek}:
\begin{equation}
 S = \frac{\int_{0}^{\alpha}N_{2-jet}(\Delta\phi^{*},x,Q^2)d\Delta\phi^{*}}
            {\int_{0}^{180^{\dg}}N_{2-jet}(\Delta\phi^{*},x,Q^2)d\Delta\phi^{*}}, 
                0 < \alpha < 180^{\dg} 
\end{equation}
with $\alpha$ being a parameter for the $\Delta\phi^{*}$ distribution.
Its advantage in comparison with the direct $\Delta\phi^{*}$ measurement 
(see e.g. analysis of the Tevatron data, subsection~\ref{sec:exp-tevatron}) 
is its better stability against migrations. For the data presented in 
Fig.~\ref{fig5} $\alpha = 120^{\dg}$ was chosen. This choice is mainly
dictated by the limited jet energy resolution which may result in an 
incorrect choice of the two most energetic jets. 
Fig.~\ref{fig5} presents the   
$S$-distribution as a function 
of $x$ in bins of $Q^2$. 
The measured value of $S$ is of the order of 5\% and increases with 
decreasing $x$. 
This rise is most prominent in the lowest $Q^2$ bin. On the contrary, 
the NLO -dijet QCD calculation 
predicts $S$-values of order 1\%, several
standard deviations below the data, and show no rise toward small $x$.
Here the NLO calculations are performed in the on-shell limit (see discussion in
section~\ref{coll-nlo}), neglecting the transverse momentum coming from the QCD
evolution. Therefore  only the ${\cal O}(\as^2)$ part of the matrix elements
gives a significant contribution for  $\Delta\phi^{*} \neq 180^{\dg}$.
The calculation of NLO-3jet is in much better agreement with the data
(shown in ~\cite{H1dijet})
which is NLO for the $S$ variable, but it still fails to
describe the rise towards small $x$. 
Since Monte Carlo generators, like RAPGAP, include the effects of the finite
transverse momentum of the incoming partons via parton showers, it is not
surprising, that they come much closer to the data than the naive NLO
calculation ignoring the off-shellness of the incoming partons. 
This shows, that care has to be taken by applying fixed order partonic
calculations to exclusive observables.
\par
The CCFM evolution 
as implemented in CASCADE \cite{jung-dis03}
describes the data reasonably well (Fig.~\ref{fig5}), 
but this is also true 
if a resolved component of the virtual photon is added, 
provided that a rather 
large scale $\mu_r^2 = \qsq + 4\pt^2$ is chosen 
(to get large enough resolved contribution).
\par 
To conclude this section let us summarize its main points:
\begin{itemize}
\item For the inclusive jet cross section in DIS,
the NLO-di-jet description  
starts to fail when jets 
become more and more forward
\item the worsening of the description is accompanied by increasing 
theoretical uncertainty due to 
scale dependence, indicating that the NNLO terms may be more important 
in the forward direction
\item NLO-di-jet calculations describe di-jet cross section in DIS data very 
well, down to $\xbj = 10^{-4}$ in the central region of rapidity,
if $\mu_f^2=70$  GeV$^2$ is used. For $\mu_f^2=Q^2$ the description is much 
worse.
\item the largest differences between NLO theory for the inclusive jet and 
di-jet cross section
 and the data are observed in the small-$x$, small $Q^2$ region.
\item the azimuthal jet de-correlation in di-jet DIS data is 
not described by  NLO-dijet calculation, which is effectively LO for that
observable. The NLO-3jet calculation is in better agreement with the data,
but still at small $x$ is not sufficient.
The CCFM 
evolution approach is consistent with the data, but so is LO ME + DGLAP 
parton shower provided that the
resolved photon contribution is taken into account
with a scale given by $Q^2 + 4\pt^2$.
\item As in the case of cross sections, the  largest 
discrepancies for azimuthal  de-correlation are found
in the small-$x$, small $Q^2$ region. 
\end{itemize}
\subsection{Where NLO DGLAP does not work}
\subsubsection{Forward jets in DIS}
Measurements described as ``Mueller forward jets in DIS''  
\cite{Mueller_fjets1,Mueller_fjets2,Bartels_fjets} were  
especially designed to search for non-DGLAP
evolution signatures. The following conditions were required to  
suppress the DGLAP and enhance the BFKL evolution  (with
$x_{jet} = E_{jet}/E_p$):
\begin{itemize}
\item a high energetic jet with an energy fraction
$x_{jet} \gg \xbj$ to enhance
BFKL evolution $x$-space 
\item  a high enough transverse momentum $E_{T,jet}$ of the jet to 
ensure that perturbative
calculations are valid e.g.  $E_{T,jet}> 3.5 \gev$
\item $E_{T,jet} \approx Q$ in order to suppress the DGLAP evolution 
\end{itemize}
At HERA, the requirement of $x_{jet}/\xbj$ to be large results in typical 
jet angles of a few 
degrees with respect to the 
forward (proton) direction.
Due to the 
unavoidable beam-pipe hole in the detector, the acceptance is limited to jets 
with an angle larger
than, for example, $7^{\circ}$ in the H1 detector. 
At smaller angles the jets are insufficiently 
contained in the detector and the experimental separation from proton remnant 
fragments might be difficult. 
As the jet approaches more and more the forward direction its profile 
gets thicker and more 
asymmetric and a large fraction disappears down the 
beam hole.
In fact, the observation of broadening of the jet profile
leads the ZEUS collaboration to restrict the forward jet analysis to  
pseudorapidities $\eta_{jet} < 2.6$ corresponding to the limiting angle 
$\Theta_{jet} > 8.5^{\circ}$. The criterion which determines the
minimum acceptable jet angle is a satisfactory description of the jet 
profile. Obviously, the jet profile and separation of the remnant fragmentation 
depends on the jet algorithm. 
In the ZEUS analysis \cite{ZEUSfwdjet1} the cone algorithm was employed. 
In principle  an algorithm like 
the \kt-cluster algorithm, which is not based on geometry, 
should be less sensitive to detector edges,
and the separation of remnant fragments should be easier in the Breit frame. 
\par
The condition $E^2_{T,jet}/Q^2 \approx 1$ is essential to suppress the 
DGLAP evolution in direct photon 
interactions. Due to limited statistics a compromise has to be found. 
In practice an interval is
defined around this central value of $E_{T,jet}^2/Q^2$.
The requirement $E_{T,jet}\approx Q$  leads to another
experimental challenge: to reconstruct jets of the 
smallest possibly  transverse momentum, 
forward jets at smallest possibly $Q^2$ (but still in perturbative region) 
and hence smallest possible $x$ are required. 
\par 
It should be noted, that at HERA energies the above cuts restrict 
the phase space not only for DGLAP
but for any type of evolution. At HERA the range between the hard 
scattering and the forward jet
covers about  4 rapidity units, limiting the 
number of hard emissions to about 3 -- 4. 
Therefore, it may be  
that there is not enough phase space for a BFKL-DGLAP discrimination. 
In Fig.~\ref{fig:deleta} (right) a typical
Feynman diagram for forward jets and particles is shown. 
\par
\begin{figure}[tb]
\begin{center}
\resizebox{0.52\textwidth}{!}{\includegraphics{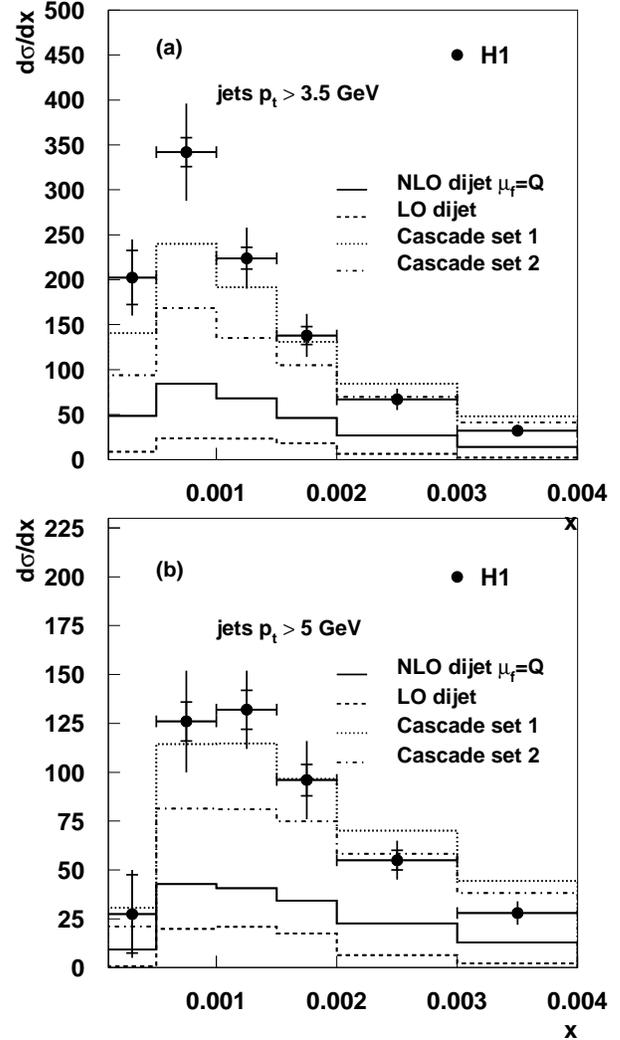}} 
\caption{{\it The cross section for 
forward jet production~\protect\cite{H1fwdjet1}
at the hadron level, as a function of $x$ for
$(a)$ $E_{T,jet}> 3.5~\gev$ and $(b)$ $E_{T,jet}> 5~\gev$.
 Also shown are the predictions from 
LO and NLO dijet calculations as well as predictions from \cascade .}}
\label{fig:jets94}
\end{center}
\end{figure}

In Tab.~\ref{tab:fjets} we present cuts
 used in the H1~\cite{H1fwdjet1} and 
ZEUS~\cite{ZEUSfwdjet1} forward jet analyses, performed  with  
the cone algorithm. In spite of small differences of the selected 
phase space, it is clear that cross sections at $E_{Tjet} > 5 \gev$ are 
compatible 
(see Fig. ~\ref{fig:jets94}$b$ and ~\ref{fig:zeusfwdjet}).
\begin{center}
\begin{table}
\begin{tabular}{|l|c|c|} \hline \hline
                                   &  H1 cuts           &  ZEUS cuts     
					     \\ \hline
  {$E'_{\sf e}$}          &  $ > 11$~GeV       &  $ > 10$~GeV   \\
  {$y_{\sf e}$}           &  $ > 0.1$          &  $ > 0.1$      \\
  {$E_{\sf{T,jet}}$}      &  $ > 3.5~(5)$~GeV     &  $ > 5$~GeV  \\
  {$\eta_{\sf{jet}}$}     &  $ 1.7$~--~$2.8 $ & $ < 2.6$     \\
  {$E_{\sf{T,jet}}/Q^2$}  &  $ 0.5$~--~$2 $    &  $ 0.5$~--~$ 2 $ \\
  {$x_{\sf{jet}}$}        &  $ > 0.035 $          &  $ > 0.036 $  \\ 
  {$p_{\sf{z,jet}}^{\sf{Breit}}$}  &       &  $ > 0$ 
  \\ \hline 
  {$x$}  & $0.0001$~--~$0.004$ & $0.00045$~--~$0.045$ \\ \hline \hline
\end{tabular}
\caption{{\it
Summary of the different selection criteria used to define 
{\it Mueller forward jets} in the H1~\protect\cite{H1fwdjet1}  and 
ZEUS~\protect\cite{ZEUSfwdjet1} measurements. }}
\label{tab:fjets}
\end{table}
\end{center}
\par
\begin{figure}[tb]
\begin{center}
\resizebox{0.52\textwidth}{!}{\includegraphics{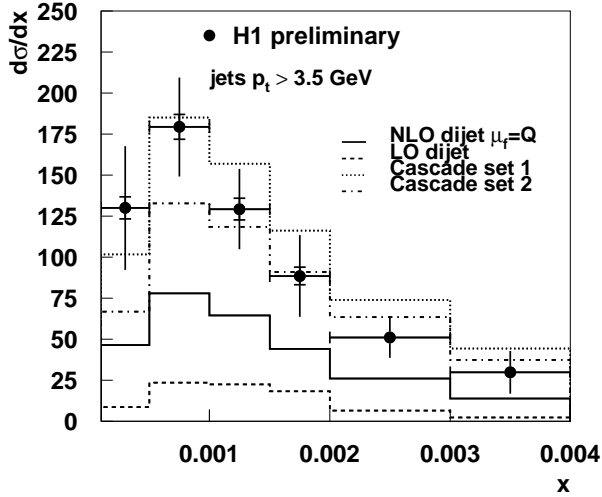}} 
\caption{{\it The cross section for 
forward jet production\protect\cite{H1fwdjet2}
at the hadron level, as a function of $x$ for
$E_{T,jet}> 3.5~\gev$. Also shown are the predictions from 
LO and NLO dijet calculations as well as predictions from \cascade .}}
\label{fig:jets97}
\end{center}
\end{figure}
\begin{figure}[tb]
\begin{center}
\resizebox{0.52\textwidth}{!}{\includegraphics{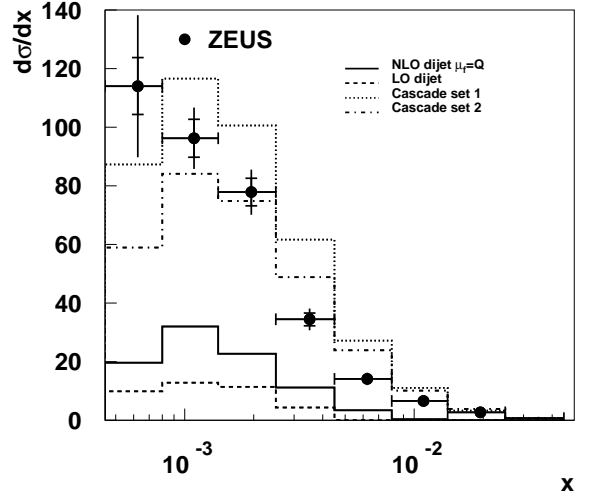}} 
\caption{{\it Forward jet cross section~\protect\cite{ZEUSfwdjet1}
as a function of $x$ 
in the kinematic region 
given in Table 2. Shown are LO and NLO dijet calculations together with
predictions from \cascade .}}
\label{fig:zeusfwdjet}
\end{center}
\end{figure}

Recently the H1 Collaboration performed a new measurement of 
the forward jet cross section \cite{H1fwdjet2} 
using much higher statistics  
and applying cuts almost identical to
those applied in~\cite{H1fwdjet1} 
($5 < \qsq < 75 ~\gevsq$, $E_{T,jet} > 3.5 ~\gev$, $7^{\circ} < 
\theta_{jet} < 20^{\circ}$, $x_{jet} = E_{jet}/E_p > 0.035$, 
$0.5 < E_{T,jet}^2/\qsq < 2$). In this analysis the jets were reconstructed
using the inclusive \kt-algorithm.
The cross section for forward jet production~\cite{H1fwdjet1,H1fwdjet2} 
as a function of $x$ is 
shown in Figs.~\ref{fig:jets94}, \ref{fig:jets97} and \ref{fig:zeusfwdjet}.
The measurements are up to a factor of two larger than the 
prediction based on 
${\cal O}(\as)$ (and also ${\cal O}(\as^2)$) QCD calculations 
in the DGLAP approach. 
Such parton level calculations are 
compared in  Figs.~\ref{fig:jets94}, \ref{fig:jets97} and  
\ref{fig:zeusfwdjet} with the measurement.
Also shown is a 
 comparison with different unintegrated gluon densities implemented in 
\cascade ~\cite{\CASCADEMC}, which shows the sensitivity of 
the predicted forward jet cross section on 
the details of the unintegrated gluon density.
It is interesting to note that also including the non-singular terms in the
CCFM splitting function ({\it J2003 set 2}) leads 
to reasonable agreement with the measurements in Figs.~\ref{fig:jets97} and  
\ref{fig:zeusfwdjet}. 
For all other distributions e.g.  ${\rm d}\sigma/{\rm d}p_{\prp\; jet}$ the 
pattern of
agreement/disagreement is similar.
The level of agreement with the measured cross section of forward jet production
is given in Tab.~\ref{chisquared}. 
\begin{figure}[tb]
\begin{center}
\resizebox{0.45\textwidth}{!}{\includegraphics{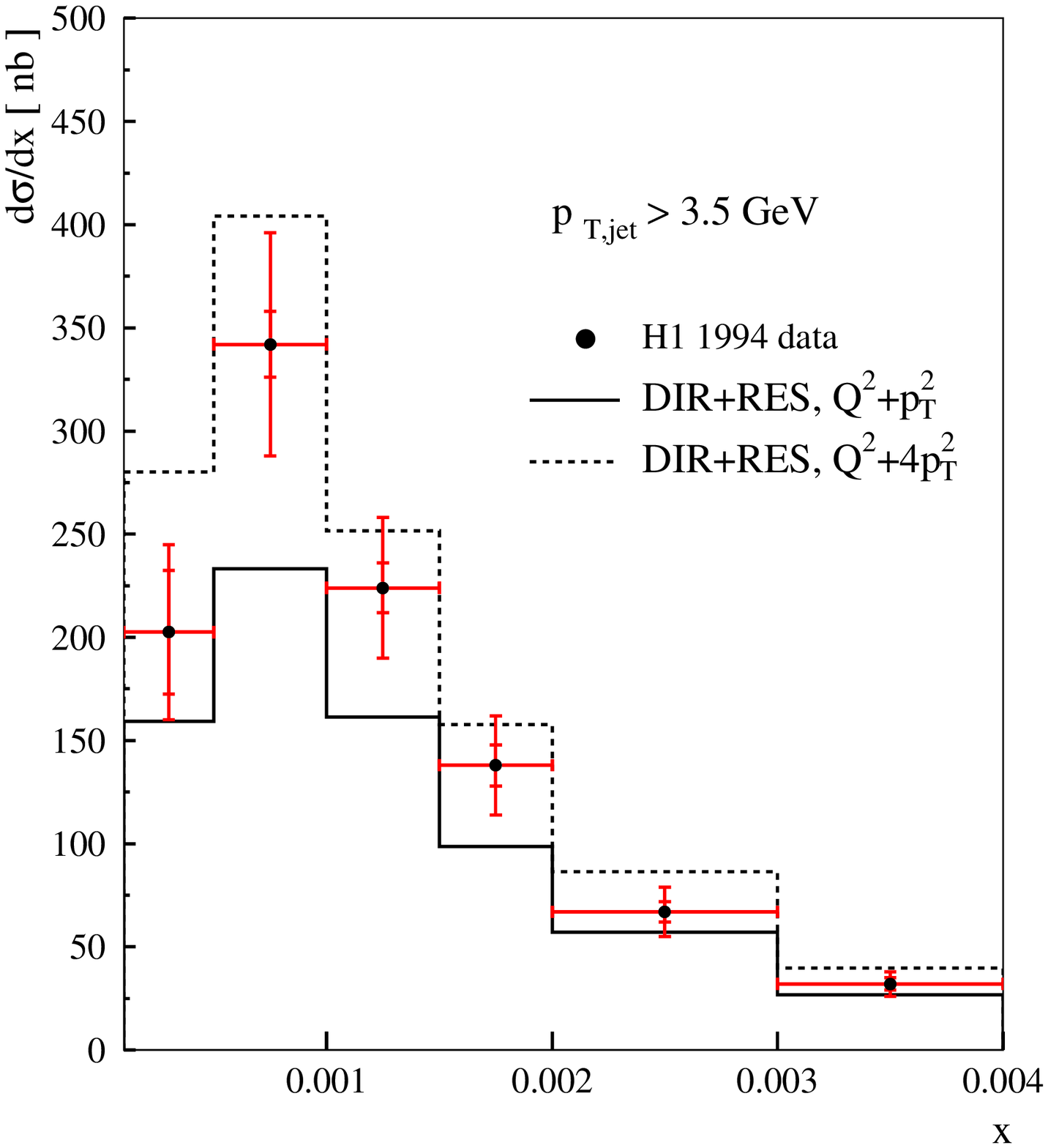}} 
\caption{{\it The predictions of the RAPGAP Monte Carlo (dir+res) at two
different scales compared to the data of \protect\cite{H1fwdjet1} 
(cone jet algorithm)}}
\label{fig:jets94rap}
\end{center}
\end{figure}
\par
The forward jet cross section was also studied
using the  BFKL formalism \cite{\BFKL}. In particular 
Kwieci\'nski, Martin and Outhwaite (KMO) in ref. \cite{Martin} used a 
modified LO BFKL equation, 
supplemented by a consistency constraint which
mimics higher orders of the perturbative expansion, to describe the 
inclusive forward jet cross section. 
The KMO  model describes the data well, however the predicted cross
section is very sensitive to the input parameters in particular 
to the infrared 
cut-off and the scale of $\as$. Thus   
the model has rather large uncertainties in the normalization of the
cross section, whereas the shape of the 
distribution in Bjorken-$x$ is expected to be more stable. 
We will come back to the KMO calculation in the next section.
\begin{figure}[tb]
\begin{center}
\resizebox{0.45\textwidth}{!}{\includegraphics{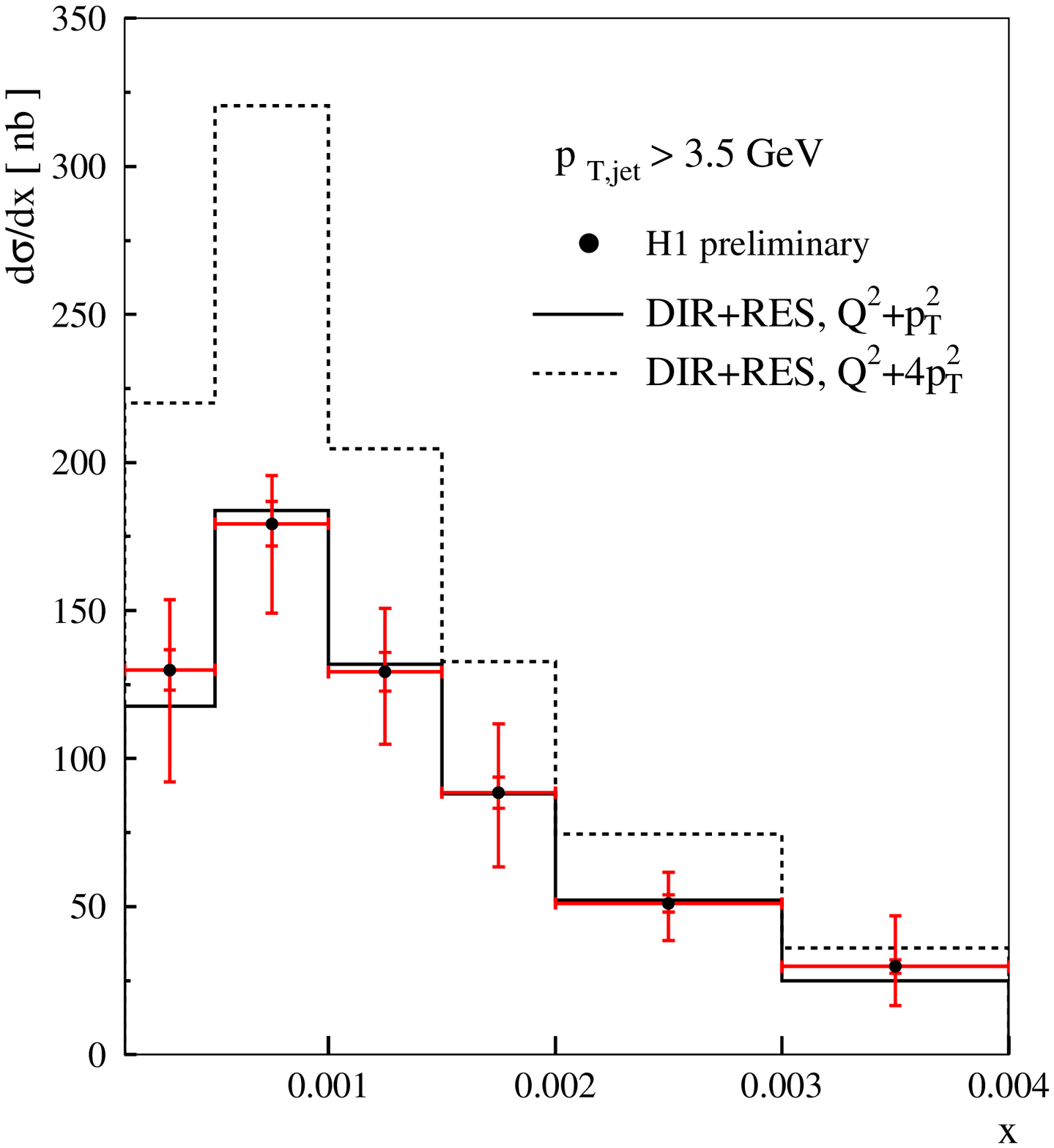}} 
\caption{{\it The predictions of the RAPGAP Monte Carlo (dir+res) at two
different scales compared to the data of \protect\cite{H1fwdjet2}
($\kt$ jet algorithm).}}
\label{fig:jets97rap}
\end{center}
\end{figure}
\par
The choice of the jet algorithm has quite an effect on the measured cross section, 
as we can see 
comparing Fig.~\ref{fig:jets97rap} (data from ~\cite{H1fwdjet2}, 
\kt\ jet algorithm in Breit frame) 
and Fig.~\ref{fig:jets94rap} ~(data from \cite{H1_fjets_data}, cone jet algorithm
in laboratory frame). 
The cross sections  come out 
different at hadron level due to the choice of the jet algorithm. 
\par
Including a contribution from resolved virtual photons (as done in 
\rapgap\  res \cite{\RAPGAPMC})
leads also to a reasonable description of the forward jet data.  
It should be noted 
however, that the predictions of the model are sensitive to the 
renormalization and factorization scales. The \rapgap\ 
package allows a choice of renormalization and factorization scale, 
and in Figs.~\ref{fig:jets94rap} and~\ref{fig:jets97rap} the predictions are  presented 
for two different choices,  $\mu^2 = Q^2 + \pt^2 $ 
and $\mu^2 = Q^2 + 4\pt^2 $, where 
$\pt$ is  the transverse momentum of the partons taking part in the 
hard scattering process. The errors 
(mainly systematic) are large and reasonable agreement with the  
data would still be achieved for a  scale of $Q^2 + 4\pt^2 $. Note, 
that for a correct description of
the azimuthal de-correlation by \rapgap\ the same 
large scale has to be employed 
(see subsection~\ref{sec:fwd-pi}).  
However it seems that the two different forward jet 
measurements prefer different choices of the scales:
in Fig. ~\ref{fig:jets97rap}
the forward jet data are well described  with a
renormalization scale
$\mu_r^2=Q^2+\pt^2$ while the  forward jet data of Fig.~\ref{fig:jets94rap}
lie between  the predictions using
$\mu_r^2=Q^2+\pt^2$  and $\mu_r^2=Q^2+4\pt^2$.
Both calculations use \rapgap\  and 
the same (CTEQ6M and SaS1d) proton and 
photon PDF's. 
\par
\begin{figure}[tb]
\begin{center}
\resizebox{0.4\textwidth}{!}{\includegraphics{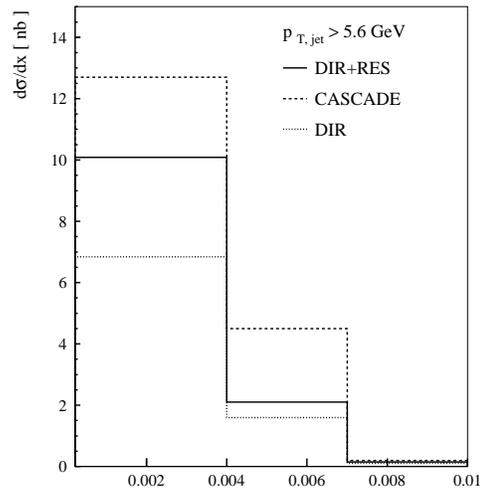}} 
\vspace{-0.5cm}
 \caption{
 {\it Forward jet cross section as a function of Bjorken $x$ in new cut 
 scenario designed for high statistics data :
$Q^2 > 16 \gevsq, p_{Tjet} > 5 \gev,  x_{jet} > 0.05, 0.1 < y < 0.7$. 
 \label{newscen}}}
\end{center}
\end{figure}

Before coming to the end of this subsection,  let us comment 
on possibilities of a new type forward jets measurements, which open with 
the advent 
of high statistics data.  
Obviously we can go to higher $Q^2$ and higher $x_{jet}$ so that the ratio 
$x_{jet}/\xbj$ 
would remain large. An example of such a scenario is 
$Q^2 > 16 ~\gevsq, p_{\prp \;jet} > 6 \gev,  x_{jet} > 0.05, 
0.1 < y < 0.7$. The cross section calculated using RAPGAP and CASCADE 
is shown in  Fig.~\ref{newscen}. 
It is 3-5 times lower than measurement with cuts presented in 
Tab.~\ref{tab:fjets}, but is
certainly measurable at the level of 100 $\rm{pb}^{-1}$ and  has several 
advantages.
The jets with higher $\pt$ and higher $x_{jet}$ are cleaner, we can 
expect smaller systematic 
errors due to the uncertainty of the calorimeter scale and detector corrections.
Furthermore in the region of higher $Q^2$ the resolved component
of the photon is suppressed, therefore the ambiguity between CASCADE-like and 
RAPGAP-like descriptions may vanish.
Another possibility was considered  by Kwieci\'nski {\it et al.}~\cite{KLM} 
who studied  deep inelastic
events containing a forward photon as a probe of small-$x$ dynamics. 
The great advantage is that  such a measurement is no longer
dependent on the hadronization mechanism. At an integrated 
luminosity of around 
1~$\rm{fb}^{-1}$ we can expect
about 300 BFKL-like events within the following phase space cuts:
$20 < Q^2 < 30, k_{\gamma \prp}^2 > 5$ \gevsq, $\Theta_{\gamma} > 5^{\circ}$ 
where $k_{\gamma \prp}$ and $\Theta_{\gamma}$  are the transverse momenta 
and the angle of the forward photon, respectively.
\footnote{It is necessary to impose an isolation cut on the photon  to
suppress background from $\pi^0$'s produced within outgoing quark jet.
Experimentally one requires that within isolation cone around photon the
energy deposit is below few percent of the photon energy.}
The DGLAP theory prediction is about 3.5 times 
lower. This process 
seems to be measurable at HERA~2, however background from $\pi^0$'s 
may turn out to be large.

\subsubsection{Forward $\pi^0$ mesons}
\label{sec:fwd-pi}
\begin{figure}[tb]
\begin{center}
\resizebox{0.5\textwidth}{!}{\includegraphics{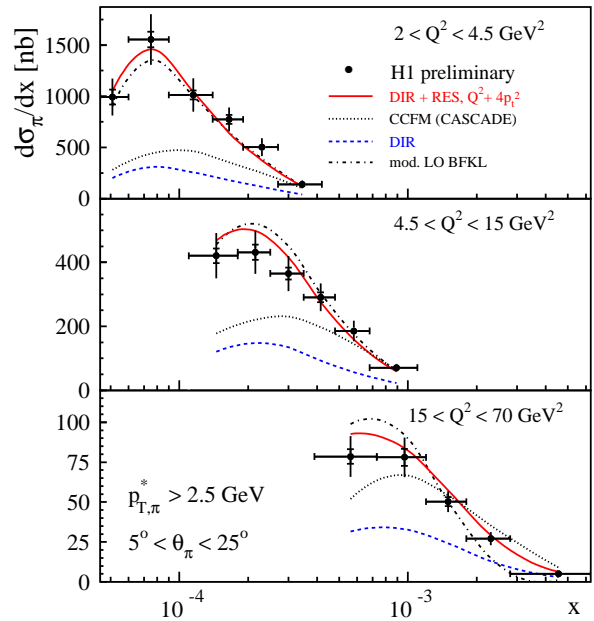}} 
\caption{{\it The cross section for forward $\pi^0$ production  
as a function of $x$ for
$p_{T,\pi^0}^{\star} > 2.5 ~\gev$.~Also shown are the predictions 
from various Monte Carlo calculations.}}
\label{fig9}
\end{center}
\end{figure}
H1 recently measured single forward $\pi^0$ meson 
production~\cite{H1pi0new}. This new measurement 
triples the number of $\pi^0$'s in comparison  to  
previously published data \cite{H1fwdjet1},  allowing the measurement 
of more differential distributions and additional final state observables. 
The analysis is restricted
to the kinematic range 
$2 < \qsq < 70 ~\gevsq$, $5^{\circ} < \theta_{\pi} < 25^{\circ}$, 
$x_{\pi}=E_{\pi}/E_p > 0.01$ (lab. system) and 
$p_{T,\pi}^{\star} > 2.5$~GeV 
(hcms). The differential cross section as a 
function of Bjorken-$x$ for 
different regions in \qsq is shown in Fig.~\ref{fig9}. It should
be noted that this measurement covers a range in $x$ down to $4\cdot 10^{-5}$. 
The prediction
of a DGLAP based Monte Carlo (\rapgap\ dir) is well below the data, 
whereas a reasonable description is obtained when the resolved virtual photon
contribution is added (\rapgap\ res). It should be also noted, 
that a rather large factorization and 
renormalization scale $\qsq + 4\pt^2$ has to be used in 
this case. Surprisingly, \cascade\  
(all sets, but only { JS2001} is shown)
falls  below the data at small $x$ values. 
The fact that \cascade , which provides good description of the forward jet
production, fails to describe the forward $\pi^0$ production at small $x$ is
interesting in itself. It may indicate that quark initiated cascades and
final state cascades (gluon splitting into quark pairs), both missing in
present \cascade\  generator code, play important role in the forward $\pi^0$
production.  In \rapgap\ both processes contribute significantly to the
forward $\pi^0$ cross section, influencing both the scale for 
string fragmentation
(string invariant mass) and the string composition (quark vs gluon
fragmentation). The final effect is such that \rapgap\ is able to produce
significantly more forward $\pi^0$'s.  It is interesting to note, that the
parton to hadron fragmentation usually viewed as a complication of the
partonic picture of deep inelastic collisions, here may serve also as the
indicator of the underlying parton dynamics.
It should be
stressed however, that there is no direct contradiction in the data: 
discrepancies in the $\pi^0$ cross section arise in the region of $x$ 
which is mostly beyond the reach 
of the forward jet measurement. It is interesting to note that the 
previously mentioned BFKL calculation of the forward jet cross section 
\cite{Martin} is consistent
with   the forward $\pi^0$ mesons measurement (dashed-dotted curve).
 The BFKL prediction for the $\pi^0$ cross section was obtained by the 
convolution of the parton distribution of KMO~\cite{Martin}
with the KKP (Kniehl, Kramer, Pötter)
fragmentation function \cite{pi0frag}. 
\par
\begin{figure}[tb]
\begin{center}
\resizebox{0.5\textwidth}{!}{\includegraphics{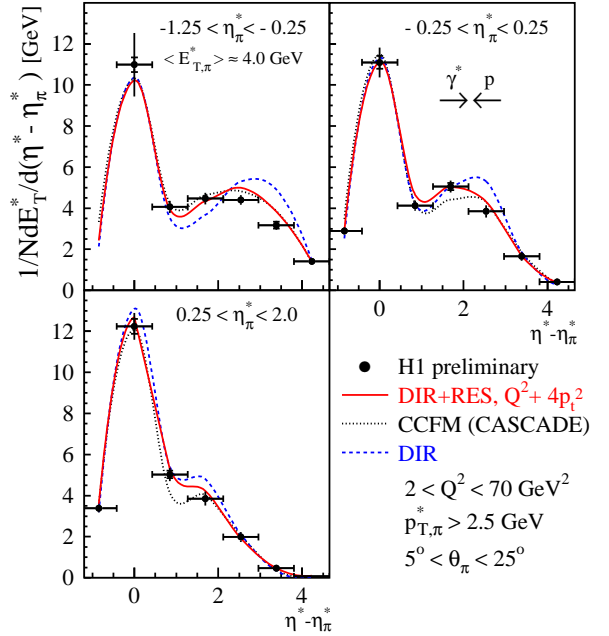}} 
\caption{{\it The distribution of transverse energy as a function of 
pseudorapidity difference in different intervals of $\pi^0$ pseudorapidity. 
Predictions
of three QCD-based models are shown.}}
\label{fig10}
\end{center}
\end{figure}

We expect  that different initial state cascade dynamics 
should lead to different 
radiation patterns and therefore to a different transverse energy flow. 
The transverse energy flow in the hadronic center of mass system, 
$\frac{1}{N}{\rm d}E_T^{\star}/{\rm d}(\eta^{\star}-\eta_{\pi}^{\star})$, 
in events
containing at least one forward $\pi^0$  is presented in Fig.~\ref{fig10}
where $\eta_{\pi}^{\star}$ gives the pseudorapidity of the pion (in the hcms). 
The spectra are presented in three intervals
of the $\pi^0$ pseudorapidity: $-1.25 < \eta_{\pi}^{\star} < -0.25 $ 
(closest to proton direction) , $-0.25 < \eta_{\pi}^{\star} < 0.25 $ 
and $0.25 < \eta_{\pi}^{\star} < 2.0$ (farthest from proton direction).
\par
The QCD-based approaches all describe the transverse energy flow around the 
$\pi^0$ but give different predictions in the {\it current} region. 
The resolved
photon picture gives a reasonable description of the spectra whilst the CCFM 
approach overestimates the transverse energy flow when the forward $\pi^0$
is closest to the proton direction (top left). The direct photon model gives 
the worst description of the data. It predicts a transverse energy flow which 
rises
strongly with increasing $\Delta \eta^{\star}$ and shows a peak at large 
values of pseudorapidity difference. This effect becomes less pronounced 
with increasing
pseudorapidity of the forward $\pi^0$. The differences between the models 
can be qualitatively understood as a consequence of the 
 ordering criteria of the parton cascade 
implemented in various Monte Carlo generators. 
\subsubsection{Di-jet production at the Tevatron}
\label{sec:exp-tevatron}
The jet production data at high energy hadron-hadron colliders can also
be used to test parton evolution dynamics.
The production of exactly two jets is described at LO by 
an $\as^2$ calculation
as being back-to-back in azimuthal angle and having their
transverse momenta balanced. 
Higher order processes
involve the
radiation of additional partons, which will upset this correlation and
additional soft radiation in higher order processes will decrease the
correlation further, leading to a smearing of the $\Delta\phi$-distribution.
Perturbative QCD has been successful in describing di-jet production up to
next-to-leading order, whereas higher order contributions have to be
accounted for by parton shower models. Since the production mechanism may
involve more than one hard interaction scale, a different treatment of the
parton radiation, such as BFKL, might be needed. 
\begin{figure}[htb]
\begin{center}
\begin{minipage}{0.5\textwidth}
\resizebox{0.9\textwidth}{!}{\includegraphics{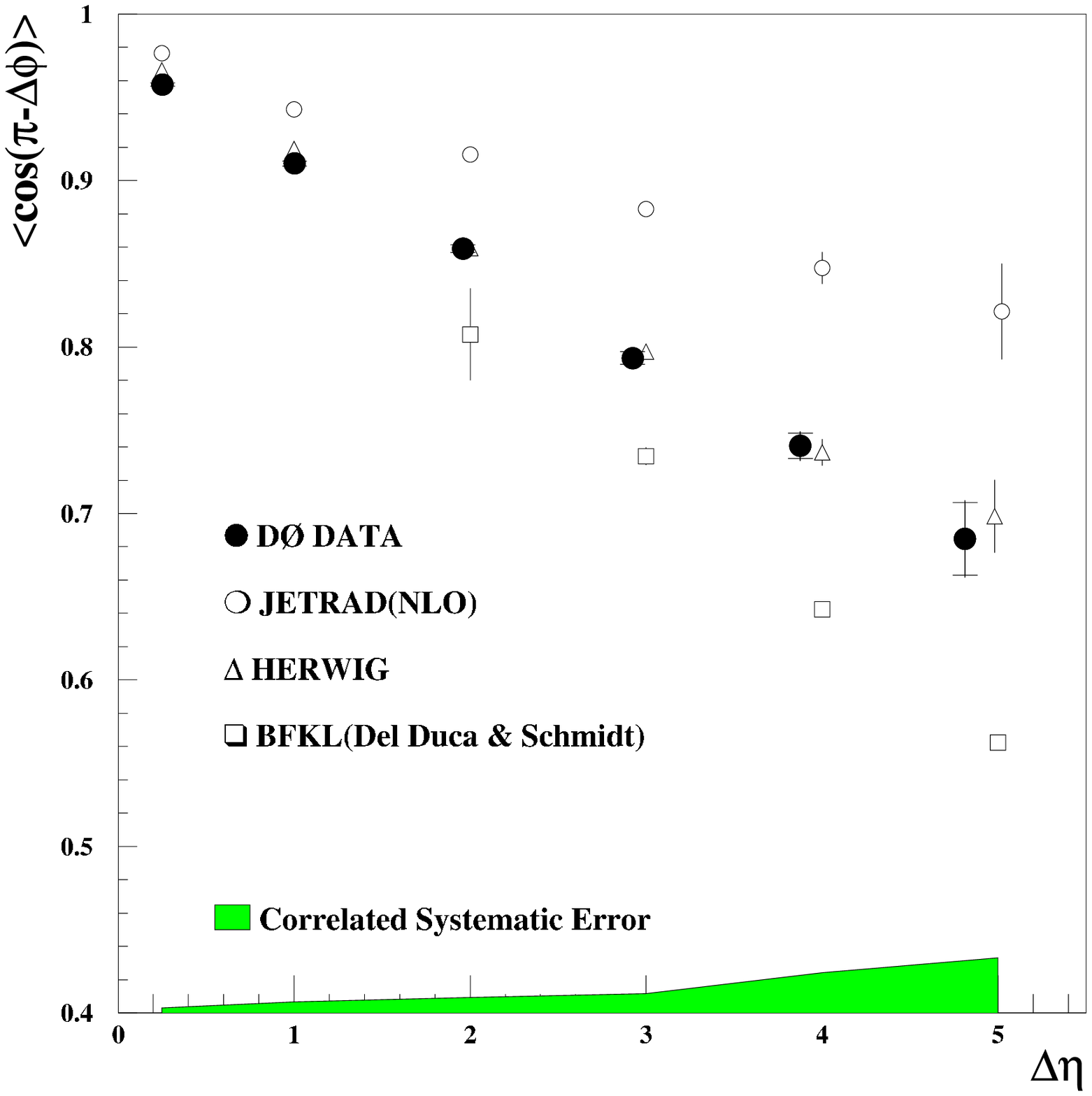}} \\
\resizebox{0.9\textwidth}{!}{\includegraphics{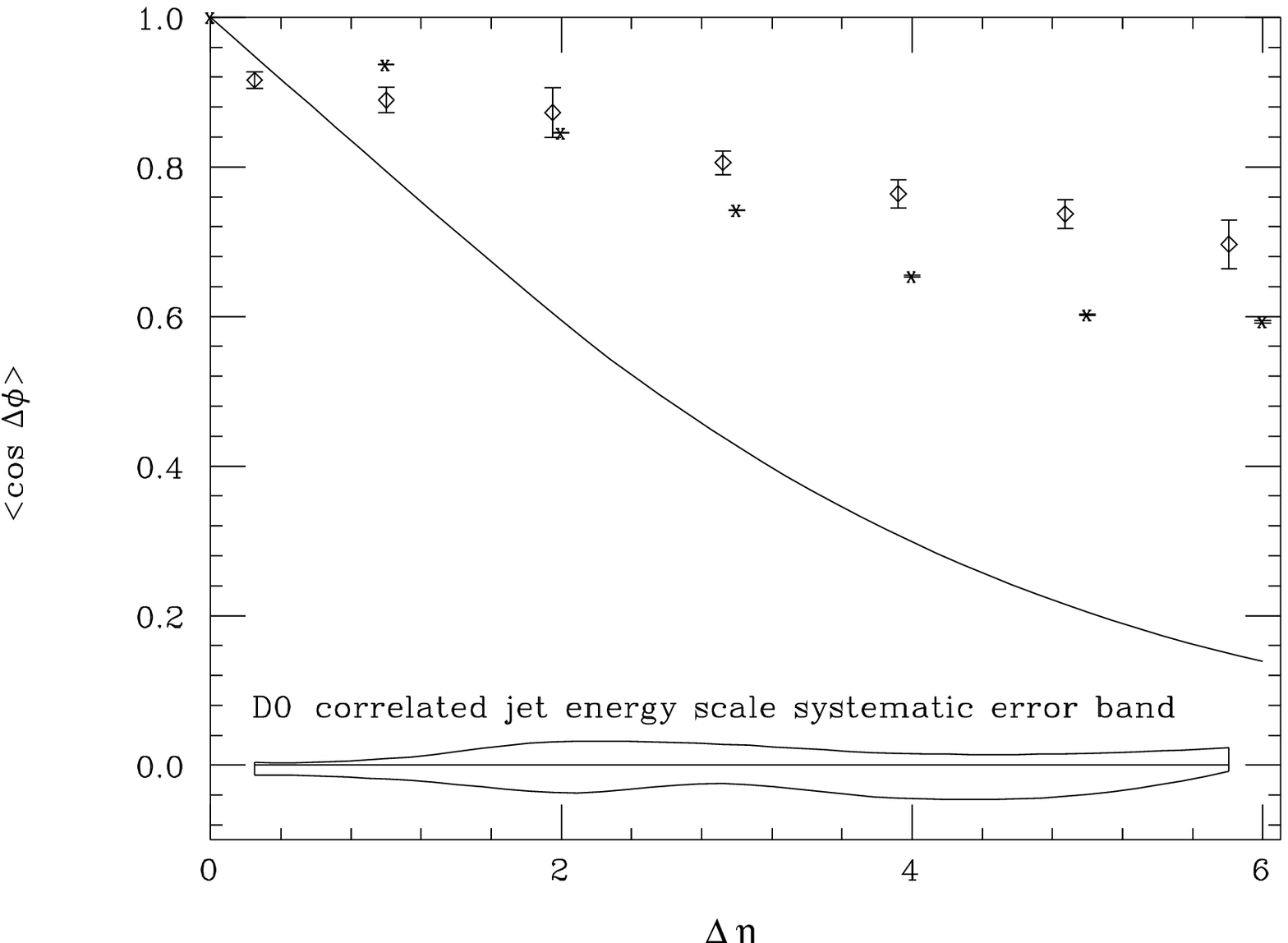}} 
\end{minipage}
\caption{{\it The average di-jet azimuthal correlation 
$cos(\pi-\Delta\phi)$ as a function $\Delta\eta$. 
$a$: Comparison with NLO, HERWIG and a BFKL calculation \protect\cite{DelDuca} 
taken from \protect\cite{D0azim}.
$b$: Comparison  with the BFKL calculation  satisfying energy/momentum
conservation \protect\cite{Orr:1997im}.
}}
\label{fig27}
\end{center}
\end{figure}
\par
The D0 experiment has studied events in which two jets, widely separated in
rapidity, have been identified. Due to their uniform calorimetric coverage
of $\pm$ 4 units in rapidity this experiment is well suited for such an
investigation. Jets were defined using a 
cone algorithm and  $E_{t\,jet}>20$ GeV. In a multi jet
event the two jets mostly separated were chosen for the analyses provided
one of them had $E_t>50$ GeV, this in order to avoid any trigger inefficiency.
\par
If $\langle\cos(\pi-\Delta\phi)\rangle$ is plotted as a function of 
the rapidity separation between the
observed jets, then one would expect to observe a decrease in this
variable as the rapidity separation increases, 
simply because the phase space for
additional radiation increases. As shown in Fig.~\ref{fig27}  
the D0 experiment \cite{D0azim} also observes a linear
decrease with the pseudorapidity interval, well described by the HERWIG
Monte Carlo. The JETRAD Monte Carlo,
which provides a NLO-dijet
calculation, predicts less de-correlation at large rapidity gaps. The BFKL
calculation by \cite{DelDuca}, valid for large $s$, on the other hand gives much
larger de-correlation effects,
although we note that in this analysis,
large effects from the constraint of energy and
momentum conservation have been ignored in the BFKL
evolution. In fact, if these are taken into account as
describe in Section~\ref{sec:momentum} the
BFKL prediction is in much better agreement with
data~\cite{Orr:1997im}.
\par 
Recently  the D0 measurements  
of Mueller-Navelet jets at the 
Tevatron  have been discussed in detail 
by Andersen et al. \cite{Andersen:2001kt}, therefore we restrict ourselves to 
quoting their main conclusions:
\begin{itemize}
\item Definitions of the momentum fractions used by D0 and some of the 
acceptance cuts imposed 
spoil the correctness of the procedure to extract the 
effective BFKL intercept from the data. 
Especially the implemented cut on the
maximum allowed transverse momentum of jets invalidates
a BFKL analysis based on the asymptotic behavior of
the BFKL prediction. Such a cut will of course always
be implicitly implemented by the constraint in energy
at a given collider, necessitating a detailed analysis
as described in section~\ref{sec:momentum}.
\item As the cuts on the transverse momenta of trigger jets 
were chosen equal, the fixed NLO QCD
calculations of both the total dijet rates and the
azimuthal de--correlations are plagued with large
logarithms of perturbative, non--BFKL origin.
\end{itemize}
The constrained phase space at the Tevatron for dijets
with large rapidity separation puts severe limits on
the phase space for mini-jets (contributing to the BFKL
evolution). The phase space constraint prohibits the
rise in cross section with increasing rapidity
separation (simply because the decrease in the PDFs is
faster than the increase in the partonic cross
section), but other observables, like the angular
correlation of dijets, still get large BFKL
corrections. The LHC promises to be very well suited
for a study of effects from the BFKL evolution.
\subsection{Experimental Conclusions and Outlook}
The measurements of forward jets and particles are sensitive to the 
dynamics of parton evolution. Several QCD-based approaches 
have been confronted
with the data. It has been shown that NLO-dijet DGLAP calculations fall well 
below forward jet data. The forward jet cross section is, however, 
 well described 
by a DGLAP based
Monte Carlo which includes a resolved photon component. Similarly, results 
obtained using the BFKL and CCFM evolution schemes are compatible with the 
data. The measurement
of the forward $\pi^0$ cross section leads essentially to the same conclusions 
for the region of Bjorken-$x$ covered by the forward jet measurements. 
For the lowest
values of $x$ i.e. those beyond the reach of the forward jet measurements, 
the \cascade\ Monte Carlo generator  
fails to describe the data, possibly due to missing quark initiated cascades. 
The comparison of results 
from various models
seems to indicate some sensitivity to the fragmentation method used to connect 
the parton and hadron levels. Study of the transverse energy flow associated
with forward $\pi^0$'s seems to favor the DGLAP direct + resolved approach. 
\par
However, 
the present measurements at HERA were mainly restricted by two factors: 
the available center-of-mass energy
and the geometrical acceptance of the detectors, requiring the forward jet to 
lie between: $ 2 \lap \eta \lap 3$. 
The dijet measurements are described best with a 
different scale (\rapgap ) or unintegrated gluon density  (\cascade ) than the
forward jet measurements (cf. see Tab.~\ref{chisquared}). This shows that indeed
new effects are seen: 
if the forward jet cross section is extended to a range of
$\eta_{jet}$ up to 6 units
(as proposed in the proposals for a continued HERA3
program~\cite{hera3-allen,hera3-h1})  the difference compared to DGLAP
becomes even more significant.

\section{Conclusions}
\label{sec:conclusions}
On the theoretical side, significant progress in understanding \sx\ effects has
been made. The soft region ($\kt \lap 1$ GeV) has been clearly identified to
have significant influence on hadronic final state observables. With the
consistent treatment of the scale in $\alpha_s$ and including the
non-singular terms into the CCFM splitting function, a necessary step forward
to  a serious application of the CCFM \sx\ evolution equation has been
taken.
\par
The question of gauge invariance of the whole \kt - factorization approach in
general and also the question of gauge invariance of (integrated or
unintegrated) PDFs has been clarified further.
\par
Many new measurements in the area of \sx\ physics have been made public, and
the interest in a better understanding of \sx\ effects is very clear. New
measurements indicate the need to go even beyond ${\cal O}(\alpha_s^3)$ if
calculations are performed in the collinear factorization approach. On the
other hand, these effects are automatically included in \kt - factorization,
which makes \kt-factorization  an important tool for studying higher order
corrections.
It was shown that, irrespective of the particular choice of
the non-perturbative CO matrix elements, the production
of $J/\psi$ mesons at HERA can be reasonably well described within the
color-singlet production mechanism (within \kt-factorization)
and color-octet contributions are not at all needed. More accurate
measurements of polarization properties of the $J/\psi$ mesons will
be an interesting test of the \kt-factorization predictions.
\par
Still, results from the experiments at HERA and the Tevatron have not yet
provided unambiguous evidence for  new \sx\ effects.
Including higher orders in the calculation according to the collinear
approach and/or including the concept resolved (virtual) photons seems to mimic
also new \sx\ effects. In order to unambiguously identify \sx\ effects at 
e.g. HERA, it is necessary to increase the angular coverage of the
experimental setup towards the proton direction, as has been shown in the
proposal for an extended HERA running beyond 2006, 
the so-called HERA~3 scenario. 
Since a correct description of the \sx\ dynamics is essential for the
understanding of QCD at high energies, and also for any asymptotically
 free field theory
it is of great importance to continue and extend the experimental and
theoretical efforts.
\section*{Acknowledgments}
We are grateful to M.~Ciafaloni for his permission to include 
section~\ref{sec:gauge} in this paper, which resulted from a
discussion with J.~Collins and Y.~Dokshitzer.
We are grateful to the DESY directorate
to the  Royal Swedish Physiographic Society and the 
Royal Swedish Academy of Science
 for financial support.
\raggedright

\end{document}